\documentclass[12pt]{article}
\usepackage{latexsym}
\usepackage{amssymb}
\catcode`\@=11
\newcount\hour
\newcount\minute
\newtoks\amorpm \hour=\time\divide\hour by 60\minute
=\time{\multiply\hour by 60 \global\advance\minute by-\hour}
\edef\standardtime{{\ifnum\hour<12 \global\amorpm={am}%
        \else\global\amorpm={pm}\advance\hour by-12 \fi
        \ifnum\hour=0 \hour=12 \fi
        \number\hour:\ifnum\minute<10
        0\fi\number\minute\the\amorpm}}
\edef\militarytime{\number\hour:\ifnum\minute<10
0\fi\number\minute}
\def\draftlabel#1{{\@bsphack\if@filesw {\let\thepage\relax
   \xdef\@gtempa{\write\@auxout{\string
      \newlabel{#1}{{\@currentlabel}{\thepage}}}}}\@gtempa
   \if@nobreak \ifvmode\nobreak\fi\fi\fi\@esphack}
        \gdef\@eqnlabel{#1}}
\def\@eqnlabel{}
\def\@vacuum{}
\def\marginnote#1{}
\def\draftmarginnote#1{\marginpar{\raggedright\scriptsize\tt#1}}
\def\draft{
        \pagestyle{plain}
        \overfullrule=2pt
        \oddsidemargin -.1truein
        \def\@oddhead{\sl \phantom{\today\quad\militarytime} \hfil
        \smash{\Large\sl DRAFT} \hfil \today\quad\militarytime}
        \let\@evenhead\@oddhead
        \let\label=\draftlabel
        \let\marginnote=\draftmarginnote
        \def\ps@empty{\let\@mkboth\@gobbletwo
        \def\@oddfoot{\hfil \smash{\Large\sl DRAFT} \hfil}
        \let\@evenfoot\@oddhead}
        \def\@eqnnum{(\theequation)\rlap{\kern\marginparsep\tt\@eqnlabel}%
        \global\let\@eqnlabel\@vacuum}  }

\renewcommand{\theequation}{\thesection.\arabic{equation}}
\renewcommand{\thefootnote}{\fnsymbol{footnote}}
\newcommand{\newsection}{    
\setcounter{equation}{0}\section}
\def\appendix#1{\addtocounter{section}{1}\setcounter{equation}{0}
\renewcommand{\thesection}{\Alph{section}}
\section*{Appendix \thesection\protect\indent \parbox[t]{11.15cm}{#1}}
\addcontentsline{toc}{section}{Appendix \thesection\ \ \ #1}}

\def \lc {{light-cone}}

\def \bi{\bibitem}
\def \la {\label}
\def \ov {\over}

\def \b {\beta}
\def \Om {\Omega}

\def \t {\theta}

\def \d {\partial}
\def \del{\partial}

\def\be{\begin{equation}}
\def\ee{\end{equation}}

\def\cH {{\cal H}}
\def\mfh {\mathfrak{h}}
\def\mfsl {\mathfrak{sl}(2, \bR)}
\def\mfuo {\mathfrak{u}(1)}
\def\mfu {\mathfrak{u}}
\def\mfsu {\mathfrak{su}}

\catcode`\@=11

\def\bea{\begin{eqnarray}}
\def\eea{\end{eqnarray}}
\def\beann{\begin{eqnarray*}}
\def\eeann{\end{eqnarray*}}
\def\beq{\begin{equation}}
\def\eeq{\end{equation}}
\def\ba{\begin{array}}
\def\ea{\end{array}}
\def\ben{\begin{enumerate}}
\def\een{\end{enumerate}}
 \def \l {\lambda}

 \def \la {\label}
 \def\be{\begin{equation}}
\def\ee{\end{equation}}

\def \coplus {\oplus_s}

\def \la {\label}

\font\mybb=msbm10 at 11pt
\font\mybbb=msbm10 at 17pt
\def\bb#1{\hbox{\mybb#1}}
\def\bbb#1{\hbox{\mybbb#1}}

\def\bR {\bb{R}}

\def\bC {\bb{C}}

\def\e  {\epsilon}

\def \k {\kappa}
\def \ov {\over}
\def \ha { { 1\ov 2}}

 \def\ep {\epsilon}

\def \del { \partial}
\def \t {\theta}

\def \ee {\epsilon}

\def \g {\gamma}
\def \bi{\bibitem}
\def\a{\alpha }

\def \ep {\epsilon}

\def \d {\delta}

\def \l {\lambda}

\def \g {\gamma}

\def \b {\beta}

\def\lc{\lrcorner}
\def\re{{\rm e}}

\def\be{\begin{equation}}
\def\ee{\end{equation}}

\def \bi {\bibitem}
\def \la{\label}

\def \t {\tau}

\begin{document}
\date{November 2002}
\begin{titlepage}
\begin{center}
\hfill hep-th/0510176\\
\hfill KUL-TF-05/23\\

\vspace{1.5cm}
{\Large \bf  The spinorial geometry of supersymmetric heterotic string backgrounds}\\[.2cm]

\vspace{1.5cm}
 {\large U.~Gran$^1$,  P.~Lohrmann$^2$ and  G.~Papadopoulos$^2$ }

 \vspace{0.5cm}
${}^1$ Institute for Theoretical Physics, K.U. Leuven\\
Celestijnenlaan 200D\\
B-3001 Leuven, Belgium\\
\vspace{0.5cm}

${}^2$ Department of Mathematics\\
King's College London\\
Strand\\
London WC2R 2LS
\end{center}

\vskip 1.5 cm
\begin{abstract}

We determine the geometry of supersymmetric heterotic string
backgrounds for which all parallel spinors with respect to the
connection $\hat\nabla$ with torsion $H$, the NS$\otimes$NS
three-form field strength, are Killing. We find that there are two
classes of such backgrounds, the null and the timelike. The Killing
spinors of the null backgrounds have stability subgroups
$K\ltimes\bR^8$ in $Spin(9,1)$, for $K=Spin(7)$, $SU(4)$, $Sp(2)$,
$SU(2)\times SU(2)$ and $\{1\}$, and the Killing spinors of the
timelike backgrounds have stability subgroups
 $G_2$, $SU(3)$, $SU(2)$ and $\{1\}$.
The former admit a single null $\hat\nabla$-parallel  vector field while the latter admit
a timelike and two, three, five and nine spacelike $\hat\nabla$-parallel vector fields, respectively.
The spacetime of the null backgrounds is a Lorentzian two-parameter family of Riemannian manifolds $B$
with skew-symmetric torsion. If the rotation of the null vector field vanishes, the holonomy
of the connection with torsion of $B$ is contained in $K$.
The spacetime of time-like backgrounds
is a principal bundle $P$ with fibre a Lorentzian Lie group and base space a suitable Riemannian manifold
with skew-symmetric torsion. The principal bundle is equipped with a connection $\lambda$ which determines
the non-horizontal part of the spacetime
metric and of $H$. The curvature of $\lambda$ takes values in an appropriate Lie algebra constructed from that
of $K$.
In addition $dH$ has only horizontal components and contains
 the Pontrjagin class
of $P$.
We have computed in all cases the Killing spinor bilinears, expressed the fluxes in terms of the geometry
and determine
the field equations that are implied by the Killing spinor equations.

\end{abstract}
\end{titlepage}
\newpage
\setcounter{page}{1}
\renewcommand{\thefootnote}{\arabic{footnote}}
\setcounter{footnote}{0}

\setcounter{section}{0}
\setcounter{subsection}{0}
\newsection{Introduction}

It has been known for some time that the geometry of supersymmetric
heterotic string backgrounds resembles that of Riemannian manifolds
that appear in the Berger classification list and admit parallel
spinors. This is because the gravitino Killing spinor equation  is a
parallel transport equation for a metric connection
 $\hat\nabla$ with torsion given by the NS$\otimes$NS three-form field strength $H$.
Therefore the solutions of the gravitino Killing spinor equation are
characterized by the holonomy of $\hat\nabla$, ${\rm
hol}(\hat\nabla)$. This holonomy group  is contained in the
stability subgroup $G$ of the parallel spinors in a suitable spin
group. Berger classified the irreducible Riemannian manifolds using
the holonomy of the Levi-Civita connection. Similarly,
 the holonomy of the Levi-Civita connection $\nabla$ of these Riemannian manifolds which in addition  admit parallel
 spinors
 is again contained in the stability subgroup of the spinors.
Because of this, it has been expected
that there must be a  relation between the holonomies of $\hat\nabla$ that appear
in supersymmetric heterotic string backgrounds and those of the Levi-Civita connection $\nabla$
of Berger irreducible Riemannian manifolds that admit parallel spinors as both are contained in the stability
subgroups of the parallel spinors.
It turns out that there is such a  relation but there are also differences because the spacetime of supersymmetric
heterotic backgrounds is a Lorentzian and not a Riemannian manifold.  So the stability subgroups
of the parallel spinors are in  $Spin(n-1,1)$ instead of  $Spin(n)$ which
is suitable for Riemannian manifolds.
In addition,  the heterotic string supergravity  has two more
 Killing spinor equations associated with the dilatino and gaugino
supersymmetry transformations.

The geometry of manifolds that admit a metric connection with
skew-symmetric torsion has been extensively investigated in the
literature. Such geometries appear in the context of supersymmetric
one- and two-dimensional sigma models, see e.g.~\cite{hull, howe,
howegp, coles}. They have also been explored as supersymmetric
solutions of the common sector of type II theories and heterotic
supergravity, and
 their properties have been examined using the Killing spinor bilinear forms, see e.g.~\cite{strominger, hullb, tseytlin}. Deformations of these geometries
due to higher curvature corrections of the heterotic string have
been investigated in e.g.~\cite{strominger, callan, howegpb,
tsimpis, lust}. It has been recognized some time ago that  these
geometries with torsion are closely related to the
 standard geometries, like K\"ahler, Calabi-Yau and hyper-K\"ahler, see e.g.~\cite{howegpc, poon, ivanovgp, ivanovgpb, salamon,
 goldstein, poonb}, and they have found applications
 in the geometry of black-hole moduli spaces \cite{gibbons, michelson, gutowski}.
 More recently, these geometries with torsion have been studied
 using the Gray-Hervella classification techniques \cite{grayhervella}, see e.g.~\cite{stefang2, stefanspin7, gauntlett, lustb, misra}.
 So far in the
 applications of these geometries
 in the context of ten-dimensional supergravity, it has been assumed that the spacetime  is a product, $\bR^{9-n, 1}\times X_n$,
  and the non-trivial part of the geometry is that of the Riemannian manifold $X_n$. We shall not make
  such an assumption and we shall find that the spacetime geometry of supersymmetric heterotic
   backgrounds is not always a product.

In this paper, we shall use the method developed in \cite{grangp} to
systematically investigate all possible geometries of supersymmetric
heterotic string backgrounds. The parallel transport equation ,
$\hat\nabla\epsilon=0$, implies that \be \hat R\epsilon=0~, \ee
where $\hat R$ is the curvature of $\hat\nabla$ and takes values in
$\mathfrak{spin}(9,1)$. If the Killing spinors $\epsilon$ have a
non-trivial stability subgroup $G$ in $Spin(9,1)$, $G\subset
Spin(9,1)$,  then the holonomy of $\hat\nabla$ must be a subgroup of
$G$, ${\rm hol}(\hat\nabla)\subseteq G$. The Killing spinors are the
singlets of $G$ in the decomposition of the Majorana-Weyl
representation $\Delta_{\bf 16}^+$ of $Spin(9,1)$ under $G$. On the
other hand if the stability subgroup is $\{1\}$, then the holonomy
of $\hat\nabla$ is the identity and \be \hat R=0~. \ee Therefore
either the Killing spinors are singlets of a proper  subgroup of
$G\subset Spin(9,1)$ or $\hat R=0$. In the former case, we shall
give all the spinors which are singlets of a subgroup of
$Spin(9,1)$. Some of these are  related to the parallel spinors that
exist on  the manifolds that appear in the Berger classification
list and have been presented in \cite{wang}. However in our case the
stability subgroups are somewhat different because the spacetime is
a Lorentzian manifold. In the latter case,  the spacetime is
parallelizable with respect to the $\hat\nabla$ connection. Using
this,  we shall show that the spacetime is a Lorentzian metric Lie
group and that $\hat\nabla$ is a parallelizable connection.

The investigation of the gaugino Killing spinor equations
$F\epsilon=0$ is similar to that of the curvature condition $\hat
R\epsilon=0$. This is because the Clifford element $F$ lies in the
$\mathfrak{spin}(9,1)$ subspace of the Clifford algebra. If the
spinors $\epsilon$ that satisfy $F\epsilon=0$ have a non-trivial
stability subgroup $G$ in $Spin(9,1)$, then the curvature $F$ takes
values in  the Lie algebra ${\rm g}\subseteq spin(9,1)$ of $G$. If
the stability subgroup is $\{1\}$, then $F=0$ and the gauge
connection is flat. In addition, the expression $F\epsilon$ for any
spinor $\epsilon$ can be read off from that for the gravitino
Killing spinor equation, in particular from the part that contains
the spin connection. Because of this, we shall not explore further
the supersymmetry conditions that arise from the gaugino Killing
spinor equation.

The dilatino Killing spinor equation is somewhat  different from the
gravitino and gaugino Killing spinor equations. In particular, there
is no understanding of the solutions of the dilatino Killing spinor
equation in terms of Lie subalgebras of $\mathfrak{spin}(9,1)$
similar to the one presented above for the gravitino and gaugino
Killing spinor equations.  However it can be analyzed using
representation theory. It is also known that there are
backgrounds\footnote{ Examples of such manifolds are Lorentzian
metric groups for which all spinors are parallel with respect to the
left-invariant connection but such backgrounds typically preserve
1/2 of the supersymmetry because  half of these spinors do not solve
the dilatino Killing spinor equation. } with  spinors which solve
the gravitino  but not the dilatino Killing spinor equation. {\it
Because of this, we shall restrict our attention to those
backgrounds for which all the solutions of the gravitino Killing
spinor equation are also solutions of the dilatino one}, i.e.~all
$\hat\nabla$-parallel spinors are Killing.
 In the terminology of \cite{grangpb}, these are the maximally supersymmetric
$G$-backgrounds, where $G$ is the stability subgroup of the Killing spinors.

It is convenient to characterize
the  supersymmetric heterotic string backgrounds in
terms of the number of supersymmetries
they admit, which we denote with $N$, and the stability subgroup of the Killing spinors
$G$ in $Spin(9,1)$, \cite{jose, bryant}.
We shall show that the stability subgroups $G$ of the Killing spinors
 are either compact groups $K$, $G=K$,
for $K= G_2 (N=2), SU(3) (N=4), SU(2) (N=8), \{1\} (N=16)$ or
$G=K\ltimes \bR^8$, for $K=Spin(7) (N=1), SU(4) (N=2), Sp(2) (N=3), SU(2)\times
SU(2) (N=4), \{1\} (N=8)$, where $N$ denotes the number of
supersymmetries. In the former case the stability subgroups $G$ are
those expected from the Berger classification list. The latter case
has no Riemannian analogue  and is due to the Lorentzian signature
of spacetime but the subgroups $K$  appear in the Berger
classification list. The Killing spinors  are chiral with respect to
a suitable chirality projector of a Clifford algebra ${\rm
Cliff}(\bR^8)\subset {\rm Cliff}(\bR^{9,1})$.

We shall show that the supersymmetric backgrounds for which the
Killing spinors have a compact stability subgroup admit a time-like
and at least two space-like parallel vector spinor
bilinears\footnote{Since the Killing spinors are parallel with
respect to $\hat\nabla$, all the Killing spinor form bilinears
$\alpha$ are also parallel with respect to $\hat\nabla$,
$\hat\nabla\alpha=0$.} with respect to $\hat\nabla$. Because of
this, we shall refer to them as time-like backgrounds. The
commutator of the parallel vector fields does not necessarily vanish
and the structure constants depend on the NS$\otimes$NS three-form
field strength $H$. If one imposes the condition that the algebra
$\mfh$ spanned by parallel vectors constructed from the spinor
bilinears closes under Lie brackets, then the spacetime $M$ for
$K\not=\{1\}$ is (locally) a principal bundle $M=P(\cH, B, \pi)$
equipped with the connection $\lambda$, where $\cH$ is a Lie group
with Lie algebra $\mfh$ and base space $B$ which is the space of
orbits of the parallel vector fields. {}The backgrounds with
$K=\{1\}$ are
  maximally supersymmetric and it has been shown in \cite{jfgp}
that the spacetime is locally isometric to $\bR^{9,1}$, $H=0$ and $\Phi={\rm const}$.
The spacetime metric and torsion can be written
as
\bea
&&ds^2=\eta_{ab}\, \lambda^a \lambda^b+ g^h~,
\cr
&&H={1\over3} \eta_{ab} \,\lambda^a\wedge d\lambda^b+{2\over3} \eta_{ab}\, \lambda^a\wedge {\cal F}^b+ H^h~,
\eea
where $g^h$ and $H^h$ are the horizontal components of the metric and $H$,  ${\cal F}$ is the curvature
of the connection $\lambda$ and $\eta$ is a Lorentzian invariant metric on $\cH$.
The dilaton $\Phi$ depends only on the coordinates of $B$.
In addition
\bea
dH=\eta_{ab}\,{\cal F}^a\wedge {\cal F}^b+d H^h~.
\eea
Therefore $dH$ contains a representative of the first Pontrjagin
class of $P$. The Killing spinor equations impose restrictions on
$\cH$, ${\cal F}$ and the geometry of the base space $B$. The
gravitino Killing spinor equation implies that the spacetime admits
a $K$-structure compatible with a metric connection with
skew-symmetric torsion, ${\rm hol}(\hat\nabla)\subseteq K$. There
are three kinds of conditions that arise from the dilatino Killing
spinor equation. One set of conditions imposes restrictions on the
Lie group $\cH$, another set of conditions suitably  restricts the
curvature ${\cal F}$ of the connection $\lambda$, and the third set
of conditions implies restrictions on the geometry of $B$.

In particular, for $K=G_2$,  $\cH$ has Lie algebra either $\mathfrak{sl}(2, \bR)$ or $\mfu(1)\oplus \mfu(1)\oplus\mfu(1)$;
for $K=SU(3)$, $\cH$ is a four-dimensional Lorentzian Lie group but otherwise unrestricted; for $K=SU(2)$,
$\cH$ is a six-dimensional Lorentzian metric Lie group but the dilatino Killing spinor equation imposes restrictions
of its structure constants which we determine.

The second set of conditions of the dilatino Killing spinor equation
implies that the connection $\lambda$ is a $\mathfrak{k}$ instanton,
i.e.~${\cal F}$ takes values in the Lie algebra $\mathfrak{k}$ of
$K$.  This is the case for all $K$ apart from $K=SU(3)$ and $\cH$
non-abelian, where ${\cal F}$ satisfies the Donaldson conditions and
takes values in $\mfsu(3)\oplus \mfuo$.

The base space $B$ has dimension, ${\rm dim}\, B=7, 6, 4$ for $K= G_2, SU(3), SU(2)$, respectively.  In addition $B$
admits a {\it conformally balanced and integrable $K$-structure, and a compatible
metric connection $\hat{\tilde \nabla}$ with skew-symmetric torsion $\tilde H$},
 $H^h=\pi^*\tilde H$,
i.e.~${\rm hol}\, (\hat{\tilde \nabla})\subseteq K$. This is the
case for all $K$ apart from   $K=SU(3)$ and $\cH$ non-abelian,
 where $B$
admits an $SU_c(3)=SU(3)\times_Z R$-structure, where $R=U(1)$ or
$R=\bR$ and $Z$ is a discrete group. The additional $R$ twist is due
to a one-dimensional representation $\rho$ of $\cH$ and the
associated line bundle $L=P\times_\rho \bC$. The conformally
balanced structure is due to the fact that a Lee form of the
$K$-structure of $B$ is related to the exterior derivative of the
dilaton as consequence of the conditions that arise from the
dilatino Killing spinor equation. An integrability of the
$K$-structure is also implied by the dilatino Killing spinor
equations. This is suitably defined for all $K$. For example,  if
$K=SU(3)$, then the associated almost complex structure is
integrable and $B$ is a complex manifold.  Furthermore, if $K=G_2$
and $\cH$ is abelian, then dilatino Killing spinor equation also
requires that $d\tilde\varphi$ is orthogonal to
$\star\tilde\varphi$, where $\tilde\varphi$ is a $G_2$ invariant
form on $B$. In the non-abelian case, this is not the case and the
inner product $(d\tilde\varphi, \star\tilde\varphi)$ is related to
the structure constants of $\cH$. In all the above cases, the
NS$\otimes$ NS three-form $H$ is determined by the form Killing
spinor bilinears and the metric of the spacetime. In addition, the
integrability conditions of the Killing spinor equations imply all
the field equations provided that the Bianchi identities of $H$ and
$F$ are satisfied.

Similarly, we shall show that the backgrounds for which the Killing
spinors have stability subgroup $K\ltimes \bR^8$, for $K=Spin(7)
(N=1), SU(4) (N=2), Sp(2) (N=3), SU(2)\times SU(2) (N=4), \{1\} (N=8)$, admit a
single null parallel one-form spinor bilinear $\kappa$ with respect
to $\hat\nabla$. Because of this, we shall refer to them  as null
backgrounds. If one adapts coordinates with respect to the null
Killing vector field $X$ associated to $\kappa$,
$X=\partial/\partial u$, then the metric and torsion can be written
as
\bea
&&ds^2=2 \re^- \re^+ + \delta_{ij} e_I^i e_J^j dy^I dy^J
\cr
&&H=\re^+\wedge d \re^-+({1\over 2} H_{-ij}^{\mathfrak{k}}+ \Omega_{-,ij}^{\mathfrak {k}^\perp})
 \re^-\wedge e^i\wedge e^j+ {1\over 3!} H_{ijk} e^i\wedge e^j \wedge
 e^k~,
\eea
where $\kappa=\re^-=(dv+m_I dy^I)$ and $\re^+=du+V dv+n_I dy^I$,
and $H_{-ij}^{\mathfrak{k}}$ denotes the component of $H_{-ij}$ in the subalgebra $\mathfrak{k}$ of $K$
and $\Omega_{-,ij}^{\mathfrak {k}^\perp}$
denotes the component of $\Omega_{-,ij}$ in the complement of $\mathfrak{k}$ in
$\Lambda^2(\bR^8)=\mathfrak{k}\oplus\mathfrak {k}^\perp$.
The Killing spinor equations determine
all components of the NS$\otimes$NS flux $H$ in terms of the form Killing spinor bilinears and the spacetime metric apart from the
component $H_{-ij}^{\mathfrak{k}}$. In addition, they imply that $d\kappa=d\re^-$  takes values in $\mathfrak{k}\coplus \bR^8$, where
$\coplus$ denotes semi-direct sum of Lie algebras.
A consequence of this is that the null parallel vector field leaves
invariant the $K\ltimes\bR^8$-structure
of spacetime.

In all null cases, the spacetime admits a codimension eight
integrable foliation with leaves a manifold $B$. For generic
backgrounds, $B$ admits a $K$-structure which is not compatible with
the induced metric connection $\hat{\tilde\nabla}$ with torsion.
However, if $d\kappa=0$, then $B$ is a  conformally balanced
integrable manifold with a $K$-structure and compatible connection
$\hat{\tilde\nabla}$ with torsion, i.e.~${\rm hol}(
\hat{\tilde\nabla})\subseteq K$. The conformally balanced and
integrability properties are consequences of the dilatino Killing
spinor equations and are defined in a way similar  to those of the
space $B$ in the timelike backgrounds we have mentioned above. We
have also shown that the Killing spinor equations imply all field
equations apart from the $E_{--}$ component of the Einstein
equations, the $LH_{-A}$ components of the two-form gauge potential
and the $LF_-$ component of the field equations of the gauge
connection provided that all the Bianchi identities are satisfied.

We also apply our results to investigate some properties of the Killing spinor equations
of the common sector of type II supergravities. We find that the IIA and IIB common sectors
should be treated separately because despite many similarities there are also differences. We mostly
focus on the   IIB common sector and investigate the supersymmetry conditions of backgrounds
with two supersymmetries. We show that there are five distinct cases to examine described by the
stability subgroups of the Killing spinors.

This paper is organized as follows: In section two, we state the field and
Killing spinor equations of heterotic
supergravity and describe the integrability conditions of the latter. In section
three, we find the stability subgroups $G$
of spinors in $Spin(9,1)$ and give the $G$-invariant spinors (singlets) in the Majorana-Weyl
representation $\Delta^+_{\bf 16}$ of $Spin(9,1)$. In section four, we describe the parallel spinors and forms
of supersymmetric backgrounds. We argue that there is always a basis up to a local Lorentz transformation
such that the parallel spinors are constant.
 In section five, we determine the geometry of $N=1$ backgrounds. In section six,
we give the geometry of $N=2$ $SU(4)\ltimes\bR^8$-backgrounds. In section seven, we describe the geometry
of $N=2$ $G_2$-backgrounds. In section eight, we investigate the geometry of $N=3$ backgrounds. In section nine,
we determine the geometry of $N=4$ $SU(3)$- and $(SU(2)\times SU(2))\ltimes \bR^8$-backgrounds. In section ten,
we describe the geometry of $N=8$ $SU(2)$- and $\bR^8$-backgrounds.
In section eleven, we show that $\hat\nabla$-parallelizable
backgrounds are Lorentzian metric Lie groups. In section twelve, we apply our results to examine
the supersymmetric solutions of the common sector of type II supergravities. In section thirteen, we give our conclusions.
In appendix A, we describe the spinors in terms of forms and compute the form spinor bilinears for all
singlets of a subgroup $G\subset Spin(9,1)$ in $\Delta_{\bf 16}^+$. In appendix B, we present the
 linear systems
associated with the Killing spinor equations of the heterotic supergravity.

\newsection{Fields and spinors}

\subsection{Field and Killing spinor equations}

The bosonic fields of heterotic supergravity   are the metric $g$,
the NS$\otimes$NS three-form field strength $H$, the dilaton scalar
$\Phi$,  and the gauge connection $A$ with curvature $F$. The field
and Killing spinor equations of the heterotic string receive string
$\alpha'$ corrections which
 can be computed
either from a sigma model beta function or from string amplitude
calculations. The field equations in the string frame to lowest
order in $\alpha'$  are
\bea
E_{MN}=R_{MN}-{1\over4} H_{PQM} H^{PQ}{}_N+2\nabla_M\partial_N\Phi&=&0~,
\cr
LH_{PQ}=\nabla_M (e^{-2\Phi} H^M{}_{PQ})&=&0~,
\cr
L\Phi=\nabla^2\Phi-2g^{MN} \partial_M\Phi \partial_N \Phi+{1\over12} H_{MNR} H^{MNR}&=&0~,
\cr
LF_N=\hat\nabla^M(e^{-2\Phi} F_{MN})&=&0~,
\eea
where $\nabla$ is the Levi-Civita connection of the metric $g$.
The field equation for the dilaton is implied from those of the
 metric and two-form gauge potential $B$
associated with $H$, $H=dB$, up to a constant. The Killing spinor
equations are
\bea
\hat\nabla\epsilon&=&0~,
\cr
(\Gamma^M\partial_M\Phi-{1\over12}\Gamma^{MNP} H_{MNP})\epsilon&=&0~,
\cr
F_{MN}\Gamma^{MN}\epsilon&=&0~,
\eea
where $\hat\nabla=\nabla+{1\over2} H$, $\nabla_M\epsilon=\partial_M\epsilon+{1\over4} \Omega_{M,AB} \Gamma^{AB}\epsilon$,
\be
\hat\nabla_N Y^M=\nabla_N Y^M+{1\over2} H^M{}_{NR} Y^R~,
\ee
and $\epsilon$ is a Majorana-Weyl spinor of positive chirality,
i.e.~$\epsilon$ is described by forms of even degree. In what
follows, we shall denote the spin connection of the $\hat\nabla$
 covariant derivative
with $\hat\Omega$.

\subsection{Integrability conditions}

It is well-known that some of the field equations of  supersymmetric backgrounds
can be implied by the Killing spinor equations. To find which field equations
are implied, one has to investigate the integrability conditions of the Killing spinor
equations.
In the case of
heterotic supergravity, these integrability conditions are, see also \cite{dewit},
\bea
[\hat\nabla_M, \hat\nabla_N]\epsilon&=&{1\over4} \hat R_{MN, AB} \Gamma^{AB}\epsilon=0~,
\cr
[\hat\nabla_M,F_{RS} \Gamma^{RS}]\epsilon&=&0~,
\cr
[\hat\nabla_M, \partial_N\Phi\Gamma^N-{1\over12} H_{NPQ} \Gamma^{NPQ}]\epsilon&=&0~,
\cr
[F_{RS} \Gamma^{RS},\partial_N\Phi\Gamma^N-{1\over12} H_{NPQ} \Gamma^{NPQ}]\epsilon&=&0~.
\eea
 Multiplying the first expression above with $\Gamma^N$, using appropriately the remaining integrability
 conditions and the identity
 \bea
 g^{MN} \partial_M\Phi \partial_N\Phi\epsilon&-&{1\over24} H_{MNR} H^{MNR}\epsilon-
 {1\over2}\partial_M\Phi H^M{}_{ST} \Gamma^{ST}\epsilon
 \cr
 &+&{1\over16} H^S{}_{MN} H_{SPQ} \Gamma^{MNPQ}\epsilon=0~,~~
\eea
one finds that
\bea
\hat R_{MA,BC} \Gamma^A \Gamma^{BC} \epsilon=-
2 E_{MN} \Gamma^N \epsilon- e^{2\Phi} LH_{MN} \Gamma^N \epsilon-{1\over6} BH_{MABC} \Gamma^{ABC}\epsilon&=&0~,
\cr
L\Phi\epsilon -{1\over4} e^{2\Phi} LH_{MN}\Gamma^{MN}\epsilon-{1\over48} BH_{MNPQ}\Gamma^{MNPQ}\epsilon&=&0~,
\cr
{1\over3} BF_{MNP}\Gamma^{MNP}\epsilon+ 2 e^{2\Phi} LF_N\Gamma^N\epsilon&=&0~,
\la{intcon}
\eea
where $BH_{MNPQ}=4 (dH)_{MNPQ}$ and $BF_{MNR}=3\nabla_{[M} F_{NR]}$.
To the order of $\alpha'$ that we have stated the field equations
above, the Bianchi identity of $H$ implies that $BH=dH=0$. However,
if the heterotic string anomaly is included and so schematically,
$BH\sim\alpha' ({\rm tr} R^2-{\rm tr} F^2)+{\cal O}(\alpha'^2)$,
then,  for consistency, one has to include the two-loop correction
to the field equations \cite{howegpb, tsimpis}.

A remarkable property of (\ref{intcon}) is that if one imposes the Bianchi identities of $H$ and $F$, $BH=0$ and $BF=0$,
respectively, then
the remaining equations are up to quadratic order in gamma matrices. As a result,
it is straightforward to construct the linear systems associated with the
integrability conditions from that of the Killing spinor equations. These linear systems are similar to those
investigated in the context of
 M-theory and IIB supergravity in \cite{grangpb}.

\newsection{Stability subgroup of spinors in $Spin(9,1)$}\la{stabgroups}

As we have mentioned in the introduction, the Killing spinors of
supersymmetric backgrounds with  $\hat R\not=0$ are singlets of the
holonomy group ${\rm hol}(\hat\nabla)$ of $\hat\nabla$. In addition, the holonomy group in every case
is a subgroup of the stability subgroup of the Killing spinors in $Spin(9,1)$.
Therefore, we have to determine all the spinor singlets of the subgroups\footnote{We only consider connected
subgroups of $Spin(9,1)$ as stability subgroups for spinors and
our computations are restricted on the Lie algebra level. However spinors can admit
disconnected stability subgroups and these
are applicable to non-simply connected  manifolds
\cite{mcinnes, amus}.}
 of
$Spin(9,1)$. This analysis closely resembles that of determining the parallel spinors
of manifolds with special holonomy which has been presented in
\cite{wang}. However, there are some differences that arise because
the spacetime is a Lorentzian manifold.

\subsection{One spinor}

There is one type of orbit of  $Spin(9,1)$ with stability subgroup
$Spin(7)\ltimes \bR^8$ in  $\Delta_{16}^+$. The proof of this has
been given in \cite{grangutowskigp} but we shall repeat the steps
here because they are useful for determining the stability subgroups
of more than one spinor. Consider the spinor
\be
1+e_{1234}~.
\ee
The stability subgroup of this spinor in $Spin(9,1)$ is $Spin(7)\ltimes \bR^8$
as it can be seen by solving the infinitesimal invariance equation
\be
\lambda_{AB} \Gamma^{AB} (1+ e_{1234})=0~,
\ee
 where $\lambda$ parameterizes  the   spinor transformations.
This computation is most easily done in the pseudo-Hermitian basis
that we have given in (\ref{hbasis}).
It is easy to see that the above condition implies
that the parameters are restricted as
\bea
\lambda_{\bar\a\bar\b}={1\over2} \epsilon_{\bar\a\bar\b}{}^{\g\d}  \lambda_{\g\d}~,~~~~~
\lambda_{\a\bar\b} g^{\a\bar\b}=\lambda_{-+}=\lambda_{+\a}=\lambda_{+\bar\a}=0~,
\eea
where $\epsilon_{\bar1\bar2\bar3\bar4}=1$. Observe that the
parameters $\lambda_{-\a}$ and $\lambda_{-\bar\a}$ are complex
conjugate to each other but otherwise unconstrained. The group that
leaves invariant $1+e_{1234}$ has Lie algebra
$\mathfrak{spin}(7)\coplus \bR^8$ and so find that the stability
subgroup is $Spin(7)\ltimes \bR^8$.

Having established this, we decompose $\Delta_{16}^+$ under the stability
subgroup $Spin(7)$ as
\be
\Delta_{16}^+= \bR<1+e_{1234}>+\Lambda^1(\bR^7)+\Delta_8~,
\la{dec}
\ee
where the singlet $\bR$ is generated by $1+e_{1234}$, $\Lambda^1(\bR^7)$
is the vector representation of $Spin(7)$ which is spanned
by the spinors associated with two-forms in the directions $e_1, \dots, e_4$ and
$i(1-e_{1234})$, and $\Delta_8$ is the spin representation of $Spin(7)$
which is spanned by the rest of spinors which are of the type
 $\Gamma^+\eta$, $\eta$ is a spinor  generated by the odd forms in
 the directions $e_1, \dots, e_4$.
Therefore the most general spinor in $\Delta_{16}^+$ can be written as
\be
\eta= a(1+e_{1234})+ \theta_1+\theta_2~,
\ee
where $\theta_1\in \Lambda^1(\bR^7)$ and $\theta_2\in \Delta_8$.
First we assume that $a\not=0$. In this case, there are
 two cases to consider depending on whether $\theta_2$ vanishes or not.
If $\theta_2=0$, since $Spin(7)$ acts with the vector representation
on $\Lambda^1(\bR^7)$, it is always possible to choose $\theta_1=i b(1-e_{1234})$.
The most general spinor in this case then is
\be
\eta= a (1+e_{1234})+ i b (1-e_{1234})~.
\ee
However, it is easy to see that this spinor is in the same orbit as
$1+e_{1234}$, e.g.~observe that
\be
\eta= h e^{\psi \Gamma_{16}} (1+e_{1234})~,
\ee
where $h^2= a^2+b^2$ and $\tan\psi= b/a$. Next suppose that
$\theta_2$ does not vanish. If $\theta_2\not=0$, there is always a
$Spin(7)$ transformation such that $\theta_2= c \Gamma^+
(e_1+e_{234})$. This is because $Spin(7)$ acts transitively on the
$S^7$ in $\Delta_8$ and the stability subgroup is $G_2$,
$Spin(7)/G_2=S^7$. In addition $G_2$ acts transitively on the $S^6$
in $\Lambda^1(\bR^7)$ with stability subgroup $SU(3)$. So it can
always be arranged such that $\theta_1=i b (1-e_{1234})$. Therefore
the most general spinor in this case is
\be
\eta=a (1+e_{1234})+ i b (1-e_{1234})+ c \Gamma^+ (e_1+ e_{234})~.
\ee
However observe that this spinor is in the same orbit of $Spin(9,1)$ as $1+e_{1234}$. Indeed
\be
\eta= e^{{b\over 2c} \Gamma^- \Gamma^6} e^{{c\over a} \Gamma^+ \Gamma^1}
a (1+e_{1234})~.
\ee
So, we find that if $a\not=0$, then there is one orbit represented by
$a (1+e_{1234})$. It remains to investigate the case where
$a=0$. In this case, it is straightforward to see that the orbit
can always be represented by $c \Gamma^+ (e_1+e_{234})$. In turn
this spinor is in the same orbit of $Spin(9,1)$ as ${c\over \sqrt{2}} (1+e_{1234})$
as it can seen
by acting on the latter with the element $\Gamma_5\Gamma_1$ of $Spin(9,1)$.
As a consequence, the stability subgroup of $c \Gamma^+ (e_1+e_{234})$
is again  $Spin(7)\ltimes \bR^8$. Therefore, there is
only one type of orbit of $Spin(9,1)$ in $\Delta^+_{16}$ which can be
represented with $a(1+e_{1234})$. To conclude, the Killing spinor of backgrounds with one supersymmetry
can be chosen, up to a Lorentz rotation for the fluxes, such that
\bea
\epsilon= f (1+e_{1234})~,
\la{ksspins}
\eea
where $f$ a spacetime function.

\subsection{Two spinors}

There are two types of $N=2$ backgrounds  distinguished  by the
stability subgroup of the Killing spinors. To see this, we choose
the first spinor to be $\epsilon_1=a_1(1+e_{1234})$ with stability
subgroup $Spin(7)\ltimes \bR^8$. Then we decompose $\Delta^+_{16}$
as in (\ref{dec}).

One option is to take the second Killing spinor $\epsilon_2\in \Lambda^1_{7}$.
It turns out that $Spin(7)$ acts transitively on the
sphere in $\Lambda^1_{7}=\Lambda^1(\bR^7)$ and so we can take  $\epsilon_2= a_2 i (1-e_{1234})$.
The stability subgroup in $Spin(9,1)$ of both $\epsilon_1$ and $\epsilon_2$  is
 $SU(4)\ltimes \bR^8$. Moreover $\Delta_8=\Lambda^1_{4}(\bC^4)\oplus \Lambda^3_{\bar 4}(\bC^4)$
 under  $SU(4)$ and so there are no additional singlets. Therefore
 one class of $N=2$ backgrounds are those for which the Killing spinors are
 \bea
 \epsilon_1&=&f (1+e_{1234})~,
 \cr
 \epsilon_2&=&g_1 (1+e_{1234})+ i g_2 (1-e_{1234})~,
 \la{kssuf}
 \eea
 with stability subgroup  $SU(4)\ltimes\bR^8$.

 Next suppose that $\epsilon_2\in \Delta_8$. $Spin(7)$ acts transitively on the sphere $S^7$
 in the spinor
 representation $\Delta_8$ with stability subgroup $G_2$. Because of this, the second Killing spinor
 can be chosen as $\epsilon_2= b_2 \Gamma^+ (e_1+e_{234})$. In addition $\Lambda_7^1(\bR^7)$
 is an irreducible representation of $G_2$ and so there are no additional singlets.
 Therefore the second Killing spinor can be chosen as $\epsilon_2= a_2 (1+e_{1234})+ b_2
 \Gamma^+ (e_1+e_{234})$. However in this case, it can be simplified further using
 the additional $\bR^8$ invariance of $1+e_{1234}$. In particular observe that
$e^{{a_2\over b_2} \Gamma^{-} \Gamma^1} b_2
 \Gamma^+ (e_1+e_{234})= a_2 (1+e_{1234})+ b_2
 \Gamma^+ (e_1+e_{234})$. Therefore, we can take as a second spinor $\epsilon_2= a_2
 \Gamma^+ (e_1+e_{234})$. To summarize, another class of $N=2$ backgrounds are those
 for which the Killing spinors are
 \bea
 \epsilon_1&=& f (1+e_{1234})~,
 \cr
 \epsilon_2&=& g \Gamma^+ (e_1+e_{234})~,
 \la{ksgt}
 \eea
which have stability subgroup $G_2$.

It remains  to take the second spinor to be an element of $\Lambda_7^1(\bR^7)\oplus
\Delta_8$. One can again use the $Spin(7)$ invariance of $\epsilon_1$ to set the component
of the second spinor $\epsilon_2$ in $\Delta_8$ to be along the direction $\Gamma^+ (e_1+e_{234})$.
As we have mentioned the stability subgroup is $G_2$. In addition $G_2$ acts transitively
on the $S^6$ in $\Lambda_7^1(\bR^7)$ with stability subgroup $SU(3)$. Because of this,
the component of $\epsilon_2$ in $\Lambda_7^1(\bR^7)$ can be set along the direction $i (1-e_{1234})$.
However, since the stability subgroup is $SU(3)$, there are two more additional singlets.
As a result, this case applies to $N=4$ backgrounds which we shall investigate below.

\subsection{Three spinors}

To find the Killing spinors of $N=3$ backgrounds, we assume that we have selected
the first two Killing spinors as it has been described above. Therefore, we have
to consider two cases. The first case is when the first two Killing spinors $\epsilon_1, \epsilon_2$
 are
the $SU(4)\ltimes \bR^8$ invariant spinors (\ref{kssuf}).
The decomposition of $\Delta^+_{16}$ under $SU(4)$ is
\be
\Delta^+_{16}=\bR<a_1(1+e_{1234})>\oplus \bR<a_2 i (1-e_{1234})>\oplus
\Lambda^2_{6}(\bC^4)\oplus \Lambda^1_{4} (\bC^4)\oplus  \Lambda^1_{\bar 4} (\bC^4)~.
\ee
The third spinor $\epsilon_3$ must be linearly independent from both $a_1(1+e_{1234})$ and
$a_2 i (1-e_{1234})$. Suppose that $\epsilon_3\in \Lambda^2_{6}(\bC^4)$. It is known that
the generic orbit of $SU(4)$ in $\Lambda^2_{6}(\bC^4)$ can be represented by
$\mu_1 e_{12}+\mu_2 e_{34}$, $\mu_1\not=\pm \mu_2$ and has stability subgroup $SU(2)\times SU(2)$.
However, there are at least two more real spinors invariant under the $SU(2)\times SU(2)$
subgroup of $SU(4)$
and so this case is suitable for backgrounds with $N>3$. However, it is well-known that
there is a special orbit of $SU(4)$ in $\Lambda^2_{6}(\bC^4)\oplus \Lambda^1_{4}$
represented by the real spinor $e_{12}-e_{34}$ which has enhanced
stability subgroup $Sp(2)$.  In addition, decomposing $\Delta_8$ under $Sp(2)$ which can be done
using $\mathfrak{sp}(2)=\mathfrak{so}(5)$,  one can find
that there are no additional singlets. To summarize, the Killing spinors
of $N=3$ backgrounds are
\bea
 \epsilon_1&=&f (1+e_{1234})~,
 \cr
 \epsilon_2&=&g_1 (1+e_{1234})+ i g_2 (1-e_{1234})~,
 \cr
 \epsilon_3&=& h_1 (1+e_{1234})+ i h_2 (1-e_{1234})+ h_3 (e_{12}-e_{34})~,
 \la{ksspt}
 \eea
 with stability subgroup  $Sp(2)\ltimes\bR^8$ in $Spin(9,1)$.
One can continue to investigate whether there are other cases of $N=3$ backgrounds.
It turns out that there are no other possibilities.

\subsection{Four spinors}
Continuing in the same way as in the above cases, one can show that
there are two cases to consider with four spinors. One case has stability subgroup $SU(3)$ and the
other has stability subgroup $(SU(2)\times SU(2))\ltimes \bR^8$.
A basis in the space of singlets in the former case is
\bea
&&\eta_1= 1+e_{1234}~,~~~\eta_2= i (1-e_{1234})~,
\cr
&&\eta_3= e_{15}+e_{2345}~,~~~\eta_4= i (e_{15}-e_{2345})~
\eea
and a basis of singlets in the latter case is
\bea
&&\eta_1= 1+e_{1234}~,~~~\eta_2= i(1-e_{1234})~,
\cr
&&\eta_3= e_{12}-e_{34}~,~~~\eta_4= i (e_{12}+e_{34})~.
\eea
The Killing spinors of supersymmetric backgrounds are linear combinations
of the (constant) spinors in the above bases. However,
we shall argue that in the case of heterotic string, one can always
find a gauge such that the Killing spinors
are constant and can be identified with the bases elements above.

\subsection{Eight spinors}
Similarly, there are two stability subgroups in $Spin(9,1)$ that leave invariant
eight spinors. One  stability subgroup is
 $SU(2)$ and a basis in the space of singlets  is
\bea
&&\eta_1= 1+e_{1234}~,~~~\eta_2= i(1-e_{1234})~,
\cr
&&\eta_3= e_{12}-e_{34}~,~~~\eta_4= i(e_{12}+e_{34})~,
\cr
&&\eta_5= e_{15}+ e_{2345}~,~~~ \eta_6= i (e_{15}- e_{2345})~,
\cr
&&\eta_7= e_{52}+ e_{1345}~,~~~\eta_8=i (e_{52}- e_{1345})~.
\eea
The other stability subgroup is $\bR^8$ and a basis in the space of singlets is
\bea
&&\eta_1= 1+e_{1234}~,~~~\eta_2= i (1-e_{1234})~,
\cr
&&\eta_3= e_{12}-e_{34}~,~~~\eta_4= i(e_{12}+e_{34})~,
\cr
&&\eta_5= e_{13}+e_{24}~,~~~\eta_6= i (e_{13}-e_{24})~,
\cr
&&\eta_7= e_{23}-e_{14}~,~~~\eta_8=  i (e_{23}+e_{14})~.
\eea
The Killing spinors of supersymmetric backgrounds with eight supersymmetries are again linear combinations
of the (constant) spinors in the above bases. As in the
previous case of four Killing  spinors,  it can always be arranged such that the Killing spinors
 are identified with the bases elements above.

Some of the results presented in this section are summarized in the table 1.

 \begin{table}[ht]
 \begin{center}
\begin{tabular}{|c|c|c|c|c|c|c|c}\hline
$\mathrm{G }$ & $\mathrm{N=1}$& $\mathrm{N=2}$ & $\mathrm{N=3}$  & $\mathrm{N=4}$
  & $\mathrm{N=8}$  &$ \mathrm{N=16}$
 \\ \hline
$Spin(7)\ltimes \bR^8$&$\surd$& -&-&- &- & -  \\
 $SU(4)\ltimes \bR^8 $&- &$\surd$ &- & -&- &- \\
$G_2$&-& $\surd$& - &- & - & - \\
$Sp(2)\ltimes \bR^8$&- & -&$\surd $& -  &  - & - \\
$(SU(2)\times SU(2))\ltimes \bR^8$&-  & -& -& $\surd$ &  - & -  \\
$SU(3)$&- & - & -& $\surd $&  - & - \\
$\bR^8$&-  &- &-  &  -& $\surd$ &-\\
$SU(2)$&- & -& -&  - & $\surd$ &-
\\
$\{1\}$&-&-&-&-&-&$\surd$
\\ \hline
\end{tabular}
\end{center}
\caption{$N$ denotes the number of parallel spinors and $G$ their
stability subgroup in $Spin(9,1)$. $\surd$ denotes the cases for
which the parallel spinors occur. $-$ denotes the cases that do not
occur.}
\end{table}

\newsection{Parallel spinors and forms}

\subsection{Holonomy, gauge symmetry and Killing spinors}\la{holspin}

As we have mentioned, the gravitino Killing spinor
equation of heterotic strings is a parallel transport
equation for  a metric connection with skew-symmetric torsion $\hat\nabla$.
Therefore, the main tool to investigate the existence of solutions
of such an equation is the holonomy of $\hat\nabla$, ${\rm hol}(\hat\nabla)$.
The bundle of parallel spinors $\hat{\cal K}$ is spanned
by the singlets of the decomposition of the Majorana-Weyl representation
$\Delta^+_{\bf 16}$ under ${\rm hol}(\hat\nabla)$. In particular, we have
\bea
0\rightarrow \hat{\cal K}\rightarrow {\cal S}\rightarrow {\cal S}/\hat{\cal K}\rightarrow 0~,
\eea
where ${\cal S}$ is the associated bundle of the principal spin bundle with typical fibre $\Delta^+_{\bf 16}$. The vector
bundle
$\hat{\cal K}$ is topologically trivial and so it is equipped with the trivial connection $\partial$.
In particular, one can introduce a basis $(\eta_i, i=1,\dots, {\rm rank}\,\hat{\cal K})$ of constant spinors in $\hat{\cal K}$.

The Killing spinor equations of the heterotic string are covariant
under (local) $Spin(9,1)$ gauge transformations as those of IIB
supergravity. However, unlike the cases of  IIB  and
eleven-dimensional supergravities, the Lie algebra
$\mathfrak{spin}(9,1)$ of the gauge group  of the Killing spinor
equations {\it coincides} with the Lie algebra that the
(super)covariant derivative
 $\hat\nabla$ takes values in.
This in particular implies that the restriction of $\hat\nabla$ on
the sections of $\hat{\cal K}$ can be trivialized with  $Spin(9,1)$
local gauge transformation. As a result, there is a gauge, up to
local $Spin(9,1)$ transformations,
 such that the
 {\it parallel spinors} of $\hat\nabla$ are {\it constant} and so they can be identified with a constant basis $\eta_i$.
 Of course the basis $\eta_i$ is defined up to a (constant) general linear transformation $GL({\rm rank}\,\hat{\cal K},\bR)$. This transformation
 can be used to simplify the expressions for the Killing spinors, for more details see \cite{grangutowskigpb}.
We remark that in IIB and eleven-dimensional supergravities, {\it there is not always a choice of a gauge for which the
solutions of the gravitino Killing spinor equations are constant}. For example, one can adapt the
results of \cite{jfgp} to show that the only maximally supersymmetric background of IIB and eleven-dimensional
supergravities with constant Killing spinors is locally isometric to Minkowski spacetime.

Given a constant basis $\eta_i$ of parallel spinors in $\hat{\cal K}$, the most general Killing spinors
can be written as
\bea
\epsilon_r=\sum_{i} f_{ri} \eta_i~,~~~~r=1,\dots, N~,~~~i=1,\dots, {\rm rank} ~{\cal K}~,
\la{kspins}
\eea
where $f=(f_{ri})$ is a real constant matrix.
In general $N< {\rm rank}~\hat{\cal K}$ because some parallel spinors may not solve the dilatino or gaugino
Killing spinor equations. The matrix $f$ can be thought of as the inclusion of the bundle of Killing spinors
${\cal K}$ in $\hat{\cal K}$.

To investigate all the supersymmetric backgrounds of the heterotic string, one has to
determine the cases for which $N<{\rm rank}~ \hat{\cal K}$ for $N>1$. It is well-known that there are such
 backgrounds  as
for example the group manifolds that have been mentioned in the
introduction. In what follows, we shall only consider the cases for
which $N={\rm rank}~ \hat{\cal K}$. These are the so called
maximally supersymmetric $G$-backgrounds in the terminology of
\cite{grangpb}. In the first few cases, we shall allow the
coefficients $f$ in (\ref{kspins}) to be spacetime functions and
show that the parallel transport equations imply that $f$ can be
taken to be the identity, up to a local $Spin(9,1)$ and constant
$GL(N,\bR)$ transformations, in agreement with the general argument
presented above.

We have mentioned in the introduction that there are null and timelike supersymmetric
backgrounds. These can be distinguished by the properties of their Killing spinors ($N={\rm rank}~ \hat{\cal K}$).
The Killing spinors of null supersymmetric backgrounds satisfy
$\Gamma^-\epsilon=0$, (see appendix A for our spinor conventions).
Since $\epsilon\in \Delta^+_{\bf 16}$, this condition implies that the
Killing spinors are also chiral with respect to the Clifford subalgebra ${\rm Cliff}(\bR^8)$
of ${\rm Cliff}(\bR^{9,1})$, where $\bR^8=\bR<e^1, \dots, e^4, e^6,\dots, e^9>$. There are
null supersymmetric backgrounds with even and odd number of Killing spinors.

The timelike supersymmetric backgrounds admit always even number of
Killing spinors. Half of these spinors satisfy the condition
$\Gamma^-\epsilon=0$ while the other half satisfies the condition
$\Gamma^+\epsilon=0$. Therefore the Killing spinors do not have a
definite chirally with respect to the above ${\rm Cliff}(\bR^8)$
subalgebra.

\subsection{Parallel forms}\la{parallelforms}

It has been known for some time that an alternative way to
characterize the geometry of supersymmetric heterotic backgrounds is
in terms of the spacetime form bilinears of the parallel spinors,
see e.g.~\cite{strominger, ivanovgp, tseytlin}. In the heterotic
case, a consequence of the Killing spinor equations is that all the
spacetime form bilinears of the parallel spinors are also  {\it
parallel} with respect to the connection $\hat\nabla$. This is
because $\hat\nabla$ is a connection that takes values in
$\mathfrak{spin}(9,1)$ and so preserves the gamma-matrices and the
spinor inner product, i.e.~$\hat\nabla \epsilon_r=0$, $r=1,\dots,
N$, implies that
\bea
\hat\nabla\alpha_{rs}=0~,~~~r,s=1,\dots, N~,
\la{parform}
\eea
where $\alpha_{rs}$ represents all the form spinor bilinears, see appendix A for the definition of $\alpha$.

A converse to the above statement has been presented in
\cite{gptsimpis}. In the case of the heterotic string a stronger
statement is valid. In particular, if the forms $\a_{rs}$ are spinor
bilinears of some spinors $\e_r$ and $\hat\nabla\a_{rs}=0$, then
$\hat\nabla\e_r=0$. This is because the stability subgroups of the
parallel spinors can also be characterized as those subgroups of
$Spin(9,1)$ that leave the forms $\a_{rs}$ invariant. Therefore, one
can use the form spinor bilinears to give an alternative description
of the geometry of spacetime of supersymmetric backgrounds.

The parallel forms of supersymmetric backgrounds generate a ring
under the wedge product. It turns out that the ring of null
supersymmetric backgrounds is nilpotent, i.e.~the wedge product of
any two forms in the ring vanishes. In all cases, there is a null
parallel one-form $\kappa=\re^-$ and all the rest of the generators
of the ring are of the form
\bea
\alpha=\re^-\wedge \phi~,
\eea
where $(\re^+, \re^-, e^i)$  is a light-cone frame adapted to the metric.
Although $\hat\nabla\alpha=0$, the form $\phi$ is not parallel with respect to the $\hat\nabla$ connection\footnote{However,
it is parallel with respect to another connection which takes values in the compact subalgebra of the
holonomy group of the null supersymmetric backgrounds.}. In particular, we have
\bea
&&\hat\nabla_A\phi_{i_1\dots i_k}=0~,~~~\hat\nabla_A\phi_{B_1\dots
B_{k-1}+}=0~,
\cr
&&\hat\nabla_A\phi_{i_1\dots i_{k-1}-}=\hat\Omega_{A,}{}^m{}_- \phi_{i_1\dots i_{k-1}m}~.
\la{paraphi}
\eea
Nevertheless  in many cases
it is convenient to use the form $\phi$ to describe the geometry of spacetime.

The timelike supersymmetric backgrounds admit {\it at least} three  parallel one forms $\kappa=\re^-$, $\kappa'=\re^+$
and $\hat\kappa=e^1$. The $N>2$ backgrounds admit more than three
parallel one-forms. The associated ring of parallel forms is not nilpotent. At the end of this section, we give the
generators of the rings of the parallel forms of all the supersymmetric backgrounds.

Some geometric properties of the spacetime follow immediately from (\ref{parform}). For example, let $\kappa$ be a one-form
parallel spinor bilinear. Then (\ref{parform}) implies that $\kappa$ is parallel, $\hat\nabla \kappa=0$.
The  associated vector
field
$X$  with respect to the spacetime metric is also parallel, $\hat\nabla X=0$. A consequence of this is that
\bea
{\cal L}_X g&=&0
\cr
d\kappa&=&i_X H~,
\eea
i.e.~$X$ is Killing and that the rotation of $\kappa$ is equal to
the particular component of the flux $H$. In addition, if $H$
satisfies the Bianchi identity, which it does at the lowest
order\footnote{It is expected that $H$ will remain invariant after
all perturbative corrections in $\alpha'$ are taken into account
provided that the classical background is invariant under the
transformations generated by $X$. This is because the corrections
are polynomials of the Riemann curvature $R$, $F$, $H$ and their
covariant derivatives which are invariant under $X$.} in $\alpha'$,
then
\bea
{\cal L}_XH=d i_X H+ i_Xd H= d i_X H=0~,
\eea
and so $H$ is also invariant under the one-parameter family of diffeomorphisms generated by $X$.

Next suppose that $X,Y$ are $\hat\nabla$-parallel vector fields and denote with $\kappa_X$ and $\kappa_Y$ the associated
one-forms. The commutator of such two Killing vector fields is Killing because ${\cal L}_{[X,Y]}
={\cal L}_X {\cal L}_Y-{\cal L}_Y {\cal L}_X$.
In addition, it is known that $i_{[X,Y]}={\cal L}_X i_Y-i_Y {\cal L}_X$ and so
\bea
i_{[X,Y]} H={\cal L}_X i_Y H={\cal L}_X d\kappa_Y=d {\cal L}_X \kappa_Y=d\kappa_{[X,Y]}~.
 \la{commh}
\eea
Therefore, the commutator $[X,Y]$ is also parallel with respect $\hat\nabla$. However $\kappa_{[X,Y]}$ {\it may not} be associated
with a one-form parallel spinor bilinear.

Another aspect of the form spinor bilinears that arise in the
context of supersymmetric heterotic string backgrounds is whether or
not they are invariant under the Killing vectors of these
backgrounds. Let  $X$ be a Killing vector associated with a one-form
spinor bilinear $\kappa$ and $\alpha$ be a $k$-form spinor bilinear.
Using $\hat\nabla\a=\hat\nabla\kappa=0$, one can show that
\bea
({\cal L}_X\alpha)_{A_1\dots A_k}=k (-1)^k (i_XH)^B{}_{[A_1} \alpha_{A_2\dots A_k]B}~,
\la{calxa}
\eea
where $A_1,\dots, A_k, B=-,+, i$.
Therefore ${\cal L}_X\alpha=0$, iff the rotation of $X$,   $i_XH$, leaves invariant the form $\a$. We shall
find that the dilatino Killing spinor equation implies such conditions.

It also turns out that the geometry of the spacetime of supersymmetric backgrounds
can be described using a minimal set of parallel forms. This particularly applies
to the conditions that arise from the gravitino Killing spinor equation. This is similar
to the characterization of K\"ahler manifolds as the Riemannian manifolds that admit
a parallel almost complex structure. The generators of the ring of  parallel forms or the rings themselves for the
 supersymmetric backgrounds, up to Hodge duality,  are summarized in table 2 below.

 \begin{table}[ht]
 \hskip-0.9cm
\begin{tabular}{|c|c|c|}\hline
$\mathrm{Supersymmetry}$ & $\mathrm{Killing~ Vectors}$& $\mathrm{Parallel~ Forms}$
 \\ \hline
$N=1$~~$Spin(7)\ltimes \bR^8$&$1$& $\re^-, \re^-\wedge \phi$\\
$N=2$~~ $SU(4)\ltimes \bR^8 $&$1$ &$\re^-~, \re^-\wedge \chi~, \re^-\wedge\omega$ \\
$N=2$~~$G_2$&$3$& $\re^-~,~ \re^+~,~e^1~,~\varphi$ \\
$N=3$~~$Sp(2)\ltimes \bR^8$&$1$ & $\re^-~, \re^-\wedge \omega_I~, \re^-\wedge \omega_J~, \re^-\wedge \omega_K $\\
$N=4$~~$(SU(2)\times SU(2))\ltimes \bR^8$& $1 $ & $\re^-~, -\re^-\wedge(e^1\wedge e^6+ e^2\wedge e^7)$
\cr
&&$ -\re^-\wedge (e^3\wedge e^8+ e^4\wedge e^9)$
\cr
&&
$\re^-\wedge (e^1+ie^6)\wedge (e^2+ie^7)$
\cr
&&
$\re^-\wedge (e^3+i e^8)\wedge (e^4+i e^9)$ \\
$N=4$~~$SU(3)$& $4$ & $\re^-~, \re^+~,~ e^1~,~ e^6~,~\hat\omega~,~\hat\chi $\\
$N=8$~~$\bR^8$& $1 $ &$\re^-\wedge \psi\,,~~~~\psi\in \Lambda^{{\rm ev}+}(\bR^8)$ \\
$N=8$~~$SU(2)$& $6$ & $\re^-~,~ \re^+~,~ e^1~,~ e^6~,~ e^2~,~ e^7~,$
\cr
&&$~-e^3\wedge e^8- e^4\wedge e^9~,~ (e^3+ie^8)\wedge (e^4+i e^9)$
\\
$N=16$~~~$\{1\}$&$ 10$& $e^A~,~~~A=0,\dots, 9$\\ \hline
\end{tabular}
\vskip 0.2cm {\small Table 2: The first column gives the number of
Killing vectors that are constructed from Killing spinor bilinears
of a supersymmetric background. The second column gives a minimal
set of $\hat\nabla$-parallel forms which characterizes the geometry
of the supersymmetric background, where $\Lambda^{{\rm
even}+}(\bR^8)=\Lambda^0(\bR^8)\oplus \Lambda^2(\bR^8)\oplus
\Lambda^{4+}(\bR^8)$ and $\Lambda^{4+}(\bR^8)$ is the space of
self-dual four-forms in $\bR^8$, and
\bea
&&\chi=(e^1+i e^6)\wedge (e^2+i e^7)\wedge (e^3+i e^8)\wedge (e^4+i e^9)~,\nonumber
\cr
&&\omega=-e^1\wedge e^6-e^2\wedge e^7-e^3\wedge e^8-e^4\wedge e^9~,~~~
\phi={\rm Re}\, \chi-{1\over 2}\omega\wedge \omega~,
\cr
&&\hat\omega=-e^2\wedge e^7-e^3\wedge e^8-e^4\wedge e^9~, ~~~\hat\chi=(e^2+i e^7)\wedge (e^3+i e^8)\wedge (e^4+i e^9)~,
\cr
&&\varphi={\rm Re}\hat\chi+ e^6\wedge \hat\omega~, ~~~\omega_I=\omega~,
\cr
&&\omega_J={\rm Re}[(e^1+i e^6)\wedge (e^2+i e^7)]+ (e^3+i e^8)\wedge (e^4+i e^9)] ~,
\cr
&&\omega_K=-{\rm Im}[(e^1+i e^6)\wedge (e^2+i e^7)+ (e^3+i e^8)\wedge (e^4+i e^9)]~.
\eea
}
\end{table}

\newpage

\newsection {$N=1$ backgrounds}

\subsection{Supersymmetry conditions}

In section \ref{stabgroups}, we have shown that
  the Killing spinor can be chosen  as  $\epsilon= f (1+e_{1234})$ and has stability
  subgroup $Spin(7)\ltimes \bR^8$, where
$f$ is a real function of the spacetime. Substituting this into the gravitino Killing spinor
equation, we find
\bea
&&\partial_A\log f (1+e_{1234})-{1\over8}
\hat\Omega_{A,\g\d} \epsilon^{\g\d}{}_{\bar\a\bar\b}\Gamma^{{\bar\a\bar\b}} 1
+{1\over4}
\hat\Omega_{A,\bar\a\bar\b}\Gamma^{{\bar\a\bar\b}} 1+ {1\over2}
\hat\Omega_{A,\a}{}^\a 1
\cr
&&
-{1\over2} \hat\Omega_{A,\a}{}^\a e_{1234}+{1\over2} \hat\Omega_{A,+\a} \Gamma^{+\a} e_{1234}
+{1\over2} \hat\Omega_{A,+\bar\a} \Gamma^{+\bar\a} 1+{1\over2} \hat\Omega_{A,-+} (1+e_{1234})=0~.
\eea
The above equation can be expanded in the basis (\ref{hbasis}). Setting every component
in this basis to zero, we find the conditions
\bea
\partial_A\log f+{1\over2}\hat\Omega_{A,-+}=0~,
\la{gspsone}
\eea
\bea
\hat\Omega_{A,\a}{}^\a=0~,~~~~~~~
\hat\Omega_{A,\bar\a\bar\b}-{1\over2} \hat\Omega_{A,\g\d} \epsilon^{\g\d}{}_{\bar\a\bar\b}=0~,
\la{gspstwo}
\eea
\bea
\hat\Omega_{A,+\bar\a}=\hat\Omega_{A,+\a}=0~.
\la{gspsthree}
\eea
The components $\hat\Omega_{A,-\a}$ and $\hat\Omega_{A,-\bar\a}$ are
unconstrained.

Similarly, one substitutes $\epsilon= f (1+e_{1234})$ into the    dilatino
Killing spinor equation to find
\be
(\Gamma^A \partial_A\Phi-{1\over12} \Gamma^{ABC} H_{ABC})(1+e_{1234})=0~.
\ee
Expanding this in the basis (\ref{hbasis}), we get that
\bea
\partial_{\bar\a}\Phi+{1\over6} H_{\b_1\b_2\b_3} \epsilon^{\b_1\b_2\b_3}{}_{\bar\a}
-{1\over2} H_{\bar\a\b}{}^\b-{1\over2} H_{-+\bar\a}=0~,
\la{dspsone}
\eea
\bea
\partial_+\Phi=0~,
\la{dspstwo}
\eea
\bea
H_{+\a}{}^\a=0~,~~~~
-H_{+\bar\a_1\bar\a_2}+{1\over2} H_{+\b_1\b_2} \epsilon^{\b_1\b_2}{}_{\bar\a_1\bar\a_2}=0~.
\la{dspsthree}
\eea
The components $\partial_-\Phi$ and $H_{-ij}$ remain undetermined
by the dilatino Killing spinor equation, where $i=\a, \bar\a$ and similarly $j$.

\subsection{The geometry of spacetime}

\subsubsection{The holonomy of $\hat\nabla$ and supersymmetry}

The gravitino Killing spinor equation implies that the holonomy of the $\hat\nabla$ connection is
contained in $Spin(7)\ltimes \bR^8$. This may have been expected on general grounds because
the Killing spinor $\epsilon$ is parallel with respect to $\hat\nabla$ and
so the holonomy of $\hat\nabla$ should be contained in the stability subgroup of the Killing spinor $\epsilon$ in $Spin(9,1)$.

One can also see this explicitly in the gauge $f=1$. This gauge can be attained
by the  spinorial transformation $e^{b \Gamma^{05}}$, which induces a Lorentz gauge transformation on $\hat\nabla$ and a
Lorentz rotation on the fluxes.
The action of $e^{b \Gamma^{05}}$ on the Killing spinor $\epsilon$ is to scale it with $e^b$.
Therefore setting $b=-\log |f|$, the spacetime dependence of the Killing spinor can be gauged away and so
the Killing spinor can be written as $\epsilon=1+e_{1234}$.
In this gauge
\bea
\hat\Omega_{A,+-}=0~,
\eea
which together with (\ref{gspsthree}) imply that all the components of $\hat\Omega_{A,+B}=0$.
It is then easy to see that the remaining components of the connection one-form, $\hat\Omega=\hat\Omega_{A} e^A$,
take values in $\mathfrak{spin}(7)\coplus\bR^8$. Note however that for generic $N=1$ backgrounds, the Levi-Civita connection
{\it does not}
have $Spin(7)\ltimes \bR^8$ holonomy.

The converse is also valid. If ${\rm hol}(\hat\nabla)\subseteq
Spin(7)\ltimes \bR^8$, there is a spinor $\epsilon$ which is
parallel with respect to $\hat\nabla$ and so $\epsilon$ satisfies
the gravitino Killing spinor equation. Thus the existence of a
solution for the gravitino Killing spinor equation can be entirely
characterized by the holonomy of $\hat\nabla$.

To investigate further the geometry of spacetime, it is convenient to introduce
the $\hat\nabla$-parallel forms associated with the parallel spinor bilinears. It turns out that
most of the fluxes and geometry can be expressed in terms of these bilinears.

\subsubsection{Spacetime forms}

Using the formulae
that we have collected
in appendix A, one can find that the non-vanishing  Killing spinor bilinears\footnote{We have normalized
the Killing spinor $\epsilon$ with an additional factor of $1/\sqrt{2}$.} are a one-form
\bea
\kappa=\kappa(\epsilon, \epsilon)=f^2 (e^0-e^5)~,~~~
\eea
and a five-form
\bea
\tau=\tau(\epsilon, \epsilon)= f^2 (e^0-e^5)\wedge \phi~,
\eea
where
\bea
\phi={\rm Re} \chi-{1\over2} \omega\wedge\omega~,
\eea
and $\chi$ and $\omega$ are defined in appendix A, see also table 2. It is easy to recognize that $\phi$ is the
usual $Spin(7)$-invariant four-form on eight-dimensional manifolds.
The forms $\omega$ and $\chi$ are not individually  well-defined on the spacetime.

To proceed, we introduce a frame $\re^+, \re^-, \re^\a, \re^{\bar\a}$,
where $\re^-=(1/\sqrt{2})\, (-e^0+e^5)$, $\re^+=(1/\sqrt{2})\, (e^0+e^5)$,  and $\re^\a=(1/\sqrt{2})\, (e^\a+ie^{\a+5}), \re^{\bar\a}=
(1/\sqrt{2})\, (e^\a-ie^{\a+5})$, and $(e^0, \dots, e^9)$ is the orthonormal frame  in appendix A. The spacetime
metric can be rewritten as
\bea
ds^2=2 \re^+ \re^-+ 2\delta_{\a\bar\b} \re^\a \re^{\bar\b}~.
\la{metrherm}
\eea
In this new frame\footnote{We have normalized the forms with a
further factor of $1/\sqrt{2}$.} $\kappa=f^2\, \re^-$ and $\tau=
f^2\, \re^-\wedge \phi$. Therefore the ring of form spinor bilinears
under the wedge product is nilpotent, i.e.~the wedge product of any
two forms vanishes. As we have explained in section
\ref{parallelforms}, $\kappa$ and $\tau$ are $\hat\nabla$-parallel.
Therefore the vector field $X=f^2\, {\rm e}_+$ associated with the
one-form $\kappa$ with respect to the spacetime metric is also
$\hat\nabla$-parallel, i.e.~$\hat\nabla\,X=0$,
 where $e^A(e_B)=\delta^A_B$ and $e_B$ is the co-frame.
This in turn implies that $X$ is Killing and $d\kappa=i_XH$.
Consequently, $di_X H=0$ and so the Bianchi  identity, $dH=0$, implies that ${\cal L}_X H=0$.
The three-form field strength $H$ is invariant under the isometries generated by $X$.
In addition (\ref{dspstwo}) implies that ${\cal L}_X \Phi=0$ as well. Therefore the metric and both fluxes
$H$ and $\Phi$ are invariant under $X$.
Furthermore as we shall explain in detail in the next section, (\ref{dspsthree})
implies that $H_{+AB}$ takes values in $\mathfrak{spin}(7)\coplus\bR^8$.
Using (\ref{calxa}), one finds  that
\bea
{\cal L}_X\tau=0~.
\eea
Therefore, the parallel vector field $X$ {\it leaves invariant} the $Spin(7)\ltimes \bR^8$-structure
of spacetime. It turns out that this is a generic property of all null supersymmetric heterotic string backgrounds
that we investigate. The null parallel vector field preserves the $K\ltimes \bR^8$-structure
of the spacetime.

\subsubsection{The solution of the Killing spinor equations}

To further investigate the Killing spinor equations, we decompose the space of two-, three- and four-forms
under $Spin(7)$ as as $\Lambda^2(\bR^8)=\Lambda^2_{\bf 7}\oplus \Lambda^2_{\bf 21}$,
$\Lambda^3(\bR^8)=\Lambda^3_{\bf 8}\oplus \Lambda^3_{\bf 48}$,  $\Lambda^4(\bR^8)=\Lambda_+^4(\bR^8)\oplus \Lambda^4_-(\bR^8)$,
 $\Lambda^{4}_+ =\Lambda_{\bf 1}^4\oplus \Lambda^4_{\bf 7}\oplus \Lambda^4_{\bf 27}$ and $\Lambda^4_-=\Lambda^4_{\bf 35}$, where
\bea
&&\Lambda^2_{\bf 7}=\{\a\in \Lambda^2(\bR^8)|\, *(\a\wedge\phi)=-3\a\}~,~~~\Lambda^2_{\bf 21}=\{\a\in \Lambda^2(\bR^8)|\,
*(\a\wedge\phi)=\a\}
\cr
&&\Lambda^3_{\bf 8}=\{ *(\a\wedge\phi)|\, \a\in \Lambda^1(\bR^8)\}~,~~~\Lambda^3_{\bf 48}=\{\a\in\Lambda^3(\bR^8)|\, \a\wedge\phi=0\}~,~~~
\cr
&&\Lambda_{\bf 1}^4=\{ r\,\phi|\, r\in \bR\}~.
\eea
 The representation
$\Lambda^2_{\bf 21}$ can be identified with the adjoint representation of $\mathfrak{spin}(7)$,
so $\mathfrak{spin}(7)=\mathfrak{so}(7)=\Lambda^2_{\bf 21}$.
Using the above decompositions,
the conditions that arise from the gravitino Killing spinor equation (\ref{dspsone})-(\ref{dspsthree}) in the gauge $f=1$
can be written as
\bea
\hat\Omega_{A,+B}=0~,~~~\hat\Omega_{A, ij}^{\bf 7}=0~,
\la{sumgsps}
\eea
where the projection to the seven-dimensional representation is done in the indices $i,j=1,\dots4, 6, \dots,9$.
In addition, the conditions that arise from the dilatino Killing spinor equation can be rewritten as
\bea
\partial_i\Phi+{1\over12} H_{jkl} \phi^{jkl}{}_{i}
-{1\over2} H_{-+i}=0~,~~~\partial_+\Phi=0~,~~~H_{+ij}^{\bf 7}=0.
\la{sumdsps}
\eea

The conditions (\ref{sumgsps}) and  (\ref{sumdsps})  can be solved to determine most of the
components of the flux in terms of the geometry. In particular, the first equation in (\ref{sumgsps}) implies that
$\kappa=\re^-$ is parallel and so $i_X H=d\kappa=d \re^-$. The second equation is equivalent to $\hat\nabla_A\phi_{ijkl}=0$ and
 so in particular
implies  that
\bea
\hat\nabla_-\phi_{ijkl}=0~,
\cr
\hat\nabla_m\phi_{ijkl}=0~.
\eea
These equations can be solved for the fluxes to give
\bea
H^{\bf 7}_{-ij}=H_{-ij}-{1\over2} H_{-kl} \phi^{kl}{}_{ij}=-{1\over12} \phi^{pqr}{}_i\nabla_-\phi_{pqrj}~,
\cr
H_{ijk}=-{1\over 4!} \nabla_{[m_1} \phi_{m_2\dots m_5]} \epsilon^{m_1 m_2\dots m_5}{}_{ijk}+\theta^m \phi_{mijk}~,~~
\la{compsps}
\eea
where
\bea
\theta_i=-{1\over 36} \nabla^p\phi_{pj_1j_2j_3} \phi^{j_1j_2j_3}{}_i~.
\eea
Observe that $\theta$ is analogous to the Lee form  of  eight-dimensional Riemannian manifolds with a $Spin(7)$-structure.
To derive the second equation in (\ref{compsps}), we have use the results of \cite{stefanspin7}.
In addition the first condition in (\ref{sumdsps}) implies that the $H_{ijk}^{\bf 8}$ component
of $H$ which is determined by  $\theta$ can be  expressed in terms of the derivative of the dilaton and
the $H_{+-i}=(d\kappa)_{-i}$ component of the flux. If $H_{+-i}\not=\partial_if$,
then the spacetime is not conformally balanced.

Therefore the metric and three-form flux of the supersymmetric spacetime can be written as
\bea
&&ds^2=2 \re^+ \re^-+\delta_{\a\bar\b} \re^\a \re^{\bar\b}
\cr
&&H= \re^+\wedge d\re^-+ \Omega^{\bf 7}_{-, ij}~ \re^-\wedge e^i\wedge e^j+{1\over2}
H^{\bf 21}_{-ij} \re^-\wedge e^i\wedge e^j
\cr
&&~~~~~~~~~~~~~~~~~~~~~~~~~~~~~~~~~~~+ {1\over3!}H_{ijk} e^i\wedge e^j\wedge e^k~,
\eea
where $H_{ijk}$ is given in (\ref{compsps}). The component $H^{\bf 21}_{-ij}$ of the fluxes is not determined by the
Killing spinor equations.

\subsubsection{Local coordinates}

One can introduce local coordinates on the spacetime $M$ by adapting a coordinate $u$ along the
null Killing vector field $X$, $
X={\partial\over\partial u}$. The spacetime metric can be written as
\bea
ds^2=2 U (dv+m_I dy^I) ( du+ V dv+  n_I dy^I )+ \gamma_{IJ} dy^I dy^J~,
\la{coormetr}
\eea
where $ U, V, m_I, n_I$ and $\gamma_{IJ}$ are functions of  $v, y^I$ coordinates,
$I,J=1,\dots,8$. All the components of the metric are independent of $u$ because $X$ is Killing.
In addition $U=f^2$. To see this, we adapt  the frame
\bea
\re^-=dv+m_I dy^I~,~~~ \re^+=U (du+ V dv+  n_I dy^I)~,~~~e^i= e^i_J dy^J~,
\la{coorframe}
\eea
where $\gamma_{IJ}=\delta_{ij}  e^i_I e^j_J$.
The Killing vector field in this frame is
\be
X=f^2e_+={\partial\over\partial u}~,
\ee
where $e_B$ is
\bea
e_+= U^{-1} {\partial\over\partial u}~,~~~ e_-= {\partial \over\partial v}-
 V{\partial \over\partial u}~,~~~
\cr
e_i= e_i^J {\partial\over\partial y^J}+(-n_i+V m_i)
{\partial\over\partial u}
- m_i {\partial\over\partial v}~,
\la{coorcoframe}
\eea
 $e^i_I e^I_j=\delta^i{}_j$, $m_i= m_I e^I_i$ and $n_i= n_I e^I_i$.
Using the above expression for the co-frame,  the Killing vector field $X$ can be written as
\bea
X=f^2e_+=f^2 U^{-1}
{\partial\over\partial u}={\partial\over\partial u}~.
\eea
Therefore $U=f^2$. In particular, we can set $U=1$ in  the gauge $f=1$.

A consequence of the torsion free condition for the Levi-Civita connection
  and $\hat\Omega_{A,+B}=0$ is that
\bea
i_XH=d m~.
\la{fluxrot}
\eea
So using $i_X H=d\kappa$, one finds that
\bea
d\kappa=dm~.
\la{kappamu}
\eea
As it may have been expected the off-diagonal part of the metric (\ref{coormetr})
proportional to $m$, which is responsible
for the deviation from Penrose coordinates, is due to the rotation of the null geodesic
congruence generated by $\kappa$.
In addition (\ref{fluxrot}) relates  this term to the
presence of non-vanishing $H$ fluxes. Furthermore,
the coordinate $v$
of the spacetime can be specified by applying the Poincar\'e lemma
on the closure relation $d(\kappa-m)=0$.

\subsubsection{A deformation family of $Spin(7)$-structures}

The spacetime $M$ of $N=1$ supersymmetric heterotic string backgrounds
 can be interpreted as a two parameter Lorentzian deformation
family\footnote{The family is trivial with respect to one of the two parameters.} of an eight-dimensional
manifold $B$ with an $Spin(7)$-structure. To see this, observe that the metric  (\ref{coormetr})
can be rewritten as
\bea
ds^2= g_{ab} du^a du^b+ g_{IJ} (dy^I+A^I_a du^a) (dy^J+A^J_b du^b)~,
\la{fammetr}
\eea
where
\bea
g_{vu}+g_{IJ} A^I_v A^J_u=U~,~~~g_{vv}+g_{IJ} A^I_v A^J_v=2UV~,~~~g_{uu}+g_{IJ} A^I_u A^J_u=0
\cr
g_{JI} A^J_u=U m_I ~,~~~g_{JI} A^J_v=U n_I+U V m_I~,
~~~
g_{IJ}=\gamma_{IJ}+ 2U n_{(I} m_{J)}~,
\la{famcoor}
\eea
 and $g_{ab}, g_{IJ}, A^I_a$ depend on all coordinates $u^a, y^I$, ($(u^a)=(u,v)$). The components $A^I_a$ can be thought of as
 the non-linear connection of the family.

The spacetime admits an integrable distribution of co-dimension eight. To see this, we adapt a frame
\bea
E^+~,~~~ E^-~,~~~E^i= \tilde e^i{}_J (dy^J+A^J_b du^b)~,
\la{eframe}
\eea
 to the
metric (\ref{fammetr}), where $E^+, E^-$ is a light-cone frame adapted to the
two-dimensional part of the metric, $g_{ab} du^a du^b=2 E^- E^+$, and
$\delta_{ij} \tilde e^i{}_I \tilde e^j{}_J=g_{IJ}$. Applying the Frobenius theorem to the one-forms $E^+, E^-$,
one can easily show that the spacetime is
 an integrable foliation of co-dimension eight with leaves the deformed manifold $B$ given by $u,v={\rm const}$.

It remains to determine the geometry of $B$ that gets deformed. It
is clear that $B$ is a Riemannian manifold with metric $d\tilde s^2=
g_{IJ} dy^I dy^J$ equipped with a three-form $\tilde H= H|_B$.  So
one can construct a Riemannian connection $\hat{\tilde \nabla}$ on
$B$ with torsion $\tilde H$. In addition $B$ admits a
$Spin(7)$-invariant form $\tilde\phi=\phi|_B$. However these data
are not compatible, i.e.~in general $\hat{\tilde
\nabla}\tilde\phi\not=0$. To see this, observe that
\bea
e^i=\ell^i{}_j (E^i+ p^j E^++ q^i E^-)
\eea
for some non-vanishing $p$ and $q$, ${\rm det} \ell\not=0$, and
similarly for the rest components of the frame. This in particular
implies that the self-dual four form $\phi$ in the $(E^i, E^+, E^-)$
frame has components in the $E^+$ and $E^-$ directions. Taking the
covariant derivative of $\phi$, i.e.~$\hat\nabla_i\phi$, one get
contributions from $\hat\nabla_i E^+=-\hat\Omega_i{}^+{}_j E^j$, and
similarly from $\hat\nabla_i E^-$, which can be identified with the
second fundamental form of $B$ with respect to the connection
$\hat\nabla$. Since these contributions do not apparently vanish,
$\hat\nabla_i\phi=0$, see (\ref{paraphi}), after restriction to $B$
does not imply that $\hat{\tilde \nabla}_i\tilde\phi=0$. Therefore
$B$ does not have a $Spin(7)$ structure compatible with the
connection $\hat{\tilde \nabla}$. Nevertheless, $B$ admits a
$Spin(7)$-structure.

There is though a {\it special case} where the $Spin(7)$-structure
of $B$ is compatible with the $\hat{\tilde \nabla}$ connection. This
is whenever the rotation of the null $\hat\nabla$-parallel vector
field vanishes, $d\kappa=d \re^-=0$, i.e.~the metric is written in
terms of Penrose coordinates. In this case, the $(\re^-, \re^+,
e^i)$ and $(E^-, E^+, E^i)$ frames are related as
\bea
\re^-= E^-~,~~~\re^+-n_ie^i=E^+~,~~~e^i=E^i-n^i E^-~.
\la{twoframes}
\eea
Then $\phi={1\over4!}\phi_{ijkl} e^i\wedge e^j\wedge e^k\wedge e^l$ can be written
 as $\phi=\psi+ E^-\wedge \tau$, where $\psi={1\over4!} \phi_{ijkl} E^i\wedge E^j \wedge E^k\wedge E^l$.
Thus we have
\bea
(\hat\nabla_i\phi)|_B= \hat{\tilde\nabla}_i\tilde\phi+(\hat\nabla_i E^-)|_B\wedge \tau|_B=\hat{\tilde\nabla}_i\tilde\phi=0
\eea
because $\nabla_i E^-=\nabla_i \re^-=-\Omega_{i,+ A} E^A=0$, since $\hat\Omega_{A,+B}=0$, $E^+|_B=E^-|_B=0$,
and $\phi$ is parallel with respect to $\hat\nabla$ along the $B$ directions, see (\ref{paraphi}), where
$\tilde\phi=\phi|_B=\psi|_B$. In addition $B$ is conformally balanced. This is because
the dilatino Killing spinors equation (\ref{sumdsps}) when restricted on $B$ gives
\bea
\tilde\theta_i=2 \partial_i\tilde\Phi~,
\la{sumdspsx}
\eea
since the rotation of the vector field vanishes. Eight-dimensional Riemannian manifolds
with a conformally balanced $Spin(7)$-structure compatible with
a connection with skew-symmetric torsion have been investigated in \cite{stefanspin7}.
Any eight-dimensional Riemannian manifold with a $Spin(7)$-structure admits a connection
with skew-symmetric torsion
\bea
\tilde H=-\star d\tilde\phi+ \star(\tilde\theta\wedge \tilde \phi)~,
\eea
where the Lee form can also be written as $\tilde\theta=-{1\over6}\star (\star d\phi\wedge \phi)$ and $\star$ is the
Hodge duality operator of $B$ for $d{\rm vol}(B)=\tilde e^1\wedge\dots\wedge \tilde e^4\wedge \tilde e^6\wedge\dots\wedge
\tilde e^9$. Note that $\tilde e^i=e^i|_B=E^i|_B$.
Our form conventions are summarized in appendix A. Of course the torsion is required
 to satisfy the (generalized) Bianchi identity for applications
to the heterotic string.

The geometry of $B$  can also be given in terms of G-structures. It is known that there are four classes
of $Spin(7)$-structures obtained by decomposing $\tilde\nabla\tilde\phi$ in terms of $Spin(7)$
representations \cite{fernandez}. These classes can be described in terms of the Lee form \cite{cabrerab} as follows:
  $W_0$  $(d\tilde\phi=0)$, $W_1$ $(\tilde\theta=0)$, $W_2$ $(d\tilde\phi={6\over7}\,\tilde\theta\wedge\tilde\phi)$
  and $W=W_1\oplus W_2$.
 The  only restriction that we find on the
$Spin(7)$-structure of $B$ arising from supersymmetry is that it is
conformally  balanced, i.e.~$\tilde\theta=2d\tilde\Phi$. These
geometries are in the same conformal class as those of the  $W_1$
$Spin(7)$-structure. To see this,
 observe that under the conformal transformation
$d\tilde s^2_{\Omega}= e^{2\Omega} d\tilde s^2$, the four-forms changes as $\tilde\phi^{\Omega}= e^{4\Omega} \tilde\phi$. Then
the Lee form of $(d\tilde s^2_{\Omega}, \phi^{\Omega})$ can be written in terms of the Lee form
of $(d\tilde s^2, \tilde\phi)$ as $\tilde\theta^{\Omega}=\tilde\theta+{14\over3} d\Omega$.
 Thus
  $\tilde\theta^{\Omega}=0$  for $\Omega=-{3\over7}\tilde\Phi$ and so $(d\tilde s^2_{\Omega}, \tilde\phi^{\Omega})$ ,
is in the $W_1$ class.

\subsection{Field equations}

It is straightforward to derive the  field
equations that follow as the integrability conditions of the
Killing spinor equations. In this way, we find the minimal set of
field equations that need to be solved in addition to solving the
Killing spinor equations. For the case at hand, we find that the integrability
conditions of the Killing spinor equations give
\bea
&&E_{++}=E_{+\a}=0~,
\cr
&&E_{-+}=-\frac{1}{2}e^{2\Phi}LH_{-+}~,
\cr
&&E_{-\a}=-\frac{1}{2}e^{2\Phi}LH_{-\a}+\frac{1}{2}BH_{-\a\g}{}^\g
-\frac{1}{6}\ep_{\a}{}^{\bar\g_1\bar\g_2\bar\g_3}BH_{-\bar\g_1\bar\g_2\bar\g_3}~,
\cr
&&E_{\a\b}=-\frac{1}{12}\ep_{\a}{}^{\bar\g_1\bar\g_2\bar\g_3}BH_{\b\bar\g_1\bar\g_2\bar\g_3}
-\frac{1}{12}\ep_{\b}{}^{\bar\g_1\bar\g_2\bar\g_3}BH_{\a\bar\g_1\bar\g_2\bar\g_3}~,
\cr
&&E_{\a\bar\b}=-\frac{1}{2}BH_{\a\bar\b\g}{}^\g-\frac{1}{12}\ep_{\a}{}^{\bar\g_1\bar\g_2\bar\g_3}BH_{\bar\b\bar\g_1\bar\g_2\bar\g_3}
-\frac{1}{12}\ep_{\bar\b}{}^{\g_1\g_2\g_3}BH_{\a\g_1\g_2\g_3}~,
\cr
&&LH_{+\a}=0~,
\cr
&&e^{2\Phi}LH_{\a_1\a_2}=-\frac{1}{2}\ep_{\a_1\a_2}{}^{\bar\b_1\bar\b_2}BH_{-+\bar\b_1\bar\b_2}~,
\cr
&&e^{2\Phi}LH_{\a\bar\b}=-BH_{-+\a\bar\b}+\frac{1}{6}\ep_{\a}{}^{\bar\g_1\bar\g_2\bar\g_3}BH_{\bar\b\bar\g_1\bar\g_2\bar\g_3}
-\frac{1}{6}\ep_{\bar\b}{}^{\g_1\g_2\g_3}BH_{\a\g_1\g_2\g_3}~,
\cr
&&LF_+=0~,
\cr
&&e^{2\Phi}LF_{\a}=-BF_{-+\a}+BF_{\a\g}{}^\g-\frac{1}{3}\ep_{\a}{}^{\bar\g_1\bar\g_2\bar\g_3}BF_{\bar\g_1\bar\g_2\bar\g_3}~,
\cr
&&L\Phi=\frac{1}{2}e^{2\Phi}LH_{-+}+\frac{1}{4}BH_{\g}{}^\g{}_\d{}^\d
+\frac{1}{24}\ep^{\bar\g_1\cdots\bar\g_4}BH_{\bar\g_1\cdots\bar\g_4}+\frac{1}{24}\ep^{\g_1\cdots\g_4}BH_{\g_1\cdots\g_4}~.
\eea
In addition, the Bianchi identities satisfy
\bea
BF_{+\a}{}^\a=0~,~~~
BF_{+\a_1\a_2}=\frac{1}{2}\ep_{\a_1\a_2}{}^{\bar\b_1\bar\b_2}BF_{+\bar\b_1
\bar\b_2}
\eea
and
\bea
&&BH_{+\a_1\a_2\a_3}=BH_{+\a_1\a_2\bar\b}=BH_{-+\g}{}^\g=0~,
~~~BH_{-+\a_1\a_2}=\frac{1}{2}\ep_{\a_1\a_2}{}^{\bar\b_1\bar\b_2}BH_{-+\bar\b_1
\bar\b_2}~,
\cr
&&BH_{\a_1\a_2\g}{}^\g=\frac{1}{2}\ep_{\a_1\a_2}{}^{\bar\b_1\bar\b_2}BH_{\bar\b_1
\bar\b_2\g}{}^\g~,
~~~\ep^{\bar\g_1\cdots\bar\g_4}BH_{\bar\g_1\cdots\bar\g_4}-\ep^{\g_1\cdots\g_4}BH_{\g_1\cdots\g_4}=0~.
\eea
It is significant to see that the Bianchi identities are restricted.
This effects the consistency of the theory when the heterotic
anomaly and the higher order corrections are considered. However, if
one works at the lowest order, one can impose the Bianchi
identities, $BH=BF=0$. In such a case, all field equations are
implied provided that in addition one imposes\footnote{The set of
field equations that should be imposed in addition to the Killing
spinor equations is not uniquely defined.}  $E_{--}=0$, $LH_{-A}=0$
and $LF_-=0$.

\newsection{ $N=2$ backgrounds with $SU(4)\ltimes \bbb{R}^8$ invariant spinors}

\subsection{Supersymmetry conditions} \la{sufsc}

We have shown in section \ref{stabgroups}\, that the $SU(4)\ltimes
\bR^8$-invariant  Killing spinors  can be written as
\bea
\epsilon_1&=&f (1+e_{1234})~,
\cr
\epsilon_2&=& g_1 (1+e_{1234})+i g_2 (1-e_{1234})~.
\eea
The Killing spinors equations for the first spinor have been investigated in the
previous section. The gravitino Killing spinor equation for the second spinor
can be written as
\bea
g_2^{-1} g_1 \partial_A \log (g_1 f^{-1}) (1+e_{1234})
+ i \partial_A\log g_2 (1-e_{1234})+ i\hat\nabla_A (1-e_{1234})=0~.
\eea
This equation can  be expanded in the basis given in (\ref{hbasis}). In particular, the components along the
 $1$ and $e_{1234}$ directions
are
\bea
g_2^{-1} g_1 \partial_A \log (g_1 f^{-1})+i \partial_A\log g_2
+{i\over 2} \hat\Omega_{A,\a}{}^\a+{i\over2} \hat\Omega_{A,-+}=0~,
\cr
g_2^{-1} g_1 \partial_A \log (g_1 f^{-1})-i \partial_A\log g_2 +
{i\over 2} \hat\Omega_{A,\a}{}^\a-{i\over2} \hat\Omega_{A,-+}=0~.
\eea
These in turn imply that
\bea
\partial_A\log(g_1 f^{-1})&=&0~,
\cr
\partial_A\log(g_2 f^{-1})&=&0~.
\la{gofgtf}
\eea
To derive  the latter, we have also used the equation that we have obtained for $f$
in the $N=1$ case. Since the Killing spinors
are specified up to a constant scale, they can be written as
\bea
\epsilon_1&=& f(1+e_{1234})
\cr
\epsilon_2&=& f [\cos\varphi (1+e_{1234})+ i\sin\varphi (1-e_{1234})]~,
\la{kspinconst}
\eea
where $\varphi$ is a constant angle.
The spinors $\epsilon_1$ and $\epsilon_2$ must be linearly independent and so
the angle $\varphi$ should  satisfy $\sin\varphi\not=0$.
The remaining conditions for the second Killing spinor  are as those we have derived  for the $N=1$ case with
the difference that  the terms
proportional to the Levi-Civita tensor epsilon  have an additional
relative minus sign. Combining, the conditions we have derived for the $\epsilon_1$
Killing spinors with those of the $\epsilon_2$ Killing spinors, we find  that the independent conditions
associated with the gravitino Killing spinor equation are
\bea
\partial_A \log f+{1\over2} \hat\Omega_{A,-+}=0~,
\la{bone}
\eea
\bea
\hat\Omega_{A, \bar\a\bar\b}=\hat\Omega_{A,\a}{}^\a=0~,
\la{btwo}
\eea
\bea
\hat\Omega_{A,+\bar\a}=\hat\Omega_{A,+\a}=0~.
\la{bthree}
\eea
It remains to find the conditions that arise from the dilatino Killing spinor equation.
We have already computed the dilatino Killing spinor equation on the spinor $\epsilon_1$ in the previous section.
So it remains to
find the conditions for $\epsilon_2$.
It is straightforward to observe using the results we have derived for the dilatino Killing spinor
equation of $\epsilon_1$ that it suffices to compute the dilatino Killing spinor equation on
$1-e_{1234}$. In turn the conditions that arise can be easily
read from those on $\epsilon_1$. The only  difference is a relative minus sign for the
terms proportional to the Levi-Civita tensor epsilon.  Combining the conditions
for the dilatino Killing spinor equation for both $\epsilon_1$ and $\epsilon_2$ spinors,
we find that
\bea
\partial_{\bar\a}\Phi-{1\over2} H_{\bar\a\b}{}^\b-{1\over2} H_{-+\bar\a}=0~,
\la{hbone}
\eea
\bea
H_{\bar\b_1\bar\b_2\bar\b_3}=0~,
\la{hbtwo}
\eea
\bea
\partial_+\Phi=0~,
\la{hbthree}
\eea
\bea
H_{+\a}{}^\a=0~, ~~~~~~H_{+\bar\a_1\bar\a_2}=0~.
\la{hbfour}
\eea
This concludes the analysis of the Killing spinor equations.

\subsection{Geometry}

 \subsubsection{Holonomy of $\hat\nabla$ connection and supersymmetry}

Applying the general arguments presented in \ref{holspin} to this case, one expects that  the gravitino Killing spinor
equation implies that the holonomy of the
 $\hat\nabla$ connection is
contained in $SU(4)\ltimes \bR^8$. This can be explicitly  seen  in the gauge $f=1$. This gauge can be attained
by using the $Spin(9,1)$ gauge transformation  $e^{b \Gamma^{05}}$ for $b=\log|f|$ as in the
$Spin(7)\ltimes \bR^8$ case that we have already investigated.
 In the gauge $f=1$, one has
\bea
\hat\Omega_{A,+-}=0~,
\eea
which together with (\ref{bthree}) imply that all the components of $\hat\Omega_{A,+B}=0$.
It is then easy to see that the remaining components of the connection one-form, $\hat\Omega=\hat\Omega_{A} e^A$,
take values in $su(4)\coplus\bR^8$. In the presence of fluxes, the Levi-Civita connection of these  backgrounds does not
have $SU(4)\ltimes \bR^8$ holonomy.

We have seen above that we can choose $f=1$. In addition, the angle that the  Killing spinors (\ref{kspinconst}) depend on can be
eliminated with a constant $GL(2,\bR)$ transformation. So the Killing spinors can be written as
$\epsilon_1=1+e_{1234}$ and
$\epsilon_2=i (1-e_{1234})$. This is in agreement with the general arguments we have
presented in  section \ref{holspin} that in the heterotic supergravity
 the Killing spinors can always be chosen to be constant.

A converse statement is also valid. If ${\rm
hol}(\hat\nabla)\subseteq SU(4)\ltimes \bR^8$, there are  spinors
$\epsilon_1, \e_2$ which are parallel with respect to $\hat\nabla$
and so they satisfy the gravitino Killing spinor equation. Thus the
existence of a solution for the gravitino Killing spinor equation
can be entirely characterized by the holonomy of $\hat\nabla$.

To investigate further the geometry of spacetime, it is convenient to introduce
the $\hat\nabla$-parallel forms associated with the parallel spinor bilinears. It turns out that
most of the fluxes and geometry can be expressed in terms of these bilinears.

\subsubsection{Geometry and spacetime forms bilinears }

The $\hat\nabla$-parallel forms associated with the spinor pair $(\epsilon_1, \epsilon_1)$ have already
been computed and can be found in the previous section. To compute the forms associated
with  the spinor pairs $(\epsilon_2, \epsilon_2)$ and $(\epsilon_1, \epsilon_2)$, we write the metric
as in  (\ref{metrherm}), i.e.
\bea
ds^2=2 \re^+ \re^-+ 2\delta_{\a\bar\b} \re^\a \re^{\bar\b}~.
\la{metrhermsu}
\eea
 Then,
after a normalization of the spinors (\ref{kspinconst})  with $1/\sqrt { 2}$,
we find
the one-forms\footnote{We have made an additional normalization of the spinor bilinears with a factor of $\sqrt{2}$.}
\bea
\kappa(\epsilon_2, \epsilon_2)&=&- f^2 \re^-
\cr
\kappa(\epsilon_1, \epsilon_2)&=&- f^2 \cos\varphi\, \re^-~,
\eea
a three-form
\bea
\xi(\epsilon_1, \epsilon_2)=-f^2\,\sin\varphi\, \re^-\wedge \omega~,
\eea
and two five-forms
\bea
\tau(\epsilon_2, \epsilon_2)&=& -f^2\, \re^-\wedge
 [{\rm Re}\,( e^{2i\varphi} \chi)-{1\over2}
\omega\wedge\omega]
\cr
\tau(\epsilon_1, \epsilon_2)&=&- f^2\, \re^-\wedge {\rm Re}\,[ e^{i\varphi} (\chi-{1\over2}
\omega\wedge\omega)]~,
\eea
where $\omega=-i \delta_{\a\bar\b} \re^\a \wedge \re^{\bar\b}$ and
$\chi=4\, \re^1\wedge \re^2\wedge \re^3\wedge \re^4$. The above
forms can be simplified in the gauge $f=1$, $\cos\varphi=0$,
$\sin\varphi=1$. It can be easily seen that if $\epsilon_1$ and
$\epsilon_2$ are linearly independent, i.e.~$\sin\varphi\not=0$, the
ring of spacetime form bilinears is generated by $\kappa=f^2 \re^-$,
$\xi=\kappa\wedge\omega$, $\tau_1=\kappa\wedge\omega\wedge\omega$
 and $\tau_2=\kappa\wedge\chi$. This ring is nilpotent as in the $Spin(7)\ltimes\bR^8$ case and
\bea
\hat\nabla\kappa=\hat\nabla\xi=\hat\nabla\tau_1=\hat\nabla\tau_2=0~,
\eea
i.e.~$\kappa$, $\xi$, $\tau_1$ and $\tau_2$ are parallel with
respect to the connection $\hat\nabla$. As we have already
mentioned, the condition $\hat\nabla \kappa=0$ implies that the
one-form $\kappa$ is associated to a {\it null Killing vector field}
$X$ and $d\kappa-i_X H=0$. The condition (\ref{hbfour}) which arises
from the dilatino Killing spinor equation   gives that the two-form
$i_XH$ takes values in $\mathfrak{su}(4)\coplus \bR^8$. This in turn
implies that $X$ preserves the $SU(4)\ltimes \bR^8$ structure, i.e.
\bea
{\cal L}_X\xi=0~,~~~{\cal L}_X\tau_1=0~,~~~{\cal L}_X\tau_2=0~.
\eea
In addition (\ref{hbthree}) implies that the dilaton is invariant under the diffeomorphisms
generated by $X$.

\subsubsection{Solution of the Killing spinor equations}

The conditions arising from the parallel transport equation imply that $\hat\nabla$ has holonomy
contained in  $SU(4)\ltimes \bR^8$.
The decomposition of the fluxes in $SU(4)\ltimes \bR^8$ representations is manifest in this case.
Nevertheless observe that under $SU(4)$ the space of two-forms decomposes as $\Lambda^2(\bR^8)\otimes \bC=
\Lambda^{2,0}_{\bf 6}\oplus
\Lambda^{0,2}_{\bf 6}\oplus \Lambda^{1,1}_{\bf 1}\oplus \Lambda^{1,1}_{\bf 15}$ and the space of three-forms
decomposes as $\Lambda^3(\bR^8)\otimes \bC=\Lambda^{3,0}_{\bf 4}\oplus \Lambda^{0,3}_{\bf 4}\oplus \Lambda^{2,1}_{\bf 20}
\oplus \Lambda^{1,2}_{\bf 20}\oplus \Lambda^{2,1}_{\bf 4}\oplus \Lambda^{1,2}_{\bf 4}$.
Using these, in the gauge $f=1$, the conditions that arise from gravitino Killing spinor equations can be written as
\bea
\hat\Omega_{A,+B}=0~,~~~\hat\Omega^{2,0}_{A, ij}=\hat\Omega^{0,2}_{A, ij}=0~,~~~\hat\Omega^{\bf 1}_{A,ij}=0
\la{gsu4}
\eea
and, similarly,  the conditions that arise from the dilatino Killing spinor equations can be written as
\bea
&&\partial_+\Phi=0~,~~~ 2\partial_i\Phi-\theta_i-H_{-+i}=0~,~~~H_{ijk}^{3,0}=H_{ijk}^{0,3}=0~,
\cr
&&H^{\bf 1}_{+ij}=0~, ~~~~~~H^{2,0}_{+ij}=H^{0,2}_{+ij}=0~,
\la{dsu4}
\eea
where the restriction to representations of $SU(4)$ is referred to the $i,j,k$ indices, $\theta$
is the Lee form
\bea
\theta_i=-\nabla^k\omega_{kj} I^j{}_i~,
\la{leesu4}
\eea
and the endomorphism $I$ is defined by $\omega_{ij}=g_{ik} I^k{}_j$.
To rewrite the conditions that arise from the dilatino Killing spinor equation in terms of the Lee form,
we have used $\hat\nabla_i\omega_{jk}=0$.

The conditions (\ref{gsu4}) and (\ref{dsu4}) can be solved to express the fluxes in terms of the geometry.
As we have mentioned already, the first condition in (\ref{gsu4}) implies that $i_X H=d\kappa$. The remaining
conditions imply that $\hat\nabla_A\omega_{ij}=0$ and $\hat\nabla_A\chi_{ijkl}=0$. The former condition implies that
\bea
\hat\nabla_-\omega_{ij}=0~,~~~ \hat\nabla_i\omega_{jk}=0~.
\eea
These two  equations can be solved using, $H_{ijk}^{3,0}=H_{ijk}^{0,3}=0$,  to give
\bea
&&H_{-ij}-H_{-kl} I^k{}_i I^l{}_j=- 2 I^m{}_i\nabla_-\omega_{mj}~,
\cr
&&H_{ijk}=-3 I^m{}_{[i} (\nabla_{j}\omega_{k]m}+\nabla_{|m|}\omega_{jk]}-\nabla_{k}\omega_{j]m})~.
\la{solsu4}
\eea
In addition $\hat\nabla_-\chi=0$ gives
\bea
H_{-\a}{}^\a={1\over 8\cdot 4!} \bar\chi^{ijkl} \nabla_-\chi_{ijkl}~.
\la{hmtrsu4}
\eea
Therefore all the fluxes apart from $H_{-ij}^{\bf 15}$ are determined in terms of the geometry and the
form Killing spinor bilinears. Of course the remaining conditions impose additional
restrictions on the metric and torsion. In particular, the component of $H$ in $\Lambda^{2,1}_{\bf 4}$
is related to the Lee form
$\theta$ and so to the derivative of the dilaton.

Therefore, the metric and torsion can be written as
\bea
ds^2&=&2  \re^- \re^++ \delta_{ij} e^i e^j~,
\cr
H&=& \re^+\wedge d\kappa-{1\over2} I^m{}_i\nabla_-\omega_{mj}\,
\re^-\wedge e^{i}\wedge e^{j}
-{1\over 64\cdot 4!} {\rm Im} (\bar\chi^{klmn} \nabla_-\chi_{klmn})\, \omega_{ij}\, \re^-\wedge e^i\wedge e^j
\cr
&&+{1\over2}
H^{\bf 15}_{-ij}\, \re^-\wedge
e^i\wedge e^i+ {1\over3!}   H_{ijk} e^i\wedge e^j\wedge e^k~,
\la{metrtorsolsu4}
\eea
where $H_{ijk}$ is given in the second equation of (\ref{solsu4}) and $d\kappa$ takes values in $\mfsu(4)\coplus \bR^8$.

\subsubsection{Local coordinates, distributions and a deformation family}

Using similar arguments to those we have presented for the $Spin(7)\ltimes \bR^8$ case and
introducing coordinates along the null Killing vector $X={\partial\over \partial u}$, we can write the spacetime
 metric in the gauge $f=1$ as
\bea
ds^2=2 (dv+m_I dy^I) (du+ V dv+  n_I dy^I )+ \gamma_{IJ} dy^I dy^J~,
\la{coormetrsu4}
\eea
where all the  components are functions of $v, y^I$ and
$i_X H=d\re^-=dm$.
In addition the second equation in  (\ref{hbfour}) implies that $dm$ takes values in $\mfsu(4)\coplus \bR^8$.
The coordinate $v$ can also be introduced as in the $Spin(7)\ltimes \bR^8$ case.  One can adapt a frame
to the above metric as in  (\ref{coorframe}).

Another aspect of the geometry of the spacetime is that it admits two integrable distributions
of codimension five. These are spanned by the the one forms $(\re^-, \re^{\a})$ and $(\re^-, \re^{\bar\a})$.
This can be seen by using the conditions that arise from the gravitino and dilatino Killing spinor equations.
This implies that the spacetime admits a ``Lorentzian'' holomorphic structure. In fact, most of the conditions that
arise from the dilatino Killing spinor equation are implied by the integrability of these distributions.

As in the case of a $Spin(7)\ltimes \bR^8$-invariant spinor, the
spacetime can be thought of as a two parameter deformation family of
an eight-dimensional manifold $B$. The metric (\ref{coormetrsu4})
can be written as the metric on the  family by introducing a
non-linear connection $A$ whose components are related to $m$ and
$n$ as in (\ref{famcoor}). It remains to investigate the geometry of
$B$. We adapt a frame $E^A$ to the metric of the family as in
(\ref{eframe}) and define the spacetime metric $d\tilde
s^2=ds^2|_B$, $\tilde H=H|_B$, $\tilde\omega=\omega|_B$ and
$\tilde\chi=\chi|_B$ on $B$. However, as in the $Spin(7)$ case, the
forms $\tilde\omega$ and $\tilde\chi$ are not always parallel with
respect to the connection $\hat{\tilde\nabla}$ of $B$ with torsion
$\tilde H$. Therefore although $B$ has an $SU(4)$-structure, it is
not compatible with the connection $\hat{\tilde\nabla}$.

There is a special case where the $SU(4)$-structure of $B$ is
compatible with the connection $\hat{\tilde\nabla}$. This appears
whenever the rotation of the the null parallel vector field $X$
vanishes, i.e.~when $d\kappa=d \re^-=0$. Using the relation
(\ref{twoframes})
 between the frames $(\re^-, \re^+, e^i)$ and
$(E^-, E^+, E^i)$,  and arguments similar to those of the $Spin(7)$
case, one can show that $B$ is a conformally balanced KT manifold
equipped with a compatible $SU(4)$ structure, i.e.~$B$ is
complex\footnote{ Note that there are two-dimensional sigma models
with extended world-volume supersymmetry and target spaces which are
almost complex manifolds \cite{wit}.}, $\tilde\omega$ and
$\tilde\chi$ define an $SU(4)$-structure and they are parallel with
respect to $\hat{\tilde\nabla}$,
i.e.~$\hat{\tilde\nabla}\tilde\omega=0$ and
$\hat{\tilde\nabla}\tilde\chi=0$, and that
\bea
\tilde\theta=2d\tilde\Phi~,
\la{conlee}
\eea
where $\tilde\theta=-\star(\star d\tilde\omega\wedge \tilde \omega)$
is the Lee form of $B$ and $\star$ is the Hodge duality operator on
$B$ associated with the volume form $d{\rm vol}(B)=\tilde
e^1\wedge\dots\wedge \tilde e^4 \wedge \tilde e^6\wedge\dots\wedge
\tilde e^9$. Note that $\tilde e^i=e^i|_B=E^i|_B$. One can show that
$B$ is a complex submanifold of the Lorentzian holomorphic manifold
$M$ by  using the integrable distributions $(\re^-, \re^{\a})$ and
$(\re^-, \re^{\bar\a})$ mentioned above. It turns out that  all
$2n$-dimensional manifolds with an $SU(n)$-structure and
skew-symmetric Nijenhuis tensor admit a compatible connection with
skew-symmetric torsion. In particular,
 the torsion of the eight-dimensional manifold $B$
is given as
\bea
\tilde H=-i_{\tilde I} d\tilde\omega=\star(d\tilde \omega\wedge \tilde\omega)
-{1\over2} \star(\tilde\theta\wedge \tilde\omega
\wedge \tilde\omega)~.
\eea
Examples of such manifolds have been given in \cite{strominger,
ivanovgpb, goldstein, salamon}. Of course for applications to the
heterotic string one has to impose the (generalized) Bianchi
identity for $H$.

The geometry of $B$ can also be described using G-structures. The
$SU(4)$-structures on an eight-dimensional manifold can be found by
decomposing $\tilde\nabla\tilde\omega$ and $\tilde\nabla\tilde\chi$
in irreducible $SU(4)$ representations. In the decomposition of
$\tilde\nabla\tilde\omega$ and $\tilde\nabla\tilde\chi$ five
irreducible $SU(4)$ representations  appear labelled by $W_1, \dots,
W_5$, so there are $2^5$ $SU(4)$-structures. One can also recover
these representations in the decomposition of $d\tilde\omega$ and
$d\tilde\chi$.  In particular, one can show that
$d\tilde\omega^{3,0}$ and $d\chi^{3,2}$ determine
$\tilde\nabla_\a\tilde\omega_{\b\g}$ and correspond to the $W_1$ and
$W_2$ classes respectively. The traceless part of
$d\tilde\omega^{2,1}$ is associated with the $W_3$ class and
determines the traceless part of of
 $\nabla_{\bar\a}\tilde\omega_{\b\g}$. Furthermore the trace part of  $d\tilde\omega^{2,1}$ and the trace part of
 $d\chi^{4,1}$ determine the trace parts  of
 $\tilde\nabla_{\bar\a}\tilde\omega_{\b\g}$ and $\tilde\nabla_{\bar\a}\tilde\chi_{\b_1\dots\b_4}$,
  respectively, and are associated with the $W_4$ and
 $W_5$ classes. The classes $W_4$ and $W_5$ are characterized by
 the Lee forms $\tilde \theta_{\tilde \omega}$ and
$\tilde \theta_{{\rm Re}\tilde \chi}$
 of $\tilde\omega$ and ${\rm Re}(\tilde \chi)$,
respectively. The Lee form $\tilde \theta_{\tilde \omega}$ has been given below (\ref{conlee}),
$\tilde \theta_{\tilde \omega}=\tilde \theta$, and the Lee form of ${\rm Re}\tilde\chi$
is defined as $\tilde \theta_{{\rm Re}\tilde \chi}=-{1\over4}\star(\star d{\rm Re}
\tilde\chi\wedge {\rm Re} \tilde\chi)$.
The remaining components of $\tilde\nabla\tilde\omega$ and $\tilde\nabla\tilde\chi$ vanish.
 The above is a  generalization of
the results of  \cite{salamonb} for the $SU(3)$ case, see also \cite{cabrera}. The further generalization
to all $SU(n)$-structures is straightforward. Returning to the geometry of the deformed
manifold $B$, since $B$ is complex, $W_1=W_2=0$.
In addition, one can show that
\bea
\tilde \theta_{\tilde \omega}=\tilde \theta_{{\rm Re}\tilde \chi}= 2d\tilde \Phi~.
\eea
This condition is reminiscent to a condition found  in \cite{lust} in the context of $\bR^{3,1}\times X_6$
heterotic string backgrounds, where $X_6$ has an $SU(3)$-structure.

\subsection{Field equations}

As was explained for the $N=1$ case,  it is straightforward to
derive the field equations that are implied from the
integrability conditions of the Killing spinor equations. In particular,
 we find for the case with $SU(4)\ltimes \bR^8$ invariant spinors that
\bea
&&E_{++}=E_{+\a}=E_{\a\b}=0~,
\cr
&&E_{-+}=-\frac{1}{2}e^{2\Phi}LH_{-+}~,
\cr
&&E_{-\a}=-\frac{1}{2}e^{2\Phi}LH_{-\a}+\frac{1}{2}BH_{-\a\g}{}^\g~,
\cr
&&E_{\a\bar\b}=-\frac{1}{2}BH_{\a\bar\b\g}{}^\g~,
\cr
&&LH_{+\a}=LH_{\a_1\a_2}=0~,
\cr
&&e^{2\Phi}LH_{\a\bar\b}=-BH_{-+\a\bar\b}~,
\cr
&&LF_+=0~,
\cr
&&e^{2\Phi}LF_{\a}=-BF_{-+\a}+BF_{\a\g}{}^\g~,
\cr
&&L\Phi=\frac{1}{2}e^{2\Phi}LH_{-+}+\frac{1}{4}BH_{\g}{}^\g{}_\d{}^\d~.
\eea
In addition, the Bianchi identities satisfy
\be
BF_{+\a}{}^\a=BF_{+\a_1\a_2}=BF_{\a_1\a_2\a_3}=0~,
\ee
and
\bea
&BH_{-\a_1\a_2\a_3}=BH_{+\a_1\a_2\a_3}=BH_{+\a_1\a_2\bar\b}=BH_{-+\a_1\a_2}=BH_{-+\g}{}^\g=0~,
\cr
&BH_{\a_1\a_2\a_3\a_4}=BH_{\a_1\a_2\a_3\bar\b}=BH_{\a_1\a_2\bar\b_1\bar\b_2}=0~.
\eea
As can be seen from the conditions above, if we choose to impose the
Bianchi identities $BF=BH=0$, the only field equations that
remain to be solved are $E_{--}=0$, $LH_{-A}=0$ and $LF_-=0$.

\newsection{$N=2$ with $G_2$ invariant spinors}

\subsection{Supersymmetry conditions}

The two Killing spinors can be chosen as, see section \ref{stabgroups},
\bea
\epsilon_1&=&f (1+e_{1234})~,~~~
\epsilon_2= g (e_{15}+ e_{2345})~.
\la{kspingtwo}
\eea
We have already derived the conditions required for $\epsilon_1$ to
 be a Killing spinor when we investigated the backgrounds with one
 supersymmetry. Therefore it remains to
derive the conditions for $\epsilon_2$ to be a Killing spinor.
After some computation, the gravitino Killing spinor  equation gives
\bea
 \hat \Omega_{A,-1}=0~,~~~~~~
\hat\Omega_{A,-\bar n}=0~,
\eea
\bea
\partial_A \log g-{1\over2} \hat\Omega_{A,1\bar1}+{1\over2}
\hat\Omega_{A,n}{}^n-{1\over2}\hat\Omega_{A, -+}=0~,
\eea
\bea
\hat\Omega_{A,\bar n 1}-{1\over2}\hat\Omega_{A, pm} \epsilon^{pm}{}_{\bar n}=0~,
\eea
\bea
\partial_A\log g+{1\over2} \hat\Omega_{A, 1\bar 1}-{1\over2}
 \hat\Omega_{A, n}{}^n-{1\over2}\hat\Omega_{A, -+}=0~,
\eea
where $m, n, p, q, \dots=2,3,4$ and $\epsilon_{mnp}=\epsilon_{1mnp}$.
In addition, the dilatino Killing spinor equation gives
\bea
2\partial_-\Phi+H_{-1\bar 1}-H_{-n}{}^n=0~,
\eea
\bea
2\partial_-\Phi-H_{-1\bar 1}+H_{-n}{}^n=0~,
\eea
\bea
2H_{-1\bar m}+ H_{-np} \epsilon^{np}{}_{\bar m}=0~,
\eea
\bea
-2\partial_1\Phi+ H_{1n}{}^n- H_{-+1}-{1\over3} H_{npm} \epsilon^{npm}=0~,
\eea
\bea
\partial_{\bar n}\Phi+{1\over2} H_{1\bar 1\bar n}-{1\over2} H_{\bar n p}{}^p
+{1\over2} H_{-+\bar n}
-{1\over2} H_{pm \bar 1} \epsilon^{pm}{}_{\bar n}=0~.
\eea
Comparing the above equations with those derived for the $\epsilon_1$ Killing spinor,
we find that the parallel transport equation gives

\be
 \partial_A\log f+{1\over2}\hat\Omega_{A,-+}=0~,
\la{gtparaone}
\ee
\bea
 \partial_A\log fg=0~,
 \la{gtparatwo}
\eea
\bea
 \hat\Omega_{A,1}{}^1=0~,
 \la{gtparathree}
\eea
\bea
 \hat\Omega_{A,n}{}^n=0~,
 \la{gtparafour}
\eea
\bea
 \hat\Omega_{A,+\alpha}=\hat\Omega_{A,-\alpha}=0~,~~~\a,\b=1,2,3,4~,
 \la{gtparafive}
\eea
\bea
 \hat\Omega_{A,\bar\alpha\bar\beta}={1\over2}\hat\Omega_{A,\gamma\delta}\epsilon^{\gamma\delta}{}_{\bar\alpha\bar\beta}~,
 \la{gtparasix}
\eea
\be
 \hat\Omega_{A,1\bar n}=-\hat\Omega_{A,\bar1\bar n}~,
\la{gtparaseven}
\ee

and the dilatino Killing spinor equation gives

\bea
 \partial_+\Phi=\partial_-\Phi=(\partial_1+\partial_{\bar 1})\Phi=0~,
 \la{gtdone}
\eea
\bea
 \partial_{\bar 1}\Phi=-{1\over12}H_{\bar n\bar p\bar m}\epsilon^{\bar n\bar p\bar m}
 +{1\over12}H_{npm}\epsilon^{npm}~,
  \la{gtdtwo}
\eea
\bea
 \partial_{\bar n}\Phi=-{1\over2}H_{\bar n}{}^p{}_p
 +{1\over4}H_{\bar1 pm}\epsilon^{pm}{}_{\bar n}-{1\over4}H_{1pm}\epsilon^{pm}{}_{\bar n}~,
  \la{gtdthree}
\eea
\bea
 H_{+\alpha}{}^\alpha=0~,
  \la{gtdfour}
\eea
\bea
 H_{+\bar\alpha\bar\beta}={1\over2}H_{+\gamma\delta}\epsilon^{\gamma\delta}{}_{\bar\alpha\bar\beta}~,
  \la{gtdfive}
\eea
\bea
 H_{-1\bar1}=H_{-n}{}^n~,
  \la{gtdsix}
\eea
\bea
 H_{-1\bar n}=-{1\over2}H_{-pm}\epsilon^{pm}{}_{\bar n}~,
  \la{gtdseven}
\eea
\bea
 H_{-+\bar1}=H_{\bar1}{}^n{}_n-{1\over6}H_{npm}\epsilon^{npm}-{1\over6}
 H_{\bar n\bar p\bar m}\epsilon^{\bar n\bar p\bar m}~,
  \la{gtdeight}
\eea
\bea
 H_{-+\bar n}={1\over2}H_{1pm}\epsilon^{pm}{}_{\bar n}+{1\over2}H_{\bar1pm}\epsilon^{pm}{}_{\bar n}-H_{\bar n1\bar 1}~.
  \la{gtdnine}
\eea
This concludes the analysis of the Killing spinor equations. In the
remainder of the section, we shall investigate the geometry of the
backgrounds with $G_2$ invariant spinors.

\subsubsection{Holonomy of $\hat\nabla$ connection}

The gravitino Killing spinor equation implies that the holonomy of
the connection $\hat\nabla$ is contained in $G_2$, which is the
stability subgroup of the spinors (\ref{kspingtwo})
 in $Spin(9,1)$.
One can also see this explicitly. This is easily done in the gauge $f=1$. As in the
previous cases we have already investigated, this gauge can be attained
by the $Spin(9,1)$ gauge transformation  $e^{b \Gamma^{05}}$ for $b=\log|f|$.
Then (\ref{gtparatwo}) implies that $g$ is also constant and so it can be chosen as $g=1$.
 In the gauge $f=g=1$, one has
\bea
\hat\Omega_{A,+-}=0~,
\eea
which together with (\ref{gtparafive}) imply that all the components
of $\hat\Omega_{A,+B}=0$. It is then easy to see that
(\ref{gtparatwo})-(\ref{gtparaseven}) imply that
 the remaining components of the connection one-form, $\hat\Omega=\hat\Omega_{A} e^A$,
take values in $\mathfrak{g}_2$. The Levi-Civita connection does not
have $G_2$ holonomy.
The  analysis of the geometry of supersymmetric backgrounds with $G_2$ invariant spinors
simplifies in the gauge $f=g=1$.

Conversely, if the connection $\hat\nabla$ has holonomy contained in $G_2$, then
there are spinors $\epsilon_1=1+e_{1234}$ and $\epsilon_2=
(e_{15}+e_{2345})$, up to a $Spin(9,1)$ gauge transformation,
which are parallel with respect to $\hat\nabla$. Therefore the holonomy of
 $\hat\nabla$ completely characterizes the
solution of the gravitino Killing spinor equation.

\subsubsection{Spacetime  form bilinears  }

To proceed further in the investigation of the geometry, we compute the spacetime forms associated
with the Killing spinor bilinears.
The spacetime forms of $\epsilon_1$ have already been described
in the previous sections. It remains to compute the forms associated
with the spinor pairs $(\epsilon_2, \epsilon_2)$ and $(\epsilon_1, \epsilon_2)$.
In particular after an additional normalization of the spinors, we find the one-forms

\bea
\kappa(\epsilon_1, \epsilon_2)=-e^{1}~,~~~\kappa(\epsilon_2, \epsilon_2)=e^0+e^5~,
\eea
the three-form
\bea
\xi(\epsilon_1, \epsilon_2)={\rm Re}\hat\chi+ e^6\wedge \hat\omega- e^0\wedge e^1\wedge e^5~,
\eea
and the five-forms
\bea
\tau(\epsilon_1, \epsilon_2)=-{\rm Re}\hat\chi\wedge e^0\wedge e^5+ {\rm Im}\hat\chi\wedge e^1\wedge e^6
+{1\over2} e^1\wedge\hat\omega\wedge\hat\omega- \hat\omega\wedge e^0\wedge e^5 \wedge e^6
\cr
\tau(\epsilon_2, \epsilon_2)=-(e^0+e^5)\wedge [ e^1\wedge {\rm Re}\hat\chi+  e^6\wedge {\rm Im}\hat\chi
+{1\over2}\hat\omega\wedge\hat\omega+ \hat\omega\wedge e^1\wedge e^6]~,
\eea
in the gauge $f=g=1$. Unlike the previous cases we have investigated, the ring of invariant forms
is not nilpotent. It is generated by the one-forms $\kappa=\re^-, \kappa'=\re^+$
and $\hat\kappa=e^1={1\over \sqrt{2}} (\re^1+\re^{\bar 1})$, the $G_2$ invariant form
\bea
\varphi={\rm Re}\hat\chi+ e^6\wedge \hat\omega~,
\eea
and its dual $\star\varphi$, where the Hodge operator is taken with respect to the
volume form  $e^2\wedge e^3\wedge e^4\wedge e^6\wedge e^7\wedge e^8\wedge e^9$.

The one-forms
 $\kappa=\re^-, \kappa'=\re^+$ and $\hat\kappa=e^1={1\over \sqrt{2}} (\re^1+\re^{\bar 1})$, in the gauge $f=g=1$,
 are associated
with the  Killing vector fields $X=\re_+,Y=\re_-$ and $Z=e_1$, respectively.
This follows from the conditions $\hat\Omega_{A,+B}=\hat\Omega_{A,-B}=\hat\Omega_{A, 1 B}+\hat\Omega_{A, \bar 1 B}=0$
which summarize (\ref{gtparaone}), (\ref{gtparafive}) and (\ref{gtparaseven}), in the gauge $f=g=1$,
 and the skew-symmetry of the torsion $H$.
 The commutators  of these Killing vector fields are
 \bea
 [X,Y]&=&-H^A{}_{+-} e_A
 \cr
 [X,Z]&=&-H^A{}_{+1} e_A
 \cr
 [Y,Z]&=&-H^A{}_{-1} e_A~.
 \la{commut}
 \eea
The components of the torsion which appear in (\ref{commut})  are not required to vanish by the Killing spinor equations.
So the
above commutators do not vanish and therefore the Killing vector fields do not necessarily commute. As we have shown in
\ref{parallelforms},
if two vector fields $X,Y$ are $\hat\nabla$-parallel  their commutator
 $[X,Y]$ is $\hat\nabla$-parallel as well.
So if the commutators of $X,Y,Z$ are independent vector fields, then the spacetime admits up to six parallel
vector fields not counting the further commutators that one can construct. So there is a large class of geometries
that can occur ranging from a spacetime with three commuting Killing vector fields $X,Y,Z$ to a spacetime that is
a Lorentzian Lie group of dimension ten equipped with a left-invariant metric $g$ and   a left-invariant
closed three form $H$. We shall not attempt to investigate the full range of possibilities. Instead, we shall focus
on the case for which the vector fields $X$, $Y$ and $Z$ span a Lie algebra under  the commutators (\ref{commut}).

\subsubsection{Backgrounds with three isometries and supersymmetry conditions}

Let $\mfh$ be the Lie algebra spanned by $X,Y$ and $Z$. Then  $[h,h]\subset h$ implies that
\bea
H_{-+ i}=H_{-1 i}=H_{+1i}=0~,~~~i,j,k,l=2,3,4,6,7,8,9~.
\la{closeh}
\eea
Therefore the structure constants of the Lie algebra are given by
the $H_{-+1}$ component of the torsion. Since $H$ is a three-form,
$\mfh$ can be either isomorphic to $\mfuo\oplus \mfuo\oplus \mfuo$,
if $H_{-+1}=0$, or isomorphic to $\mfsl$, if $H_{-+1}\not=0$. The
analysis can be done for both cases simultaneously.

First consider the consequences of (\ref{closeh}) on the gravitino Killing spinor equation. It is straightforward
to find that (\ref{gtparaone})-(\ref{gtparaseven}) can be rewritten as
\bea
\hat\Omega_{A,aB}=0~,~~~\hat\Omega^{\bf 7}_{A,ij}=0~,~~~a=-,+,1~,
\la{gkscloseh}
\eea
where we have used the decomposition of $\Lambda^2(\bR^7)= \Lambda^2_{\bf 7}\oplus \Lambda^2_{\bf 14}$ under
$G_2$,
\bea
\Lambda^2_{\bf 7}=\{\star(\star\varphi\wedge \a)|\, \a\in \Lambda^1(\bR^7)\}
\cr
\Lambda^2_{\bf 14}=\{\a\in \Lambda^2(\bR^7)|\, \star(\varphi\wedge
\a)=-\a\}
\eea
and $\Lambda^2_{\bf 14}$ can be identified with the Lie algebra $\mathfrak{g}_2$ of $G_2$.
 Similarly, using (\ref{closeh}), the conditions implied by the dilatino Killing spinor equation can be rewritten
as
\bea
\partial_+\Phi=\partial_-\Phi=\partial_1\Phi=0~,
\cr
i_X H^{\bf 7}_{ij}=i_Y H^{\bf 7}_{ij}=i_Z H^{\bf 7}_{ij}=0~,
\cr
\partial_i\Phi+{1\over12} H_{jkl}\, \star\varphi^{jkl}{}_i=0~,
\cr
H_{-+1}+{1\over6} H_{ijk} \varphi^{ijk}=0~.
\la{dkscloseh}
\eea
 The
second equation in (\ref{dkscloseh}) implies that $(i_XH)_{ij}$, $(i_YH)_{ij}$  and $(i_ZH)_{ij}$ take values
in $\mathfrak{g}_2$. This together with (\ref{closeh}) imply that $X,Y$ and $Z$ leave invariant the forms $\varphi$
and its dual $\star\varphi$, i.e.
\bea
{\cal L}_X\varphi={\cal L}_Y \varphi={\cal L}_Z\varphi=0~,
\eea
and similarly for $\star\varphi$.
The last equation in (\ref{dkscloseh}) implies that the structure constants of $\mfh$
can be identified with the singlet of $H$ under the $G_2$ decomposition
$\Lambda^3(\bR^7)=\Lambda^3_{\bf 1}\oplus\Lambda^3_{\bf 7}\oplus \Lambda^3_{\bf 27}$, where
\bea
&&\Lambda^3_{\bf 1}=\{r\,\varphi|\, r\in \bR\}~,
\cr
&&\Lambda^3_{\bf 7}=\{\star(\varphi\wedge\a)|\,\a\in \Lambda^1(\bR^7)\}~,
\cr
&&\Lambda^3_{\bf 27}=\{\a\in \Lambda^3(\bR^7)|\, \a\wedge\varphi=0,~\a\wedge\star\varphi=0\}=\{s\in S^2(\bR^7)|\, {\rm tr}( s)=0\}~.
\eea
In addition
the seven-dimensional component of $H$ in the above decomposition is identified with the
exterior derivative of the dilaton.

\subsubsection{The solution of the Killing spinor equations}\la{prinbundle}

The space of supersymmetric backgrounds with $G_2$ invariant spinors is (locally) a principal bundle $P$ equipped
with a connection $\l$. To see this, we again assume that the algebra $\mfh$ spanned by the vector fields\footnote{These
vector fields do not have fixed points because they are $\hat\nabla$-parallel and so they cannot vanish.} $X,Y$ and $Z$ closes under
Lie brackets and consider a Lie group $\cH$ with Lie algebra $\mfh$. Then the spacetime $M=P(\cH, B, \pi)$,
where the base space $B$ is the space of orbits of $\cH$ in $M$ and $\pi$ is the projection of $P$ onto $B$. It remains to determine
the connection $\lambda$. This is identified with the components of the frame $e$ along the $X,Y$ and $Z$ directions, i.e.
\bea
\lambda^a= e^a~.
\la{pconn}
\eea
One can immediately see that $\lambda$ satisfies the requirements of
a connection, i.e.~$\lambda^a(X_b)=e^a(X_b)=\delta^a{}_b$ where
$\{X_b, b=+,- 1\}=\{X,Y,Z\}$, and ${\cal L}_{X_b} \lambda^a=
H^a{}_{bc} \lambda^c$, where $H_{abc}$ are interpreted as the
structure constants of $\mfh$. The latter is the infinitesimal
expression of the requirement that $R_g^*\lambda=Ad_{g^{-1}}
\lambda$, $g\in \cH$, of a principal bundle connection, see
e.g.~\cite{nomizu}. Then the tangent bundle decomposes into the
vertical and horizonal subspaces, $TM=TP=T^vP\oplus T^hP$, where
$T^vP$ is spanned by the vector field $X,Y$ and $Z$ and $T^hP$ is
(locally) spanned  by the dual vector fields of the $e^i$ components
of the frame because $\lambda^a(e_i)=g^{MN} e^a{}_M e_{iN}=0$.

To determine the Cartan structure equations for this connection, we
use the conditions (\ref{dkscloseh}) to write  (\ref{gkscloseh}) as
\bea
&\Omega_{a,bc}-{1\over2}
H_{abc}=0~,~~~\Omega_{i,ab}=0~,~~~\Omega_{a,bi}=0~,~~~\Omega_{i,aj}^{\bf
7}=0~, ~~~\Omega_{(i,j)a}=0~,
\cr
&\Omega_{a,ij}^{\bf 7}=0~,~~~\hat\Omega_{k,ij}^{\bf 7}=0~,
\eea
where the restriction to the seven-dimensional representation is referred to the $i,j$ indices. These
in turn give rise to the torsion free conditions
\bea
de^a+\Omega_{b,}{}^a{}_c e^b\wedge e^c+\Omega_{i,}{}^a{}_j e^i\wedge e^j=0~,
\cr
de^i+\Omega_{j,}{}^i{}_k e^j\wedge e^k+\Omega_{a,}{}^i{}_j e^a\wedge
e^j+\Omega_{j,}{}^i{}_a e^j\wedge e^a=0~. \la{torfreeg2}
\eea
The first torsion free condition rewritten in terms of $H$  can be interpreted as the Cartan
structure equation for the connection $\lambda$. In particular,  we have
\bea
d\lambda^a-{1\over2} H^a{}_{bc}  \lambda^b\wedge \lambda^c-{1\over2} H^a{}_{ij} e^i\wedge e^j=0~.
\la{pcartan}
\eea
Since the curvature ${\cal F}$ of a principal bundle connection $\l$ is identified with the horizontal part of $d\l$, we find
that
\bea
{\cal F}^a= {1\over2} H^a{}_{ij} e^i\wedge e^j~.
\la{pcurv}
\eea
In addition the condition (\ref{dkscloseh}), part of which can also
be written as $H^{\bf 7}_{aij}=0$, implies that the curvature ${\cal
F}^a$ is that of a $\mathfrak{g}_2$ type of instanton on $B$ with
gauge Lie algebra  $\mfsl$ or $\mfuo\oplus \mfuo\oplus \mfuo$. In
terms of these principal bundle data, the metric $ds^2$ and torsion
$H$ of spacetime can be written as
\bea
ds^2&=& \eta_{ab} \lambda^a \lambda^b+ \pi^* d\tilde s^2
\cr
H&=&{1\over3} \eta_{ab} \lambda^a \wedge d\lambda^b+{2\over3} \eta_{ab} \lambda^a\wedge {\cal F}^b+ \pi^*\tilde H~,
\la{ph}
\eea
where $d\tilde s^2$ and  $\tilde H$ is a metric and a three-form on $B$ and horizontally lifted to $P$ with $\pi$, respectively.

It remains to determine the geometry of the base space $B$ of the
principal bundle. $B$ is equipped with a metric $d\tilde
s^2=\delta_{ij} e^i e^j|_B$ and a three-form $\tilde H$ which is the
horizontal part of $H$. Note that $d\tilde H\not=0$. In addition  it
is equipped with a $G_2$-invariant three-form $\tilde \varphi$ such
that $\varphi=\pi^*\tilde \varphi$. This is because $\varphi$ is
horizontal and ${\cal L}_a\varphi=0$. Furthermore
$\hat{\tilde\nabla}\tilde\varphi=0$ which follows from
$\hat\nabla\varphi=0$. Therefore $B$ is a Riemannian manifold
equipped with a metric connection with skew-symmetric torsion and
${\rm hol}(\hat{\tilde\nabla})\subseteq G_2$ and thus admits a
$G_2$-structure. It has been shown in \cite{stefang2} that any
seven-dimensional manifold with an integrable
 $G_2$-structure admits a unique
metric connection $\hat{\tilde\nabla}$ with torsion a three-form
\bea
\tilde H=-{1\over6} (d\tilde\varphi, \star \tilde\varphi)\, \tilde\varphi+ \star d\tilde\varphi-\star (\tilde\theta\wedge\tilde\varphi)
\eea
such that ${\rm hol}({\hat{\tilde\nabla}})\subseteq G_2$, where
\bea
\tilde\theta=-{1\over3} \star(\star d\tilde\varphi\wedge \tilde\varphi)
\eea
is the Lee-form and $d{\rm vol}(B)=e^2\wedge e^3\wedge e^4\wedge e^6\wedge e^7\wedge e^8\wedge e^9$.
An integrable $G_2$-structure satisfies $d\star\tilde \varphi=-\tilde\theta\wedge \star\tilde\varphi$.
In addition the third condition in (\ref{dkscloseh}) can be rewritten as
\bea
\tilde\theta=2d\tilde\Phi
\eea
and so $B$ is conformally balanced. If $\mfh=\mfuo\oplus \mfuo\oplus \mfuo$, then the last condition implies that
the singlet $\tilde H^{\bf 1}$ of $\tilde H$ in the decomposition
$\Lambda^3(\bR^7)=\Lambda^3_{\bf 1}\oplus\Lambda^3_{\bf 7}\oplus \Lambda^3_{\bf 27}$
vanishes, $\tilde H^{\bf 1}=0$. This implies that $d\tilde\varphi$ is orthogonal to $\star\tilde\varphi$.
These are precisely the manifolds with $G_2$-structures investigated in the context of supersymmetric backgrounds
in \cite{stefang2}. Moreover it can be shown that  these $G_2$-structures are
 conformally equivalent to  cocalibrated $G_2$-structures of pure $W_3$ type\footnote{The covariant derivative
 $\tilde\nabla\tilde\varphi$ can be decomposed into four irreducible $G_2$ representations $W_1, W_2, W_3$ and $W_4$ which
 are determined by $d\tilde\varphi$ and $d\star\tilde\varphi$ \cite{grayfern}, see also \cite{salamonb}. So there are sixteen
 $G_2$-structures on a seven-dimensional manifold.}. However, if $\mfh=\mfsl$, then $\tilde H^{\bf 1}$ is
 identified with the structure constants of $\mfsl$
and so $\tilde H^{\bf 1}\not=0$.

To summarize, the solution of the Killing spinor equations for the
backgrounds that we have investigated above can be described as
follows: The spacetime is (locally) a principal bundle $P(\cH, B,
\pi)$. The group $\cH$ of the fibre has Lie algebra either
$\mfuo\oplus \mfuo\oplus \mfuo$ or $\mfsl$, and $P$ is equipped with
a connection $\lambda$ whose curvature ${\cal F}$ is a
$\mathfrak{g}_2$ instanton. The base space $B$ is a
seven-dimensional manifold equipped with a metric connection with
skew-symmetric torsion $\hat{\tilde\nabla}$ and ${\rm
hol}(\hat{\tilde\nabla})\subseteq G_2$ and the associated $G_2$
structure is conformally balanced, i.e.~it satisfies the conditions
$d\star\tilde\varphi=-\tilde\theta\wedge \star\tilde\varphi$ and
$\tilde\theta=2 d\Phi$. The metric and torsion are given by
(\ref{ph}), and the dilaton $\Phi$ is a function of $B$. If in
addition $\cH$ is abelian, then $(d\varphi, \star\varphi)=0$.

Using the description of  the geometry of spacetime in terms of principal bundle data,
we can write the exterior derivative of the torsion
\bea
dH=\eta_{ab}\, {\cal F}^a\wedge {\cal F}^b+\pi^*\, d \tilde H~.
\la{pdh}
\eea
The first term in the right-hand-side of $dH$ can be recognized as a representative of the first Pontrjagin class
 of the principal bundle $P$. Therefore the non-horizontal part of $H$ is the form that trivializes the first Pontrjagin
 class of $P$ on the bundle space.
If one requires that $dH=0$, then the representative of the first
Pontrjagin class of $P$ should cancel against the contribution from
the base space $B$. Of course if $P$ is a globally defined principal
bundle over $B$, then the condition $dH=0$ implies that the first
Pontrjagin form is exact and therefore the first  Pontrjagin class
of the principal bundle should vanish. Observe that it is {\it not}
required that $d\tilde H=0$.

\subsection{Field equations}

\subsubsection{Integrability conditions}

We shall demonstrate that if the Bianchi identities of $H$ and $F$ are satisfied, then the Killing spinor
equations imply all the field
equations. To see this, one can show that the
 integrability conditions of the Killing spinor equations imply
\bea
&E_{-A}=E_{+A}=0 ~,\quad E_{n1}=-E_{n\bar
1}=-\frac{1}{2}BH_{n1m}{}^m~,
\cr
&E_{11}=-E_{1\bar 1}=\frac{1}{6}\ep^{\bar n\bar p\bar m}BH_{\bar
1\bar n\bar p\bar m}~,\quad E_{np}=-\frac{1}{2}\ep_{(n}{}^{\bar
m\bar q}BH_{p)1 \bar m\bar q}~,
\cr
&E_{n\bar p}=-\frac{1}{2}BH_{-+n \bar p}-\frac{1}{2}BH_{n\bar p
m}{}^m-\frac{1}{4}\ep_{n}{}^{\bar m\bar q}BH_{\bar p\bar m\bar q
1}+\frac{1}{4}\ep_{\bar p}{}^{mq}BH_{nmq1}~,
\cr
&LH_{-A}=LH_{+A}=LH_{1A}=0~,\quad
e^{2\Phi}LH_{np}=BH_{npm}{}^m-\ep_{np}{}^{\bar m}BH_{\bar m\bar1
q}{}^q~,
\cr
&e^{2\Phi}LH_{n\bar p}=BH_{-+n\bar p}-\frac{1}{2}\ep_{n}{}^{\bar
m\bar q}BH_{\bar p \bar m \bar q 1}-\frac{1}{2}\ep_{\bar
p}{}^{mq}BH_{n mq \bar 1}~,
\cr
&LF_{-}=LF_{+}=0~,\quad
e^{2\Phi}LF_{n}=BF_{-+n}+BF_{np}{}^p-BF_{n1\bar1}-\ep_{n}{}^{\bar p\bar q}BF_{1\bar p\bar q}~,
\cr
&e^{2\Phi}LF_{1}=BF_{-+\bar1}-BF_{\bar1n}{}^n-\frac{1}{3}\ep^{\bar n\bar m\bar p}BF_{\bar n\bar m\bar p}~,
\cr
&L\Phi=\frac{1}{4}BH_{p}{}^p{}_m{}^m+\frac{1}{3}\ep^{npm}BH_{1npm}
\eea
and
\bea
&BF_{-1\bar1}=BF_{-n}{}^n~,\quad BF_{+1\bar1}=-BF_{+n}{}^n~,
\cr
&BF_{-+1}-BF_{-+\bar1}-BF_{1n}{}^n-BF_{\bar1 n}{}^n=0~,
\cr
&BF_{-+1}-BF_{1n}{}^n+\frac{1}{6}\ep^{npm}BF_{npm}+\frac{1}{6}\ep^{\bar
n\bar p\bar m}BF_{\bar n\bar p\bar m}=0~,
\cr
&BF_{-n\bar1}=\frac{1}{2}\ep_{n}{}^{\bar p\bar m}BF_{-\bar p\bar
m}~,\quad BF_{+n1}=-\frac{1}{2}\ep_{n}{}^{\bar p\bar m}BF_{+\bar
p\bar m}~,
\cr
&BF_{-+n}-BF_{n1\bar 1}-\frac{1}{2}\ep_{n}{}^{\bar p\bar m}BF_{\bar
p\bar m\bar 1}-\frac{1}{2}\ep_{n}{}^{\bar p\bar m}BF_{\bar p\bar m
1}=0~,
\cr
&BH_{-ABC}=BH_{+ABC}=BH_{np1\bar 1}=BH_{1\bar1 n}{}^n=0~,
\cr
&\ep^{\bar n\bar p\bar m}BH_{\bar 1 \bar n\bar p\bar
m}=\ep^{npm}BH_{1npm}=-\ep^{\bar n\bar p\bar m}BH_{1 \bar n\bar
p\bar m}~,
\cr
&\frac{1}{6}\ep^{\bar p\bar m\bar q}BH_{n\bar p\bar m\bar
q}=BH_{n1p}{}^p=-BH_{n\bar1 p}{}^p~,
\cr
&BH_{-+n\bar p}+\frac{1}{4}\ep_{n}{}^{\bar m\bar q}BH_{\bar p\bar
1\bar m\bar q}+\frac{1}{4}\ep_{n}{}^{\bar m\bar q}BH_{\bar p 1\bar
m\bar q}-\frac{1}{4}\ep_{\bar p}{}^{mq}BH_{nmq\bar
1}-\frac{1}{4}\ep_{\bar p}{}^{mq}BH_{nmq1}=0~,
\cr
&BH_{-+n\bar p}+BH_{n\bar p 1\bar1}-\frac{1}{2}\ep_{\bar
p}{}^{mq}BH_{nmq\bar 1}-\frac{1}{2}\ep_{\bar p}{}^{mq}BH_{nmq1}=0~.
\eea
In the above conditions, we have not imposed  the Bianchi identity
$BH$ of $H$. In the order of $\alpha'$ that we are working $BH=dH=0$
and so there is no contribution from the Bianchi identities. But in
the next order up in $\alpha'$, the above integrability conditions
are believed to hold but $dH\not=0$. As a result some of the field
equations that are derived in the one-loop sigma model approximation
 can be expressed in terms of $dH$. This has been used in
\cite{tsimpis} to investigate heterotic backgrounds taking into account
the two-loop and higher order corrections to the field equations.

If we take $BH=BF=0$, the integrability conditions above imply all
field equations. In the absence of the gauge field $A$, the only
Bianchi identity that has to be imposed is that of $H$. This has
been computed in (\ref{pdh}). We shall explore this to give examples
of some supersymmetric backgrounds.

\subsubsection{Examples}

As an example, let us consider the case where $\mfh=\mfuo\oplus \mfuo\oplus \mfuo$. Then
\bea
\lambda^a= dx^a+ A^a
\eea
and so
\bea
ds^2&=&\eta_{ab} (dx^a+A^a) (dx^b+A^b)+ \delta_{ij} e^i e^j~,
\cr
H&=& \eta_{ab} (dx^a+A^a) \wedge dA^b+ \pi^*\tilde H~.
\eea
If one requires closure of $H$ and choose $\tilde H=-\eta_{ab} A^a\wedge dA^b+ H_B$, then
\bea
H=\eta_{ab}\, dx^a\wedge dA^b+ H_B~,
\eea
where $H_B$ is a three-form on $B$ such that $dH_B=0$.
Clearly $dH=0$. Within a brane interpretation of these solutions, the connection $A^0$ along the time direction is thought of
as rotation while the remaining connections are thought of as wrapping.

A special case of this example is whenever the only non-vanishing rotation and wrapping is a along a null direction. In this case,
the Chern-Simons
form contribution vanishes. Thus one can set $\tilde H=H_B$. Such kind of solutions have been consider before\footnote{Our current results
correct some of the fractions of supersymmetry that have appeared in \cite{gpapas}.} in \cite{gpapas}.
The metric and torsion are
\bea
ds^2&=&2 dv (du+A)+dx^2+ \delta_{ij} e^i e^j~,
\cr
H&=&2\, dv\wedge dA+ H_B~.
\eea
In such a case, the base space $B$ is a conformally balanced Riemannian manifold equipped with a connection $\hat{\tilde \nabla}$
with torsion a three-form $\tilde H$ such that ${\rm hol}(\hat{\tilde \nabla})\subseteq G_2$ and $\tilde H$ is closed.

The form of $dH$ in (\ref{pdh}) raises the possibility of canceling
the representative of the first Pontrjagin class of $P=M$ against
the representatives first Pontrjagin classes of the tangent bundle
of $M$ and that of the gauge bundle. This will solve the generalized
Bianchi identity of $H$, schematically $dH\sim\alpha' ({\rm tr}
R^2-{\rm tr} F^2)$, where $dH$ is given in (\ref{pdh}). As we have
already mentioned consistency in this case requires that the two
loop correction to the
 field equations should be taken into account.
Nevertheless the integrability conditions we have derived are still valid because the gravitino, dilatino and
gaugino supersymmetry transformations are not believed to receive corrections to this order but $dH\not=0$.
A systematic investigation of such solutions
will be given elsewhere.

\newsection{N=3 backgrounds}

\subsection{Supersymmetry conditions}

We have shown in section \ref{stabgroups}  that the three Killing spinors can be written as

\bea
 \e_1 &=& f\left(1+e_{1234}\right)~,
 \cr
 \e_2 &=& g_1(1+e_{1234})+ig_2(1-e_{1234})~,
 \cr
 \e_3 &=& h_1(1+e_{1234})+ih_2(1-e_{1234})+ih_3(e_{12}+e_{34})~,
\eea
where $f, g_1, g_2, h_1, h_2, h_3$ are spacetime functions.

Using the results we have derived for backgrounds with two supersymmetries, we can write
the gravitino  Killing spinor equation of $\e_3$ as

\bea
 h_1\del_A\log(h_1f^{-1})(1+e_{1234})+ih_2\del_A\log(h_2g_2^{-1})(1-e_{1234})&&
  \cr
    +i\del_A h_3(e_{12}+e_{34})+ih_3\hat{\nabla}_A(e_{12}+e_{34})&=&0~.
\eea
Evaluating this equation along $1$ and $e_{1234}$, we find that
\bea
 h_1\del_A\log(h_1f^{-1})+ih_2\del\log(h_2g_2^{-1})-ih_3\hat{\Om}_{A, 12}-ih_3\hat{\Om}_{A,34}&=&0~,
 \cr
 h_1\del_A\log(h_1f^{-1})-ih_2\del\log(h_2g_2^{-1})+ih_3\hat{\Om}_{A, \bar{1}\bar{2}}
 +ih_3\hat{\Om}_{A,\bar{3}\bar{4}}&=&0~,
\eea
and using (\ref{gofgtf}) and (\ref{btwo}), we get
\bea
 \del_A\log(h_1f^{-1})&=&0~,
 \cr
 \del_A\log(h_2f^{-1})&=&0~.
\eea
The remaining conditions of the gravitino Killing spinor equations on $\epsilon_3$ are
\be
 \del_A\log h_3+{1\over 2}\hat{\Om}_{A,-+}=0~,
\ee
\be
 \hat{\Om}_{A,1\bar{1}}+\hat{\Om}_{A,2\bar{2}}-\hat{\Om}_{A,3\bar{3}}-\hat{\Om}_{A,4\bar{4}}=0~,
\ee
\be
 \hat{\Om}_{A,+\alpha}=0~,
\ee
\be
 \hat{\Om}_{A,4\bar{2}}=-\hat{\Om}_{A,1\bar{3}}~,
\ee
\be
 \hat{\Om}_{A,3\bar{2}}=\hat{\Om}_{A,1\bar{4}}~.
\ee
The dilatino Killing spinor equation for $\e_3$ implies the conditions

\bea
&& \del_+\Phi=0~,
\cr
&& \del_{\bar{1}}\Phi=-H_{2\bar{3}\bar{4}}+{1\over 2} H_{2\bar{2}\bar{1}}
 -{1\over 2} H_{3\bar{3}\bar{1}}-{1\over 2} H_{4\bar{4}\bar{1}}-{1\over 2} H_{+-\bar{1}}~,
\cr
 &&\del_{\bar{2}}\Phi=H_{1\bar{3}\bar{4}}+ {1\over 2}H_{1\bar{1}\bar{2}}
 -{1\over 2} H_{3\bar{3}\bar{2}}-{1\over 2} H_{4\bar{4}\bar{2}}-{1\over 2} H_{+-\bar{2}}~,
\cr
&& \del_{\bar{3}}\Phi=-H_{4\bar{1}\bar{2}}-{1\over 2} H_{1\bar{1}\bar{3}}
 -{1\over 2} H_{2\bar{2}\bar{3}}+{1\over 2} H_{4\bar{4}\bar{3}}-{1\over 2} H_{+-\bar{3}}~,
\cr
&& \del_{\bar{4}}\Phi=H_{3\bar{1}\bar{2}}-{1\over 2} H_{1\bar{1}\bar{4}}
 -{1\over 2} H_{2\bar{2}\bar{4}}+{1\over 2} H_{3\bar{3}\bar{4}}-{1\over 2} H_{+-\bar{4}}~,
\cr
&& H_{+\bar{1}1}+H_{+\bar{2}2}-H_{+\bar{3}3}-H_{+\bar{4}4}=0~,
\cr
&& H_{+\bar{3}\bar{4}}=-H_{+\bar{1}\bar{2}}~,
\cr
&& H_{+\bar{4}2}=-H_{+\bar{1}3}~,
\cr
&& H_{+\bar{3}2}=H_{+\bar{1}4}~.
\eea
Combining the above results with the conditions we have derived for the first two  Killing spinors
 $\epsilon_1, \epsilon_2$, in section \ref{sufsc}, the gravitino Killing spinor equations implies the conditions

\be
 \del_A\log f+{1\over 2}\hat{\Omega}_{A,-+}=0~,
 \la{spgone}
\ee
\be
 \del_A\log (g_r f^{-1})=\del_A\log (h_p f^{-1})=0, \quad r=1,2\quad p=1,2,3~,
 \la{spgtwo}
\ee
\be
 \hat{\Omega}_{A,+\bar{\alpha}}=0~,~~~\a,\b=1,2,3,4~,
 \la{spgthree}
\ee
\be
 \hat{\Omega}_{A,\bar{\alpha}\bar{\beta}}=0~,
 \la{spgfour}
\ee
\be
 \hat{\Omega}_{A,1\bar{1}}+\hat{\Omega}_{A,2\bar{2}}=0~,
 \la{spgfive}
\ee
\be
 \hat{\Omega}_{A,3\bar{3}}+\hat{\Omega}_{A,4\bar{4}}=0~,
 \la{spgsix}
\ee
\be
 \hat{\Omega}_{A,4\bar{2}}=-\hat{\Omega}_{A,1\bar{3}}~,
 \la{spgseven}
\ee
\be
  \hat{\Omega}_{A,3\bar{2}}=\hat{\Omega}_{A,1\bar{4}}~,
  \la{spgeight}
\ee

and  the dilatino Killing spinor gives

\be
 \del_+\Phi=0~,
 \la{spdone}
\ee
\be
 \del_{\bar{1}}\Phi=-{1\over 2} H_{2\bar{3}\bar{4}}+{1\over 2} H_{2\bar{2}\bar{1}}-{1\over 2} H_{+-\bar{1}}~,
 \la{spdtwo}
\ee
\be
 \del_{\bar{2}}\Phi={1\over 2} H_{1\bar{3}\bar{4}}+{1\over 2} H_{1\bar{1}\bar{2}}-{1\over 2} H_{+-\bar{2}}~,
 \la{spdthree}
\ee
\be
 \del_{\bar{3}}\Phi=-{1\over 2} H_{4\bar{1}\bar{2}}+{1\over 2} H_{4\bar{4}\bar{3}}-{1\over 2} H_{+-\bar{3}}~,
 \la{spdfour}
\ee
\be
 \del_{\bar{4}}\Phi={1\over 2} H_{3\bar{1}\bar{2}}+{1\over 2} H_{3\bar{3}\bar{4}}-{1\over 2} H_{+-\bar{4}}~,
 \la{spdfive}
\ee
\be
 H_{\bar{\alpha}\bar{\beta}\bar{\gamma}}=0~,
 \la{spdsix}
\ee
\be
 H_{2\bar{3}\bar{4}}+H_{3\bar{3}\bar{1}}+H_{4\bar{4}\bar{1}}=0~,
 \la{spdseven}
\ee
\be
 -H_{1\bar{3}\bar{4}}+H_{3\bar{3}\bar{2}}+H_{4\bar{4}\bar{2}}=0~,
 \la{spdeight}
\ee
\be
 H_{4\bar{1}\bar{2}}+H_{1\bar{1}\bar{3}}+H_{2\bar{2}\bar{3}}=0~,
 \la{spdnine}
\ee
\be
 -H_{3\bar{1}\bar{2}}+H_{1\bar{1}\bar{4}}+H_{2\bar{2}\bar{4}}=0~,
 \la{spdten}
\ee
\be
 H_{+\bar{\alpha}\bar{\beta}}=0~,
 \la{spdeleven}
\ee
\be
 H_{+1\bar{1}}+H_{+2\bar{2}}=0~,
 \la{spdtwelve}
\ee
\be
 H_{+3\bar{3}}+H_{+4\bar{4}}=0~,
 \la{spdthirteen}
\ee
\be
 H_{+\bar{4}2}=-H_{+\bar{1}3}~,
 \la{spdfourteen}
\ee
\be
 H_{+\bar{3}2}=H_{+\bar{1}4}~.
 \la{spdfifteen}
\ee
It remains to investigate the geometric properties of $N=3$ backgrounds which are
implied by the above conditions.

\subsection{Geometry}
\subsubsection{The holonomy of $\hat\nabla$ connection}

As we have explained in previous cases, the holonomy of $\hat\nabla$
is contained in the stability subgroup of the Killing spinors in
$Spin(9,1)$, which in this case is $Sp(2)\ltimes\bR^8$. This can be
seen explicitly in the gauge where the spacetime functions $f$,
$g_r$ and $h_p$ in the Killing spinors are constant. It is clear
from the supersymmetry conditions (\ref{spgtwo}) that for this it
suffices to find a gauge $Spin(9,1)$ transformation to set $f=1$. As
in previous cases, this gauge can be attained with a $Spin(9,1)$
gauge transformation
 in the direction $\Gamma_{05}$. In this gauge, (\ref{spgone}) implies that $\hat\Omega_{A,-+}=0$ and so
 with (\ref{spgthree}), we have
 \bea
 \hat\Omega_{A,+B}=0~.
 \eea
Then the remaining conditions of the gravitino Killing spinor equations imply that the connection $\hat\nabla$
takes values in $\mathfrak{sp}(2)\coplus\bR^8$ and so the holonomy of the connection is contained in $Sp(2)\ltimes \bR^8$.
Moreover, we can set $\e_1= 1+e_{1234}$, $\e_2=i (1-e_{1234})$
and $\e_3=i(e_{12}+e_{34})$ using a constant $GL(3,\bR)$ transformation. This
is in agreement with the general arguments we have presented in section  \ref{holspin}. Conversely, if ${\rm hol}(\hat\nabla)
\subseteq Sp(2)\ltimes \bR^8$, then there are three parallel spinors which solve the gravitino Killing spinor equation.

\subsubsection{Spacetime forms and the geometry of spacetime}

We have shown that in  the gauge $f=1$ the Killing spinors can be chosen as $\e_1= 1+e_{1234}$, $\e_2=i (1-e_{1234})$
and $\e_3=i(e_{12}+e_{34})$. The spacetime form bilinears associated with the spinors $(\e_1, \e_1)$,
$(\e_1, \e_2)$ and $(\e_2,\e_2)$ have already been computed in previous sections.
 After an additional normalization of the spinors with  $1/\sqrt{2}$, we find the non-vanishing spacetime form bilinears
 of the pairs $(\e_1, \e_3)$, $(\e_2, \e_3)$, $(\e_3, \e_3)$ are the
one-forms
\bea
 \kappa(\e_3,\e_3)&=&e^0-e^5~~,
\eea
the three-forms
\bea
 \xi(\e_1,\e_3)&=&(e^0-e^5)\wedge \omega_K~,
  \cr
 \xi(\e_2,\e_3)&=&-(e^0-e^5)\wedge \omega_J~,
 \la{N3threeforms}
\eea
and the five-forms
\bea
 \tau(\e_1,\e_3)&=&-(e^0-e^5)\wedge\omega_I\wedge \omega_J~,
  \cr
 \tau(\e_2,\e_3)&=&-(e^0-e^5)\wedge\omega_I\wedge \omega_K~,
  \cr
 \tau(\e_3,\e_3)&=&-(e^0-e^5)\wedge[4\,{\rm Re}\,(\re^{\bar1}\wedge \re^{\bar2}\wedge \re^3\wedge \re^4)
 +\ha \check\omega\wedge \check\omega]~,
\eea
where we have set $\omega_I=\omega$,
\bea
\omega_J&=&\re^1\wedge \re^2+\re^{\bar{1}}\wedge \re^{\bar{2}}+\re^3\wedge \re^4 +\re^{\bar{3}}\wedge \re^{\bar{4}}~,
\cr
\omega_K&=&i(\re^1\wedge \re^2-\re^{\bar{1}}\wedge \re^{\bar{2}}+\re^3\wedge \re^4 -\re^{\bar{3}}\wedge \re^{\bar{4}})~,
\eea
and
\bea
\check\omega=i(\re^1\wedge \re^{\bar{1}}+\re^2\wedge \re^{\bar{2}}-\re^3\wedge \re^{\bar{3}}-\re^4\wedge \re^{\bar{4}})~.
\eea
The forms $\omega_I$, $\omega_J$ and $\omega_K$ are the familiar two-forms that appear on manifolds
with an $Sp(2)$-structure and $I,J$ and $K$ are the associated endomorphisms.
The two-form $\check\omega$ does not have an invariant meaning but it is necessary to write
the five-form of the $Spin(7)\ltimes\bR^8$-structure associated with the spinor $\eta_3$.

As in all the previous null supersymmetric backgrounds, the one-form
$\kappa=\re^-$ is associated with a null parallel vector field $X$.
In addition the conditions  (\ref{spdeleven})-(\ref{spdfourteen}) of
the dilatino Killing spinor equations imply that $i_X H$ takes
values in $\mathfrak{sp}(2)\coplus \bR^8$. This in turn implies that
$X$ leaves invariant the $Sp(2)\ltimes \bR^8$-structure of the
spacetime, i.e.
\bea
{\cal L}_X \alpha=0~,
\eea
where $\alpha$ are all the form bilinears constructed from the parallel spinors.

\subsubsection{Solution of the Killing spinor equations}

The solution of the Killing spinor equations in this case is similar to that of
the $SU(4)\ltimes \bR^8$ case. This is because the conditions conditions that we get for $Sp(2)\ltimes \bR^8$ are those
of $SU(4)\ltimes \bR^8$ but with respect to each $I,J$ and $K$ endomorphisms.

The supersymmetry conditions that arise  from the gravitino and
dilatino Killing spinor equations can be decomposed in
representations of $\mathfrak {sp}(2)$. This is easily done using
$\mathfrak {sp}(2)=\mathfrak{so}(5)$ but we shall not pursue this
here because of the similarity of this case with that of
$SU(4)\ltimes \bR^8$. For example, the conditions that arise from
the dilatino Killing spinor equation can be written as
\bea
&&\partial_+\Phi=0~,~~~2\partial_i\Phi-H_{-+i}=(\theta_I)_i=(\theta_J)_i=(\theta_K)_i~,
\cr
&&H_{+ij} (\delta^i{}_m\pm i (I_r)^i{}_m)
(\delta^j{}_n\pm i(I_r)^j{}_n) =0~,
\cr
&&H_{ijk} (\delta^i{}_m\pm i (I_r)^i{}_m)
(\delta^j{}_n\pm i(I_r)^j{}_n) (\delta^k{}_l\pm i (I_r)^k{}_l)=0~,~
\la{dsp2}
\eea
where $\theta_I$, $\theta_J$ and $\theta_K$ are the Lee forms of the
endomorphisms $I, J$ and $K$, see (\ref{leesu4}),
$(I_r,\,r=1,2,3)=(I,J,K)$, and $i,j,k,l, m,n=1,2,3,4,6,7,8,9$. The
last three conditions are the vanishing of the (3,0) and (0,3)
components of $H$ with respect to $I,J$ and $K$.

The gravitino Killing spinor equation implies that $\kappa$ is
parallel and so $i_X H=d\kappa$. In addition $\hat\nabla_A
(\omega_I)_{ij}=\hat\nabla_A (\omega_J)_{ij}=\hat\nabla_A
(\omega_K)_{ij}=0$ and so the torsion can be expressed in terms of
the geometry and the form spinor bilinears $\omega_I, \omega_J$ and
$\omega_K$. The expressions are those that we have given for
$SU(4)\ltimes \bR^8$ (\ref{solsu4})  but with respect to each of the
$I,J$ and $K$ endomorphisms. The only component of the torsion that
it is not specified is $H_{-ij}^{\bf 10}$, where the ten-dimensional
representation is the adjoint representation of $\mathfrak {sp}(2)$.
The metric and torsion can be written in a way similar to that of
$SU(4)\ltimes \bR^8$ in (\ref{metrtorsolsu4}).

\subsubsection{Special coordinates and a deformation family }

As in the $Spin(7)\ltimes\bR^8$ and $SU(4)\ltimes\bR^8$ cases
before, one can introduce coordinates adapted to the parallel vector
field $X$, $X=\partial/\partial u$, and write the metric as in
(\ref{coormetrsu4}). The analysis of the construction is similar to
the cases we have already investigated and so we shall not pursue
this further here. For example, one can introduce a frame $(\re^-,
\re^+, e^i)$ adapted to the special coordinates mentioned above as
in (\ref{coorframe}).

The spacetime also admits three pairs of integrable distributions,
one pair for each of the endomorphisms $I$, $J$ and $K$. This is
similar to the $SU(4)\ltimes\bR^8$ case which we have shown to admit
one pair of integrable distributions with respect to the
endomorphism $I$.

The spacetime  again has an interpretation as a two parameter family
of an eight-dimensional  manifold $B$ with an $Sp(2)$-structure.
Again for this, one has to introduce a frame $(E^-, E^+, E^i)$ as in
(\ref{eframe}), where $(E^+, E^-)$ are chosen to define an
integrable distribution of codimension eight with  typical leaf $B$.
There are two cases to consider. If the null vector has
non-vanishing rotation then, although $B$ admits an
$Sp(2)$-structure, it is not compatible with the induced connection
$\hat{\tilde \nabla}$ with torsion. The details are similar to those
of the $Spin(7)\ltimes\bR^8$ and $SU(4)\ltimes\bR^8$ cases, we have
already investigated. However, if the rotation of $X$ vanishes, then
the above data are compatible, i.e.~$B$ is a hyper-complex manifold
and all the complex structures are parallel with respect to the
induced connection $\hat{\tilde \nabla}$ with torsion. The
conditions of the dilatino Killing spinor equation (\ref{dsp2}) also
imply that $B$ is conformally balanced. Therefore $B$ is a
conformally balanced HKT manifold. The geometric properties of such
manifolds have been extensively investigated in the literature, see
e.g.~\cite{howegpc, poon, gibbons, michelson}.

\subsubsection{Field Equations}

The integrability conditions of the Killing spinor equations imply
that if the Bianchi identities of $H$ and $F$ are satisfied, $BH=0,
BF=0$, then all the field equations are satisfied provided that
$E_{--}=0$, $LH_{-A}=0$ and $LF_-=0$. This may have been expected
because these models are special cases of those with $Spin(7)\ltimes
\bR^8$- and $SU(4)\ltimes \bR^8$-invariant spinors. Examples of
backgrounds with $Sp(2)\ltimes \bR^8$-invariant Killing spinors have
been given in, e.g.~\cite{townsendgp, teschgp}.

\newsection {$N=4$ backgrounds}

\subsection{Backgrounds with $SU(3)$-invariant spinors}

As we have explained, without loss of generality the Killing spinors can be chosen as
\bea
\epsilon_1=1+e_{1234}~,~~~\epsilon_2=i(1-e_{1234})~,~~~\epsilon_3=(e_{15}+e_{2345})~,~~~\epsilon_4=i(e_{15}-e_{2345})~.
\eea
Substituting these into the gravitino Killing spinor equation, one finds  that the connection $\hat\nabla$ takes values
in $su(3)$, i.e.
\bea
&&\hat\Omega_{A,-+}=\hat\Omega_{A,-1}=\hat\Omega_{A,+1}=\hat\Omega_{A,1\bar1}=\hat\Omega_{A,n}{}^n=\hat\Omega_{A,np}=0~,
\cr
&&\hat\Omega_{A,-n}=\hat\Omega_{A,+n}=\hat\Omega_{A, 1 n}=\hat\Omega_{A,\bar 1 n}=0~,~~~~~~~m,n,p,\dots=2,3,4.
\la{gconsut}
\eea
In addition, the dilatino Killing spinor equation implies the conditions
\bea
\partial_+\Phi=\partial_-\Phi=\partial_1\Phi=\partial_{\bar 1}\Phi=0~,
\cr
\partial_{\bar n}\Phi-{1\over2} H_{\bar n p}{}^p=0~,
\la{dconsuta}
\eea
\bea
H_{+1\bar 1}+H_{+n}{}^n=0~,~~~H_{+np}=0~,~~~H_{+\bar1 \bar n} =0~,
\cr
H_{-1\bar 1}-H_{-n}{}^n=0~,~~~H_{-np}=0~,~~~H_{- 1 \bar n}=0~,
\cr
H_{1np}=H_{npm}=0~,
\cr
H_{\bar 1n}{}^n+H_{-+\bar 1}=0~,~~~H_{1\bar n\bar p}=0~,~~~H_{\bar n 1\bar 1}+H_{-+\bar n}=0~.
\eea
It remains  to investigate the restrictions on the geometry of spacetime imposed by the
above conditions.

\subsection{Spinor bilinears, backgrounds with four isometries and supersymmetry conditions}

The conditions (\ref{gconsut}) imply that the holonomy of the connection $\hat\nabla$ is contained in $SU(3)$,
${\rm hol}(\hat\nabla)\subseteq SU(3)$. The form bilinears of the first two Killing spinors have already
been computed in the context of supersymmetric backgrounds with $SU(4)\ltimes\bR^8$-invariant spinors,
 and the form bilinears of the first and
the third Killing spinor have already been computed in the context of  supersymmetric backgrounds
with $G_2$-invariant spinors. The remaining spinor pairs give the
one-forms
\bea
 \k(\e_1,\e_4)&=&-\k(\e_2,\e_3)=-e^6~,~~~
 \k(\e_2,\e_4)=-e^1~,
\cr
 \k(\e_3,\e_3)&=&\k(\e_4,\e_4)=e^0+e^5~,
\eea
the three-forms
\bea
 \xi(\e_1,\e_4)&=&e^0\wedge e^5\wedge e^6-{\rm Im}(\hat{\chi})-e^1\wedge \hat{\omega}~,
\cr
 \xi(\e_2,\e_3)&=&-e^0\wedge e^5\wedge e^6-{\rm Im}(\hat{\chi})+e^1\wedge \hat{\omega}~,
\cr
 \xi(\e_2,\e_4)&=&-e^0\wedge e^1\wedge e^5-{\rm Re}(\hat{\chi})+e^6\wedge\hat{\omega}~,
\cr
 \xi(\e_3,\e_4)&=&(e^0+e^5)\wedge(e^1\wedge e^6+\hat{\omega})~,
\eea
and the five-forms
\bea
 \tau(\e_1,\e_4)&=&e^0\wedge e^5\wedge{\rm Im}(\hat{\chi})+e^1\wedge e^6\wedge{\rm Re}(\hat{\chi})
            +\ha e^6\wedge\hat{\omega}\wedge\hat{\omega}-\hat{\omega}\wedge e^0\wedge e^1\wedge e^5~,
\cr
 \tau(\e_2,\e_3)&=&e^0\wedge e^5\wedge{\rm Im}(\hat{\chi})+e^1\wedge e^6\wedge{\rm Re}(\hat{\chi})
            -\ha e^6\wedge \hat{\omega}\wedge\hat{\omega}+\hat{\omega}\wedge e^0\wedge e^1\wedge e^5~,
\cr
 \tau(\e_2,\e_4)&=&e^0\wedge e^5\wedge{\rm Re}(\hat{\chi})-e^1\wedge e^6\wedge{\rm Im}(\hat{\chi})
            +\ha e^1\wedge \hat{\omega}\wedge\hat{\omega}-\hat{\omega}\wedge e^0\wedge e^5\wedge e^6~,
\cr
 \tau(\e_3,\e_3)&=&(e^0+e^5)\wedge[-e^1\wedge{\rm Re}(\hat{\chi})-e^6\wedge{\rm Im}(\hat{\chi})
            -\ha\hat{\omega}\wedge\hat{\omega}-\hat{\omega}\wedge e^1\wedge e^6]~,
\cr
 \tau(\e_3,\e_4)&=&(e^0+e^5)\wedge[e^1\wedge{\rm Im}(\hat{\chi})-e^6\wedge{\rm Re}(\hat{\chi})]~,
\cr
 \tau(\e_4,\e_4)&=&-(e^0+e^5)\wedge[-e^1\wedge{\rm Re}(\hat{\chi})-e^6\wedge{\rm Im}(\hat{\chi})
            +\ha\hat{\omega}\wedge\hat{\omega}+\hat{\omega}\wedge e^1\wedge e^6]~.
\eea

As we have explained all the form bilinears of the parallel spinors
are parallel with respect to the $\hat\nabla$ connection. In
particular, the one-forms $\re^-$, $\re^+$, $\re^1$ and $\re^{\bar
1}$ are $\hat\nabla$-parallel. Let us denote with $X,Y,Z$ and $\bar
Z$ the associated vector fields, respectively. In general, the
vector space spanned by $X,Y,Z$ and $\bar Z$ is not closed under Lie
brackets. To see this recall that commutator of two vector field,
say $X,Y$, is given by $i_X i_Y H$ and the supersymmetry conditions
do not imply that the component $H_{-+k}$ of $H$ vanishes, and
similarly for the other vector fields. This is reminiscent to the
situation we have encountered in the backgrounds with
$G_2$-invariant Killing spinors. Again there are many possibilities
ranging from requiring $X,Y,Z$ and $\bar Z$ to commute to taking the
spacetime to be a non-abelian Lorentzian Lie group. We shall not
investigate all cases, instead we shall require that the vector
space $\mfh$ spanned by $X,Y,Z$ and $\bar Z$ closes under Lie
brackets, i.e.~that $\mfh$ is a Lie algebra. This in particular
implies that
\bea
H_{ab n}=0~,~~~a,b=+,-, 1, \bar 1~,~~~n=2,3,4~.
\eea
Observe that some of these conditions are already implied by the
dilatino Killing spinor equation. The remaining non-vanishing
commutators of the vector fields are unconstrained by the conditions
of the dilatino Killing spinor equation and so they can span any
four-dimensional Lorentzian Lie algebra. The conditions from the
gravitino and dilatino Killing spinor equations in the case where
$\mfh$ is a Lie algebra can be summarized as
\bea
\hat\Omega_{A, ab}=\hat\Omega_{A,n}{}^n=\hat\Omega_{A,an}=\hat\Omega_{A,np}=0~,
\eea
and
\bea
\partial_a\Phi=0~,~~~\partial_{\bar n}\Phi-{1\over2} H_{\bar n p}{}^p=0
\cr
H_{mnp}=H_{abn}= H_{anp}=0~,~~~{1\over 3!}\epsilon_a{}^{bcd} H_{bcd}-i H_{ap}{}^p=0~,
\la{dsu3}
\eea
respectively, where $\epsilon=-i \re^-\wedge \re^+\wedge \re^1\wedge \re^{\bar 1}$. The above supersymmetry
conditions have been written in apparent representations
of $\mathfrak{so}(3,1)\oplus \mfsu (3)$.

We have explained that $H_{abn}=0$ is required by the closure of the Lie brackets of the vector fields. The condition
$H_{anp}=0$ implies that
\bea
{\cal L}_X \hat\omega=0~,~~{\cal L}_Y \hat\omega=0~,~~{\cal L}_Z \hat\omega=0~,~~{\cal L}_{\bar Z} \hat\omega=0~.
\eea
However, unlike the $G_2$ case, $X,Y,Z$ and $\bar Z$ do not always preserve the $SU(3)$-structure. This is because
from the last  condition of the dilatino Killing spinor equations the components $H_{ak}{}^k$ of the torsion
are not required to vanish, unless
$\mfh$ is abelian. In particular, we find that
\bea
{\cal L}_{X_a} \hat\chi= H_{an}{}^n \hat\chi~,
\la{liechisu3}
\eea
where $X_a= (X,Y, Z, \bar Z)$.
Thus  the parallel vector fields do not preserve the holomorphic volume form of the
$SU(3)$-structure.

\subsubsection{Solution of the Killing spinor equations}

As for the supersymmetric backgrounds with $G_2$-invariant spinors,
the spacetime $M$ of backgrounds with $SU(3)$-invariant
spinors can be interpreted as a principal bundle $M=P(\cH, B,\pi)$
equipped with a connection $\lambda$, where
$\cH$ is a Lorentzian group with Lie algebra $\mfh$ spanned by the four parallel
vector fields,
$B$ is the space of orbits of $\cH$ in $M=P$,  and $\pi$ is the
projection of the principal bundle.
The various formulae we have found in (\ref{prinbundle})
 can be extended to the $SU(3)$ case we are investigating here.
 In particular, the connection on the principal bundle can again be chosen as $\lambda^a=e^a$,
 where now $a=+,-, 1, \bar 1$.

Combining the conditions of the gravitino and  dilatino Killing spinor equations, we find that
the Levi-Civita connection of the spacetime satisfies the conditions,
\bea
\Omega_{n,ab}=0~,~~~2\Omega_{[a,bc]}=H_{abc}~,~~~\Omega_{a,bc}^{\bf 20}=0~,
\cr
2\Omega_{a,n}{}^n=H_{an}{}^n=-{i\over3!}\epsilon_a{}^{bcd} H_{bcd}~,~~~2\Omega_{p,n}{}^n=H_{pn}{}^n=-2\partial_p\Phi~,
\cr
\Omega_{a,bn}=0~,~~~\Omega_{n,pa}=0~,~~~2\Omega_{\bar n,a p}=H_{\bar n a p}~,
\cr
\Omega_{n,pm}=0~,~~~2\Omega_{\bar n,mq}=H_{\bar n mq}~,~~~\Omega_{a,np}=0~,~~~
\la{lcconsut}
\eea
which we have expressed in terms of $\mathfrak{so}(3,1)\oplus
\mfsu(3)$ representations. Some of above conditions can also be seen
as expressing the flux $H$ in terms of the Levi-Civita connection.
As we shall see
 $H$ is determined from the spinor bi-linears. Using the relations
(\ref{lcconsut}),  the torsion free conditions
of the frame $e^A$ can be written as
\bea
&&de^a={1\over2} H^a{}_{bc} e^b\wedge e^c+ H^a{}_{n\bar p} \re^n\wedge \re^{\bar p}~,
\cr
&&d\re^n=-\Omega_{a,}{}^n{}_p e^a\wedge \re^p+{1\over2} H^n{}_{pa} \re^p\wedge e^a
-\Omega_{\bar p,}{}^n{}_ m \re^{\bar p}\wedge \re^m
\cr
&&~~~~~~~~~~~~~~~~~~-\Omega_{p,}{}^n{}_ {\bar m} \re^{ p}\wedge
\re^{\bar m} -\Omega_{p,}{}^n{}_ m \re^{p}\wedge \re^m~.
\la{torfreesu3}
\eea

In terms of principal bundle data, the first torsion free condition above can be interpreted as the Cartan structure
equation for the connection $\lambda$, i.e.
\bea
d\lambda^a-{1\over2} H^a{}_{bc}  \lambda^b\wedge \lambda^c-{1\over2} H^a{}_{ij} e^i\wedge e^j=0~,~~~i,j,k,l=2, 3,4, 7,8,9~,
\la{pcartansu}
\eea
and so the curvature of $\lambda$ is
\bea
{\cal F}^a= {1\over2} H^a{}_{ij} e^i\wedge e^j~.
\la{pcurvsu}
\eea
In addition the condition (\ref{dsu3}) implies
that the curvature ${\cal F}^a$ satisfies the Donaldson condition. Note that ${\cal F}$ takes values in $\mfu(3)$ rather than in
$\mfsu(3)$ because if $\mfh$ is not abelian, the complex trace of ${\cal F}$ does not vanish.
The torsion $H$ of spacetime can be written as in (\ref{ph}). Therefore the metric and torsion of spacetime in terms of
principal bundle data can be written as
\bea
ds^2&=&\eta_{ab} \lambda^a \lambda^b+\delta_{ij} e^i e^j
\cr
H&=&{1\over 3}\eta_{ab} \lambda^a \wedge d\lambda^b+{2\over3}\eta_{ab} \lambda^a\wedge {\cal F}^b+\pi^*\tilde H
\la{metrtorsu3}
\eea
where $\tilde H$ is a three form of the base space $B$ horizontally
lifted to $P$.

It remains to investigate the geometry of the base space $B$.  The Riemannian manifold $B$ is equipped with a metric
$d\tilde s^2$ and a three form $\tilde H$ such that
$\pi^*d\tilde s^2=\delta_{ij} e^i e^j$. Therefore, one can define a metric connection $\hat{\tilde \nabla}$
 on $B$ with skew-symmetric torsion. In addition $B$  is equipped with a two-form
$\tilde \omega$ such that  $\hat\omega=\pi^*\tilde\omega$. This
follows from the property of $\hat\omega$ to be invariant under
$\cH$. The two-form $\tilde \omega$ is parallel with respect to
$\hat{\tilde \nabla}$. This follows from $\hat\nabla \hat\omega=0$.
The associated almost complex structure $\tilde I$ of the pair
$(d\tilde s^2, \tilde \omega)$ is integrable. This follows from the
fact that $\tilde H$ is (2,1) and (1,2) with respect to $\tilde I$.
Therefore $B$ is a KT manifold which is conformally balanced. The
latter condition follows   from the second equation in the
conditions that arise from the dilatino Killing spinor equation in
(\ref{dsu3}), i.e.~the Lee form can be written as $\tilde
\theta=2d\Phi$, where
\bea
\tilde\theta=-\star(\star d\tilde\omega\wedge \tilde\omega)~,
\la{leesu3}
\eea
and $d{\rm vol}(B)=e^2\wedge e^3\wedge e^4\wedge e^7\wedge e^8\wedge e^9$.
It is  well-known that for such manifolds the torsion three-form is uniquely determined in terms of the
complex structure and the metric as
\bea
\tilde H=-i_{\tilde I} d\tilde \omega=\star d\tilde\omega-\star(\tilde\theta\wedge \tilde\omega)~.
\eea
It remains to find whether $B$ has a compatible $SU(3)$-structure. There are two cases to consider depending
on whether or not $\cH$ is abelian.

First suppose that $\cH$ is abelian, then $H_{an}{}^n=0$ and
therefore $\hat\chi$ is invariant under $\cH$. In such a case $B$
admits a compatible $SU(3)$-structure, i.e.~there is a (3,0)-form
$\tilde \chi$ on $B$ such that it is parallel with respect to
$\hat{\tilde \nabla}$. Thus ${\rm hol}(\hat{\tilde \nabla})\subseteq
SU(3)$. Therefore $B$ is a conformally balanced KT manifold with a
compatible $SU(3)$-structure. An analysis of the geometry of these
manifolds in terms of $G$-structures can be found in \cite{salamonb,
lust}. The covariant derivatives $\tilde\nabla\tilde\omega$ and
$\tilde\nabla\tilde\chi$ can be decomposed in terms of five
irreducible $SU(3)$ representations as in the case of
eight-dimensional manifolds with $SU(4)$-structures that we have
already described. Since $B$ is complex $W_1=W_2=0$. In addition
$W_4$ and $W_5$, which can be represented by the Lee forms
$\theta_{\tilde\omega}=\tilde\theta$ given in (\ref{leesu3}) and
\bea
\theta_{ {\rm Re} \tilde\chi}=-{1\over2} \star(\star d {\rm Re} \tilde\chi\wedge {\rm Re} \tilde\chi),
\eea
respectively, are related as
\bea
\theta_{\tilde\omega}=\theta_{ {\rm Re} \tilde\chi}=2 d\Phi~.
\eea

Next suppose that $\cH$ is not abelian. There are three distinct
four-dimensional non-abelian Lorentzian Lie algebras. This is
because the structure constants of such Lie algebra are dual to a
vector in four-dimensional Minkowski space. Since the generators of
the Lie algebra are determined up to a Lorentz transformation, there
are three types of Lie algebras depending on whether the vector is
timelike, spacelike or null. These are isomorphic to $\bR\oplus
{\mathfrak su}(2)$, ${\mathfrak sl}(2,\bR)\oplus {\mathfrak u}(1)$
or ${\mathfrak so}(2)\coplus {\mathfrak h}_2(\bR)$, respectively,
where ${\mathfrak h}_2$ is the Heisenberg algebra of dimension
three. In this case, $\hat\chi$ is not invariant under $\cH$. As a
result although there still exists a $\tilde \chi$  on $B$ which
horizontally  lifts to $\hat\chi$, $\tilde \chi$ is a section of
$\Lambda^{3,0}\otimes L$, where $L=P\times_\rho \bC$ is a line
bundle over $B$ associated to the principal bundle $P$. The
representation\footnote{We assume that $\rho$ is not trivial. If it
is trivial, then the analysis reduces to that of the abelian case.}
$\rho$ is induced from (\ref{liechisu3}). Therefore $\tilde \chi$ is
a tensorial form of degree three associated with the representation
$\rho$ in the terminology of \cite{nomizu}. The structure that it is
associated with such a form is reminiscent of a $Spin_c$-structure
and so we shall say that $B$ admits an $SU_c(3)$-structure but not
an $SU(3)$ one as one might have been expecting. Therefore $B$ is a
conformally balanced  KT manifold. The geometry of these manifolds
can also be examined  using the Gray-Hervella classes
\cite{grayhervella}. It turns out that $W_1=W_2=0$, because $B$ is
complex, and $W_4$ which is represented by the Lee form
$\tilde\theta$ defined in (\ref{leesu3}).

To examine some other aspects of the geometry of spacetime, we consider
 the one-forms $\{\re^n, \re^{\bar n}\}$, $\{\re^n, \re^1, \re^-, \}$ and $\{e^a, \re^n\}$. These span
   integrable distributions of
co-dimensions four, five and three, respectively\footnote{There are
other integrable distributions, i.e.~the one spanned by the
one-forms $\{\re^n\}$.}. The first distribution is associated with
the principal bundle structure of the spacetime which we have
already investigated.  The second distribution implies that the
space admits a certain Lorentzian complex structure, i.e.~the
spacetime is a ``Lorentzian''-holomorphic manifold. Observe that
$\{\re^{\bar n}, \re^{\bar 1}, \re^-, \}$ is also an integrable
distribution. The third distribution is related to the property of
$B$ to be a complex manifold.

To summarize the geometry of backgrounds with $SU(3)$ invariant
spinors, we have found that the spacetime is (locally) a principal
bundle $M=P(\cH, B, \pi)$ equipped with a connection $\lambda$. The
fibre is a four-dimensional Lorentzian  Lie group and the curvature
${\cal F}$ of the connection $\lambda$ satisfies the Donaldson
condition. The geometry of the base space $B$ depends on whether or
not $\cH$ is abelian. If $\cH$ is abelian, then the base space $B$
is a balanced KT manifold with an $SU(3)$-structure. If $\cH$ is not
abelian, then $B$ is a balanced KT manifold equipped with a bundle
$L$ associated to $P$ and a section $\tilde \chi$ of
$\Lambda^{3,0}\otimes L$. The metric and torsion of spacetime in
terms of principal bundle data are given in (\ref{metrtorsu3}), and
the dilaton $\Phi$ is a function of $B$.

As in the $G_2$ case, one can compute $dH$ to find
\bea
dH=\eta_{ab}\, {\cal F}^a\wedge {\cal F}^b+\pi^*\, d \tilde H~.
\la{pdhsu3}
\eea
The first term in the right-hand-side of $dH$ is a representative of the first Pontrjagin class
 of the principal bundle $P$.
If one requires that $dH=0$, then the representative of the first
Pontrjagin class of $P$ must cancel against a form on $B$.   Of
course if $P$ is a globally defined principal bundle over $B$ and
one imposes the condition $dH=0$, then the first Pontrjagin form is
exact and therefore the first  Pontrjagin class of the principal
bundle should vanish.

\subsubsection{Field equations and examples}

An investigation of the integrability conditions of the Killing
spinor equations imply all the field equations are satisfied
provided that one imposes the Bianchi identities $BH=BF=0$. This may
have been expected because this class of backgrounds is a special
case of those with $N=2$ supersymmetry and $G_2$-invariant spinors.

In the absence of a gauge field, the only Bianchi identity that has to be imposed is that of $H$ which
has been computed in (\ref{pdhsu3}). As an example, one can take $\cH$ to be abelian. Then, one
can write $\lambda^a= dx^a+A^a$ and so we have
\bea
&&ds^2=\eta_{ab} (dx^a+A^a) (dx^b + A^b)+ \delta_{ij} e^i e^j~,
\cr
&&H=\eta_{ab} (dx^a+A^a)\wedge dA^b+ H'~,
\eea
where ${\cal F}=dA$ the curvature of the connection takes values in
$\mfsu(3)$, i.e.~it satisfies the Donaldson condition. If in
addition we require that $dH=0$, then
\bea
H=\eta_{ab} dx^a\wedge dA^b+H_B
\eea
after  choosing $H'=-\eta_{ab} A^a dA^b+ H_B$,
where  $H_B$ is a three-form on $B$ such that $dH_B=0$. Within a brane interpretation of these solutions, the connections
$A^a$ can be thought of as a rotation and wrapping.

A special case of this example is whenever the only non-vanishing rotation and wrapping is a along a null direction.
In this case, the Chern-Simons
form contribution vanishes. Thus one can set $\tilde H=H_B$. Such kind of solutions have been consider
 before in \cite{gpapas}.
The metric and torsion are
\bea
ds^2&=&2 dv (du+A)+dx^2+ dy^2+\delta_{ij} e^i e^j~,
\cr
H&=&\, dv\wedge dA+ H_B~.
\eea
In such a case, the base space $B$ is a integrable, conformally balanced strong KT Riemannian manifold
such that ${\rm hol}(\hat{\tilde \nabla})\subseteq SU(3)$
and $\tilde H$ is closed. An example of such manifold was found in \cite{volkov} and has been used
as gravitational dual to $N=1$ Yang-Mills theory in four-dimensions \cite{nunez}. The geometry of these models
has been investigated in \cite{tseytlin}. It is remarkable that the six-dimensional manifold is also a principal bundle.

\subsection{Backgrounds with $(SU(2)\times SU(2))\ltimes\bR^8$-invariant spinors}

\subsubsection{Supersymmetry conditions}

As we have already explained in \ref{holspin}, one can always arrange without loss
of generality such that the Killing spinors are
\bea
\epsilon_1=1+e_{1234}~,~~~\epsilon_2=i(1-e_{1234})~,~~~\epsilon_3=e_{12}-e_{34}~,~~~\epsilon_4=i(e_{12}+e_{34})~.
\eea
The gravitino Killing spinor equation implies that the connection
$\hat\nabla$ takes values in $(su(2)\oplus su(2))\coplus\bR^8$,
i.e.~that
\bea
\hat\Omega_{A,\a\b}&=&\hat\Omega_{A,np}=0~,~~~\hat\Omega_{A,\a}{}^\a=\hat\Omega_{A,n}{}^n=0~,
\cr
\hat\Omega_{A, \a n}&=&\hat\Omega_{A,\bar\a n}=0~,
\cr
\hat\Omega_{A,+B}&=&=0~,~~~~~~~~~~~~~~~~~~~~~~~~~\a,\b, \dots =1,2~,~~~n,p, \dots =3,4~.
\la{gconsusu}
\eea
The conditions that arise from the dilatino Killing spinor equation are
\bea
\partial_+\Phi=0~,~~~
\partial_{\bar\a}\Phi-{1\over2} H_{\bar\a\b}{}^\b-{1\over2}H_{-+\bar\a}=0~,~~~
\partial_{\bar n}\Phi-{1\over2} H_{\bar n p}{}^p-{1\over2} H_{-+\bar n}=0~,
\la{dconsusua}
\eea
\bea
H_{\a\b n}=H_{\a np}=0~,~~~
H_{\bar n\a\b}=H_{\bar \a np}=0~,
~~~H_{n\a}{}^\a=H_{\a n}{}^n=0~,
 \la{dconsusu}
\eea
\bea
H_{+\a\b}=H_{+\a}{}^\a=H_{+np}=H_{+n}{}^n= H_{+\a n}=H_{+\a\bar n}=0~.
\la{dconsusud}
\eea
It remains to investigate the restrictions on the geometry of the spacetime that are
implied by the above conditions.

\subsubsection{Geometry, holonomy and spinor bilinears }

As in the previous null cases, the conditions that arise from the gravitino Killing spinor
equation (\ref{gconsusu}) imply that ${\rm hol} (\hat\nabla)\subseteq (SU(2)\times SU(2))\ltimes \bR^8$.
To give a further insight into the geometry, one can compute the form bilinears of the Killing spinors.
This has already been done for the pairs of the first three Killing spinors. Therefore, it remains
to compute the forms of the pairs $(\epsilon_1, \epsilon_4)$, $(\epsilon_2, \epsilon_4)$,
$(\epsilon_3, \epsilon_4)$ and $(\epsilon_4, \epsilon_4)$. In particular, we find the
one form
\bea
 \k(\e_4,\e_4)=e^0-e^5~,
\eea
the three-forms
\bea
 \xi(\e_1,\e_4)&=&-(e^0-e^5)\wedge {\rm Im} (\chi_1+\chi_2)~,
\cr
 \xi(\e_2,\e_4)&=&-(e^0-e^5)\wedge {\rm Re} (\chi_1+\chi_2)~,
\cr
 \xi(\e_3,\e_4)&=&-(e^0-e^5)\wedge (\omega_1-\omega_2)~,
\eea
and the five-forms
\bea
 \tau(\e_1,\e_4)&=&-(e^0-e^5)\wedge(\omega_1\wedge\rm Re(\chi_2)+\omega_2\wedge\rm Re(\chi_1))~,
\cr
 \tau(\e_2,\e_4)&=&(e^0-e^5)\wedge(\omega_{1}\wedge\rm Im(\chi_{2})+\omega_{2}\wedge\rm Im(\chi_{1}))~,
\cr
 \tau(\e_3,\e_4)&=&-(e^0-e^5)\wedge\rm Im(\chi_{1}^*\wedge\chi_{2})~,
\cr
 \tau(\e_4,\e_4)&=&-(e^0-e^5)\wedge(\rm Re(\chi_{1}^*\wedge\chi_{2})+\ha(\omega_{1}
 -\omega_{2})\wedge(\omega_{1}-\omega_{2}))~,
\eea
where $\omega_1=-(e^1\wedge e^6+ e^2\wedge e^7)$, $\omega_2=-(e^3\wedge e^8+ e^4\wedge e^9)$,
$\chi_1= (e^1+ie^6)\wedge (e^2+i e^7)$ and $\chi_2=(e^3+ie^8)\wedge (e^4+i e^9)$.
All the form bilinears $\alpha$ constructed from the Killing spinors are parallel with respect
to the connection $\hat\nabla$,
\bea
\hat\nabla\alpha=0~.
\eea
Moreover as expected,  (\ref{dconsusud})
implies that $i_X H$ takes values in $(\mfsu(2)\oplus \mfsu(2))\coplus \bR^8$, where $X$ is the associated parallel
vector field to $\kappa$. This in turn
gives that
\bea
{\cal L}_X \alpha=0~,
\eea
where again $\alpha$ stands for any form Killing spinor bilinear. Since $X$ is Killing,
${\cal L}_X H=0$ ($dH=0$) and ${\cal L}_X\Phi=0$, $X$ leaves  all the geometry of spacetime
including the $(SU(2)\times SU(2))\ltimes\bR^8$-structure invariant.

\subsubsection{Solution of the Killing spinor equations}

The solution of the Killing spinor equations is similar to that of
the $SU(4)\ltimes \bR^8$ case. The supersymmetry conditions of both
the gravitino and dilatino Killing spinor equations have been
decomposed in terms of $SU(2)\times SU(2)$ representations in an
apparent way. The minimal set of covariantly constant forms that
characterizes the conditions (\ref{gconsusu}) that arise from the
gravitino Killing spinor equation are
\bea
\kappa=\re^-~,~~\xi_I=\re^-\wedge \omega_I~,~~~\xi_J=\re^-\wedge \omega_J~,~~~\xi_1=\re^-\wedge \chi_1~,~~~
\xi_2=\re^-\wedge \chi_2~.
\la{minsusu}
\eea
where
\bea
\omega_I= \omega_1+\omega_2~,~~~\omega_J=\omega_1-\omega_2~.
\eea
In particular, if we denote the forms (\ref{minsusu}) collectively with $\beta$, then the conditions that
arise from the gravitino Killing spinor equation are equivalent to
\bea
\hat\nabla\beta=0~.
\eea
Note that the endomorphisms $I,J$ of the tangent bundle of the
spacetime commute, i.e.~$IJ=JI$. The forms $(\omega_1, \chi_1)$ and
$(\omega_2,\chi_2)$ are associated with an $SU(2)\times
SU(2)$-structure.

The conditions (\ref{dconsusu}) imply that $H$ is a (2,1) and (1,2)
form with respect to both $I$ and $J$, i.e.~the (3,0) and (0,3)
components with respect to both $I,J$  vanish. The last two
conditions in (\ref{dconsusua}) can be rewritten  as
\bea
2\partial_\a\Phi=(\theta_1)_\a~,~~~2\partial_n\Phi=(\theta_2)_n~,
\eea
where $\theta_1$ and $\theta_2$ are the Lee forms of the endomorphisms $I_1$ and $I_2$
associated with $\omega_1$ and $\omega_2$, respectively, see (\ref{leesu4}).
The conditions (\ref{dconsusud}) imply that $H_{+ij}$, $i,j=1,2,3,4,6,7,8,9$ takes values in $\mfsu(2)\oplus\mfsu(2)$.

The supersymmetry conditions can be solved for the fluxes. It is
easy to see that the expressions for $H_{-\a\b}$ and
$H_{\a\b\bar\g}$, and  $H_{-np}$  and $H_{nm\bar q}$ can be given as
in (\ref{solsu4}) but now with respect to the endomorphisms $I_1$
and $I_2$ associated to the forms $\omega_1$ and $\omega_2$,
respectively. Similarly $H_{-\a}{}^\a$ and $H_{-n}{}^n$ can be
expressed as in (\ref{hmtrsu4}) but now with respect to $\chi_1$ and
$\chi_2$, respectively. Therefore all $H$ fluxes are determined in
terms of the spinor bilinears and the metric of spacetime apart from
the component $H^{\bf 6}_{-ij}$ which takes values in
$\mfsu(2)\oplus\mfsu(2)$. The metric and torsion of the spacetime
can be written as
\bea
&&ds^2=2 \re^- \re^++ 2\delta_{\a\bar\b} \re^\a \re^{\bar\b}+2 \delta_{m\bar n} \re^m \re^{\bar n}
\cr
&&H=\re^+ \wedge d\kappa -[{1\over2} (I_1)^m{}_i \nabla_-(\omega_1)_{mj}+{1\over2} (I_2)^m{}_i \nabla_-(\omega_2)_{mj}+
 (I_1)^m{}_i \nabla_-(\omega_2)_{mj}\cr
 &&~~~~~~
 + (I_2)^m{}_i \nabla_-(\omega_1)_{mj}] e^-\wedge e^i\wedge e^j
 -\, {1\over8}\, {\rm Im}((\bar\chi_1)^{\a\b}\nabla_-(\chi_1)_{\a\b})\,  e^-\wedge \omega_1
 \cr
&&~~~~~~~-\, {1\over8}\, {\rm Im}((\bar\chi_2)^{ m n}\nabla_-(\chi_2)_{m n}) e^-\wedge \omega_2
+{1\over2} H_{-ij}^{\bf 6} e^-\wedge e^i\wedge e^j
\cr
&&~~~~~~~~~~~~~~~~+
{1\over 3!} H_{ijk} e^i\wedge e^j\wedge e^k~,
\eea
where $i,j,k,l=1,2,3,4,6,7,8,9$, $H_{ijk}$ is determined in terms of
$\omega_1$, $\omega_2$, $I_1$ and $I_2$ as has been explained above.

\subsubsection{Local coordinates, distributions and a deformation family}

One can adapt coordinates along the null parallel vector field $X$ and write the metric of
the spacetime as in (\ref{coormetr}). Then one can adapt a frame
$(\re^-, \re^+, \re^\a, \re^{\bar\a}, \re^n, \re^{\bar n})$
in a way similar to that in (\ref{coorframe}). The spacetime admits various integrable distributions. Apart from the
distributions of co-dimension five spanned by $(\re^-, \re^\a, \re^n)$ and $(\re^-, \re^{\bar\a}, \re^{\bar n})$ which
are analogues to those of the $SU(4)\ltimes \bR^8$ backgrounds, there are also integrable distributions
of codimension three spanned by $(\re^-, \re^\a, \re^n, \re^{\bar n})$ and $(\re^-, \re^\a, \re^{\bar\b}, \re^n)$.
This can be easily seen
using the torsion free conditions of the frame $(\re^-, \re^+, \re^\a, \re^{\bar\a}, \re^n, \re^{\bar n})$.

The spacetime has the interpretation as a two parameter family of an
eight-dimensional manifold with an $SU(2)\times SU(2)$-structure.
This is done by adapting a frame $(E^-, E^+, E^i)$ similar to that
in (\ref{eframe}). The one-forms $(E^-, E^+)$ span a codimension
eight integrable distribution with leaves the eight-dimensional
manifold $B$. There are two cases to consider. In the generic case,
and in particular if the rotation of the null vector field does not
vanish, $d\kappa=d\re^-\not=0$, $B$ admits an $SU(2)\times
SU(2)$-structure which is {\it not} compatible with the induced
connection $\hat{\tilde \nabla}$ with torsion $\tilde H=H|_B$. The
explanation for this has been presented in detail for the
$Spin(7)\ltimes \bR^8$ case and it will not be repeated here.
However, if $d\kappa=d\re^-=0$, then $B$ admits an $SU(2)\times
SU(2)$-structure which is  compatible with the induced connection
$\hat{\tilde \nabla}$. In such a case, $B$ is a conformally balanced
eight-dimensional manifold equipped with (i) metric connection
$\hat{\tilde \nabla}$ with torsion given by a three-form  $\tilde
H$, (ii)
 two commuting complex structures $\tilde I$ and $\tilde J$, $\tilde I\tilde J=\tilde J \tilde I$,
 which are the restrictions
 of $I,J$,  such that
 $\hat{\tilde \nabla} \tilde I=\hat{\tilde \nabla} \tilde J=0$ and (iii)
 ${\rm hol} (\hat{\tilde \nabla})\subseteq SU(2)\times SU(2)$. The manifold $B$ is conformally
 balanced because the Lee forms satisfy
 $\tilde\theta_{\tilde I}=
 \tilde\theta_{\tilde J}=2d\tilde\Phi$ as required by the supersymmetry conditions of the
 dilatino Killing spinor equation
 (\ref{dconsusua}). Note that $B$ admits an integrable product
 structure $\Pi=\tilde I \tilde J$, i.e.~$\Pi^2=1$. The geometry of $B$ can be also analyzed using $G$-structures
 but this is very similar to the $SU(n)$ cases we have already examined and we shall not pursue this further here.

The geometric properties of manifolds equipped with two commuting complex structures  compatible with a metric connection
with torsion have been investigated before in the physics literature in the context of sigma models
with (2,2) worldvolume supersymmetry \cite{hull, howegp}. In particular, special
coordinates have been introduced and the local expression for the metric has been given. In addition the simultaneous
integrability properties of the complex structures and the product structure have been examined in detail.
As in the previews null cases, the integrability conditions of the Killing spinor imply that all field
equations are satisfied provided that the Bianchi identities are imposed and one requires that
$E_{--}=0$, $LH_{-A}=0$ and $LF_-=0$.

\newsection {$N=8$ backgrounds}

\subsection{Backgrounds with $SU(2)$-invariant spinors}

\subsubsection{Supersymmetry conditions}

One can choose without loss of generality the Killing spinors, see section \ref{holspin},  as
\bea
&\epsilon_1=1+e_{1234}~,~~\epsilon_2= i (1-e_{1234})~,~~
\epsilon_3= e_{12}-e_{34}~,~~\epsilon_4= i (e_{12}+e_{34})~,~~~~~~
\cr
&\epsilon_5= e_{15}+ e_{2345}~,~ \epsilon_6= i(e_{15}- e_{2345})~,~
\epsilon_7= e_{52}+ e_{1345}~,~\epsilon_8=i (e_{52}- e_{1345})\,.
\eea
Substituting them in to the gravitino Killing spinor equation, one finds that
the connection $\hat\nabla$ takes values in $su(2)$, i.e.
\bea
\hat\Omega_{A,+\a}=\hat\Omega_{A,-\a}=\hat\Omega_{A,-+}=\hat\Omega_{A,\a\b}
=\hat\Omega_{A,\a\bar\b}=
\hat\Omega_{A,\a n}=\hat\Omega_{A,\a\bar n}=0~,
\cr
\hat\Omega_{A,+n}=\hat\Omega_{A,-n}=0~,~~~\hat\Omega_{A,np}=\hat\Omega_{A,n}{}^n=0~,~~~\a, \b=1,2~,~~n,p=3,4~.
\eea
Similarly, the conditions that arise from the  dilatino Killing spinor equation equation are
\bea
\partial_+\Phi=\partial_-\Phi=\partial_\a\Phi=0~,~~~
\partial_{\bar n}\Phi-{1\over2} H_{\bar n p}{}^p=0~,
\cr
H_{\bar\a\b}{}^\b+ H_{-+\bar\a}=0~,~~~H_{\a n}{}^n=0~,~~~H_{-+n}=0~,~~~
\cr
H_{\a\b n}=H_{\a np}=0~,~~~H_{\bar n \a \b}= H_{\bar\a np}=0~,~~~ H_{n \a \bar\b}=0~, H_{n \a}{}^\a=0~,
\cr
H_{-n}{}^n=0~,~~~H_{-\a n}=0~,~~~ H_{-\bar n\bar p}=0~,~~~H_{-\a\bar n}=0~,~~~H_{-1\bar 2}=0~,~~~H_{-1\bar1}=H_{-2\bar 2}~,
\cr
H_{+\a\b}=0~,~~~H_{+\a}{}^\a=0~,~~~H_{+np}=0~,~~~H_{+n}{}^n=0~,~~~H_{+\a n}=0~,~~~H_{+\a \bar n}=0~.
\eea
We shall next investigate the restrictions that these conditions above impose on the
geometry of spacetime.

\subsubsection{Geometry and form bilinears}

The gravitino Killing spinor equation implies that the holonomy of
$\hat\nabla$ is contained in $SU(2)$, ${\rm
hol}(\hat\nabla)\subseteq SU(2)$. Several of the Killing spinor form
bilinears have been computed in previous cases. The remaining pairs
can also easily be computed. This is because we have found the form
spinor bilinears of all possible types of spinors, see
\ref{appexsut}. As a result, the forms of the remaining pairs are
given from those we have computed already by an appropriately
relabeling of indices. However, we shall not explicitly
 give the result  because it is not particularly enlightening.

All the forms that arise form the Killing spinor bilinears are parallel with respect to $\hat\nabla$.
A basis in the ring of parallel forms is $\re^-, \re^+, e^1, e^6, e^2, e^7, \omega_2$ and $\chi_2$, where
\bea
\omega_2=-( e^3\wedge e^8+ e^4\wedge e^9)~,~~~\chi_2= (e^3+i e^8)\wedge (e^4+i e^9)~.
\eea
In particular this implies that the one-forms $\re^-, \re^+, {\rm e}^\a, {\rm e}^{\bar\a}$,
${\rm e}^\a={1\over\sqrt{2}} (e^\a+i e^{5+\a})$,
 are parallel and so the associated
vectors fields $X,Y, Z_\a, Z_{\bar\a}$ are Killing. Unlike the cases of backgrounds with $G_2$-
and $SU(3)$-invariant spinors, the vector space $\mfh={\rm Span}(X,Y, Z_\a, Z_{\bar\a})$  {\it closes
under Lie brackets}. This is a consequence of the conditions of the dilatino Killing spinor equations and
in particular the vanishing of the components
\bea
H_{ab n}=0~,~~~a,b=+,-,\a, \bar\a~,~~~n=3,4
\eea
of the NS$\otimes$NS three-form field strength. The Lie algebra $\mfh$ is not arbitrary but rather constrained
by supersymmetry. In particular, the structure constants
satisfy  the conditions
\bea
&&H_{\bar\a\b}{}^\b+ H_{-+\bar\a}=0~,~~~H_{-1\bar 2}=0~,~~~H_{-1\bar1}=H_{-2\bar 2}~,~~~
\cr
&&H_{+\a\b}=0~,~~~H_{+\a}{}^\a=0~,~~~\a,\b=1,2~,
\la{reststcon}
\eea
of the dilatino Killing spinor equation. Observe that the structure constants of $Sl(2,\bR)\times SU(2)$ whose
lie algebras are spanned by $sl(2,\bR)=\bR<e_-, e_+, e_1={1\over  \sqrt2} (e_1+e_{\bar 1})>$
and $su(2)=\bR<e_2, e_{\bar 2}, e_6={1\over i \sqrt2} (e_{\bar 1}-e_{1})>$ satisfy these conditions
provided one identifies their structure constants as in the first condition of (\ref{reststcon}).
The remaining conditions of the dilatino Killing spinor equation, apart from those involving the dilaton,
imply that
\bea
{\cal L}_X \omega_2=0~,~~~{\cal L}_Y \omega_2=0~,~~~{\cal L}_{Z_\a} \omega_2=0~,~~~
{\cal L}_{Z_{\bar\a}} \omega_2=0~,~~~
\eea
and
\bea
{\cal L}_X \chi_2=0~,~~~{\cal L}_Y \chi_2=0~,~~~{\cal L}_{Z_\a} \chi_2=0~,~~~
{\cal L}_{Z_{\bar\a}} \chi_2=0~.~~~
\eea
 The
conditions involving the dilaton imply that $\Phi$ is invariant
under the action of $\mfh$ on the spacetime and that
$\partial_n\Phi$ is related to the torsion $H$.

The conditions on the geometry of supersymmetric backgrounds with $SU(2)$ invariant spinors can be summarized
as follows: The gravitino Killing spinor equation gives
\bea
\hat\Omega_{A, ab}=0~,~~~\hat\Omega_{A,aB}=0~,~~~\hat\Omega_{A,np}=0~,~~~\hat\Omega_{A,n}{}^n=0~,~~~a=-,+,\a, \bar\a~,
\la{gsu2}
\eea
and the dilatino Killing spinor equation gives
\bea
&&\partial_a\Phi=0~,~~~\partial_{\bar n}\Phi-{1\over2} H_{\bar n p}{}^p=0~,
\cr
&&H_{anp}=0~,~~~H_{an}{}^n=0~,~~~H_{nab}=0
\cr
&&H_{\bar\a\b}{}^\b+ H_{-+\bar\a}=0~,~~~H_{-1\bar 2}=0~,~~~
\cr
&&H_{-1\bar1}=H_{-2\bar 2}~,~~~H_{+\a\b}=0~,~~~H_{+\a}{}^\a=0~.
\la{dsu2}
\eea
This concludes the discussion of the supersymmetry conditions.

Combining the conditions of the gravitino and  dilatino Killing spinor equations, we find that
the Levi-Civita connection of the spacetime satisfies the conditions,
\bea
\Omega_{k,ab}=0~,~~~2\Omega_{[a,bc]}=H_{abc}~,~~~\Omega_{a,bc}^{\bf 70 }=0~,
\cr
2\Omega_{a,n}{}^n=H_{an}{}^n=0~,~~~2\Omega_{p,n}{}^n=H_{pn}{}^n=-2\partial_p\Phi~,
\cr
\Omega_{a,bn}=0~,~~~\Omega_{n,pa}=0~,~~~2\Omega_{\bar n,a p}=H_{\bar n a p}~,
\cr
\Omega_{n,pm}=0~,~~~2\Omega_{\bar n,mq}=H_{\bar n mq}~,~~~\Omega_{a,pn}=0~,~~~
\la{lcconsu2}
\eea
which we have expressed in terms of $\mathfrak{so}(5,1)\oplus \mfsu(2)$ representations.
Some of above conditions can  been seen as expressing the flux $H$ in terms of the Levi-Civita connection.
In this case, we shall show that $H$ is determined by the metric and the form spinor bilinears.

\subsubsection{Solution of the Killing spinor equations}

As for the supersymmetric backgrounds with $G_2$- and
$SU(3)$-invariant spinors, the spacetime $M$ of backgrounds with
$SU(2)$-invariant spinors can be interpreted as a principal bundle
$M=P(\cH, B,\pi)$ equipped with a connection $\lambda$, where $\cH$
is a Lorentzian group with Lie algebra $\mfh$ spanned by the six
parallel vector fields and $\pi$ is the projection of the principal
bundle. Unlike the $G_2$ and $SU(3)$ cases, the group $\cH$ is not
arbitrary but its structure constants satisfy the (\ref{reststcon}).
The connection is $\lambda^a=e^a$. The Cartan structure equations
for $\lambda$ can be found by considering the torsion free
conditions of the frame. Using the relations (\ref{gsu2}) and
(\ref{dsu2}),  the torsion free conditions of the frame $e^A$ can be
written as
\bea
&&de^a={1\over2} H^a{}_{bc} e^b\wedge e^c+ H^a{}_{n\bar p}\, \re^n\wedge \re^{\bar p}~,
\cr
&&de^n=-\Omega_{a,}{}^n{}_p e^a\wedge \re^p+{1\over2} H^n{}_{pa} \re^p\wedge e^a-
\Omega_{\bar p,}{}^n{}_ m \re^{\bar p}\wedge \re^m
\cr
&&~~~~~~~~~~~~~~~~~~-\Omega_{p,}{}^n{}_ {\bar m} \re^{ p}\wedge \re^{\bar m}
-\Omega_{p,}{}^n{}_ m \re^{p}\wedge \re^m~.
\eea
The first condition is interpreted as the Cartan structure equation and so the curvature of the
connection $\lambda$
is
\bea
{\cal F}^a{}_{ij}={1\over2} H^a{}_{ij}~,~~~i,j=3,4,8,9~.
\eea
The conditions (\ref{dsu2}) imply that ${\cal F}^a$ is self-dual,
i.e.~it takes values in the $\mfsu(2)\subset \mathfrak{so}(4)$. The
metric of the spacetime and the torsion $H$ in terms of principal
bundle data can be written as
\bea
&&ds^2=\eta_{ab}\lambda^a \lambda^b +\delta_{ij} e^i e^j~,
\cr
&&H={1\over3} \eta_{ab} \lambda^a\wedge d\lambda^b+ {2\over3} \eta_{ab} \lambda^a\wedge {\cal F}^b+\pi^*\tilde H~.
\la{metrtorsu2}
\eea
It remains to investigate the geometry of the four-dimensional base
space $B$. As in the $G_2$ and $SU(3)$ cases, the base space $B$ is
a Riemannian manifold equipped with a metric $d\tilde s^2$ and a
three-form $\tilde H$, and so with a metric connection $\hat{\tilde
\nabla}$ with three-form torsion. In addition, since both $\omega_2$
and $\chi_2$ are invariant under $\cH$, the base space $B$ is also
equipped with two two-form $\tilde \omega_2$ and $\tilde \chi_2$.
Both $\tilde\omega_2$ and $\tilde \chi_2$ are parallel with respect
to $\hat{\tilde \nabla}$. The almost complex structure $\tilde I$
associated with the pair $(d\tilde s^2, \tilde\omega_2)$ is
integrable, and $\tilde \chi$ is (2,0) with respect $\tilde I$.
Therefore, $B$ is a conformally balanced HKT manifold. The conformal
balance condition follows from the second equation in (\ref{dsu2})
and $\tilde\theta=2d\Phi$, where $\tilde\theta$ is the Lee form of
$\tilde\omega_2$ defined as in (\ref{leesu4}).

As in the $SU(3)$ case, the spacetime admits various  distributions spanned by the
 one-forms $\{\re^n, \re^{\bar n}\}$, $\{\re^n, \re^-, \re^1, \re^2\}$ and $\{e^a, \re^n\}$. These are
   {\it integrable} distributions of
co-dimensions six, five and two, respectively. The first integrable distribution is that of the principal
bundle structure that we have already investigated.  To show that $\{\re^n, \re^-, \re^1, \re^2\}$ span an integrable
  distribution, one also needs the conditions (\ref{reststcon}) on the structure constants $H_{abc}$ imposed by
supersymmetry.  Note that there is another co-dimension five distribution spanned by the
one-forms   $\{\re^{\bar n}, \re^+, \re^{\bar 1}, \re^2\}$.    Both codimension five distributions imply
 that the spacetime admits a certain ``Lorentzian''
complex structure, i.e.~$M$ is a ``Lorentzian-holomorphic'' space,
in two different ways. The  distribution spanned by $\{e^a, \re^k\}$
is related to the property of  $B$ to be  a complex manifold.

To summarize the geometry of supersymmetric backgrounds with
$SU(2)$-invariant spinors, we have found that the spacetime is a
principal bundle $P(\cH, B, \pi)$ equipped with a connection
$\lambda$. The structure constants of the six-dimensional group
$\cH$ are constrained by (\ref{reststcon}). The curvature of
$\lambda$ is a self-dual two-form. The base space is a
four-dimensional, conformally balanced  HKT manifold. The metric and
torsion are given in (\ref{metrtorsu2}) and the dilaton $\Phi$ is a
function of $B$.

\subsubsection{Field equations and examples}

The integrability conditions of the Killing spinor equations imply
that all field equations are satisfied provided that $BH=BF=0$ as in
the previous cases with compact stability subgroups. In addition,
one can show that  $dH={\cal F}\wedge {\cal F}+\pi^* d\tilde H$.
This expression is similar to those in the $G_2$ and $SU(3)$ cases.
If $\cH$ is abelian, it is straightforward to introduce coordinates
and write explicit expressions for the metric and torsion. Since we
have done this for the $SU(3)$ case and the analysis is very
similar, we shall not repeat the various formulae here. One can also
easily construct the examples with non-vanishing null rotation which
have also been considered in \cite{gpapas}. An example of a
background with $SU(2)$-invariant spinors and eight supersymmetries
is the NS5-brane \cite{callan}.

\subsection{Backgrounds with $\bR^8$-invariant spinors}

\subsubsection{Supersymmetry conditions}

As we have already explained, the Killing spinors can be chosen as
\bea
\epsilon_1=1+e_{1234}~,~~~\epsilon_2=i(1-e_{1234})~,~~~\epsilon_3=e_{12}-e_{34}~,~~~\epsilon_4=i(e_{12}+e_{34})~,
\cr
\epsilon_5=e_{13}+e_{24}~,~~~\epsilon_6=i(e_{13}-e_{24})~,~~~ \epsilon_7=e_{23}-e_{14}~,~~~\epsilon_8=i (e_{23}+e_{14})~.
\eea
Observe that the above spinors can also be thought of as spanning
the real chiral representation  $\Delta^+_{\bf 8}$ of $Spin(8)$. The
gravitino Killing spinor equation implies that the connection
$\hat\nabla$ takes values in $\bR^8$, i.e.~that
\bea
\hat\Omega_{A,ij}=~,~~~
\hat\Omega_{A,+B}=0~,~~~~~~~~i,j=1,2,3,4~,6,7,8, 9
\la{gconre}
\eea
i.e.~the only non-vanishing components of the connection are
$\hat\Omega_{A,-i}$. The dilatino Killing spinor equation in
addition implies that
\bea
&&\partial_+\Phi=0~,~~~
\partial_i\Phi-{1\over2} H_{-+i}=0~,~~~
\cr
&&H_{ijk}=0~,
~~~H_{+ij}=0~.
\la{dconre}
\eea
We shall next investigate the conditions on the geometry of spacetime.

\subsubsection{Solution of the Killing spinor equations}

It is clear from (\ref{gconre}) that the connection $\hat\Omega$
takes values in $\bR^8$ and so ${\rm hol}(\hat\nabla)\subseteq
\bR^8$. The forms of the Killing spinor bilinears can be written as
\bea
\a=\re^-\wedge \phi~,~~~~~\phi\in \Lambda^{{\rm ev}+}(\bR^8)~,
\eea
where $\Lambda^{{\rm ev}+}(\bR^8)=\Lambda^0(\bR^8)\oplus
\Lambda^2(\bR^8)\oplus \Lambda^{4+}(\bR^8)$ and
$\Lambda^{4+}(\bR^8)$ denotes the subspace of self-dual four-forms
and $\bR^8=\bR<e^1, \dots, e^4, e^6, \dots, e^9>$. As in the
previous cases, the last condition in (\ref{gconre}) implies that
 $\kappa=\re^-$ is a parallel null one-form associated with a Killing vector field $X$.
In addition the last condition in (\ref{dconre}) implies that $X$
preserves the $\bR^8$-structure, i.e.
\bea
{\cal L}_X\alpha=0~,
\eea
where $\alpha$ is any form Killing spinor bilinear.

The only non-vanishing components of $H$ are $H_{-ij}$ and $H_{-+i}$. Unlike previous $K\ltimes \bR^8$ cases, $H$ is determined
in terms of the Levi-Civita connection and the dilaton. In particular, the conditions of the dilatino Killing spinor equation
  (\ref{dconre}) express $H_{-+i}$ in terms of $\Phi$ and the condition $\hat\Omega_{-,ij}=0$ of the
  gravitino Killing spinor equation gives $H_{-ij}=2\Omega_{-,ij}$. One can also adapt coordinates along the
  parallel vector field $X$,  $X={\partial/\partial u}$, and write the metric and torsion as
\bea
ds^2&=&2(dv+m_I dy^I) (du+V dv+n_I dy^I)+ \gamma_{IJ} dy^I dy^J
\cr
H&=&\re^+\wedge d\kappa+\Omega_{-,ij} \re^-\wedge e^i\wedge e^j~.
\la{solr8}
\eea
Next consider the torsion free conditions for the frame $(\re^-, \re^+, e^i)$ which we can introduce as in (\ref{coorframe}).
In particular, the torsion free conditions for $\re^-$ and $e^i$ imply that there are functions $m=m(v,y)$ and $e^i=e^i(v,y)$ such
that
\bea
m_I=\partial_I m~,~~~~e^i_J=\partial_Je^i(v,y)~.
\eea
In addition since $d\kappa=dm$, we have from the dilatino Killing spinor equation that
\bea
\partial_I(2\Phi+\partial_vm)-2m_I \partial_v\Phi=0~.
\la{dddr8}
\eea
If the rotation of $X$ does not vanish, $d\kappa=d \re^-\not=0$, it is not apparent that there is a
diffeomorphism which preserves the form of the metric in (\ref{solr8})
and transforms the ``transverse'' metric $\gamma_{IJ} dy^I dy^J$ to that of flat space.
However if $d\kappa=d \re^-=0$, the dilaton $\Phi=\Phi(v)$, and one can perform the diffeomorphism $u=u, v=v$ and $y^i=e^i(v,y^I)$
which preserves the form of the metric in (\ref{solr8}) and transforms the transverse part of the metric to that of flat space.
In such a case, the solution of the Killing spinor equations can be written as
\bea
ds^2&=&2dv (du+V dv+n_i dy^i)+ \delta_{ij} dy^i dy^j~,
\cr
H&=&\Omega_{-,ij} dv\wedge dy^i\wedge dy^j~,~~~~~\Phi=\Phi(v)~.
\la{solrz8}
\eea

As in all previous $K\ltimes\bR^8$ cases, the spacetime can be
interpreted as a two parameter family of an eight-dimensional
manifold. This can be done by introducing a frame $(E^-, E^+, E^i)$
as in (\ref{eframe}). If the rotation of $X$ does not vanish,
$d\kappa\not=0$, then the deformed submanifold $B$ although it
admits a $\{1\}$-structure it is not compatible with the induced
$\hat{\tilde\nabla}$ connection. However, if the rotation vanishes,
then $B$ is locally isomorphic to $\bR^8$ as we have shown in
(\ref{solrz8}).

\subsubsection{Field equations and examples}

One can show using the integrability conditions of the Killing
spinor equations that the only field equations that need to be
imposed in addition  to the Bianchi identities $BH=0, BF=0$ are
$E_{--}=0$, $LH_{-A}=0$ and $LF_-=0$. This is similar to all the
previous $K\ltimes \bR^8$ backgrounds as might have been expected.

An example of a background with $\bR^8$-invariant spinors and $N=8$ supersymmetries is that of the
fundamental string solution of \cite{ruiz}. The non-vanishing fields are
\bea
ds^2=2 h^{-1} dx dy + ds^2(\bR^8)~,
\cr
H=-dx\wedge dy\wedge dh^{-1}~,
\cr
e^{2\Phi}= h^{-1}~,
\eea
where $h$ is a harmonic function on $\bR^8$. To see this, one performs the coordinate transformation
\bea
u=x~,~~~v=h^{-1}y~,
\eea
and the metric and torsion can be rewritten as
\bea
ds^2=2 (dv+ v h^{-1} dh) du+ ds^2(\bR^8)~,
\cr
H=h^{-1} du\wedge dv\wedge dh= du\wedge d(v h^{-1} dh)~.
\eea
Setting $\re^-=dv+v h^{-1} dh$ and $\re^+=du$, the metric and torsion have the form of (\ref{solr8}).
In particular observe that (\ref{dddr8}) is satisfied.

\newsection{Parallelizable string backgrounds}

As we have mentioned in the introduction, if the Killing spinors
have stability subgroup $\{1\}$, then the background is
parallelizable with respect to the $\hat\nabla$ connection,
i.e.~$\hat R=0$. In addition the gauge connection is flat,
i.e.~$F=0$. In this section, we shall show that backgrounds for with
$\hat R=0$ are group manifolds provided that $H$ is closed at the
0-th order in $\alpha'$. For this, we write the expression for the
$\hat R$ curvature in terms of the Riemannian curvature $R$ as
\bea
\hat R_{MN,RS} = R_{MN,RS}-{1\over2} \nabla_{M} H_{NRS}+{1\over2} \nabla_{N} H_{MRS}
+{1\over 4} H_{RML} H^L{}_{NS}-{1\over 4} H_{RNL} H^L{}_{MS}~,
\la{hatcurv}
\eea
where $M,N,\dots=0,\dots,9$ are coordinate indices.
Skew-symmetrizing in all four indices and using  $dH=0$, we find that
\bea
H_{L[MN} H^L{}_{RS]}=0
\eea
which can be rewritten as
\bea
H_{LM[N} H^L{}_{RS]}=0~.
\eea
Skew-symmetrizing (\ref{hatcurv}) in the $N,R,S$ indices, we get
that
\be
-\nabla_{M} H_{NRS}+\nabla_{[N} H_{RS]M}=0
\ee
which together with the closure of $H$ gives
\be
\nabla_{M} H_{NRS}=0~.
\ee
Therefore the spacetime admits a parallel three-form which satisfies
the Jacobi identity and so the spacetime is a ten-dimensional
Lorentzian group manifold. Of course Lorentzian group manifolds
admit sixteen parallel spinors with respect to the $\hat\nabla$
which can be identified with the left-invariant connection. In
addition if we demand that the parallel spinors are also Killing,
then the background is maximally supersymmetric and so the spacetime
is locally isometric to Minkowski space \cite{jfgp}.

The field equations impose additional conditions on backgrounds for which $\hat R=0$. In particular, we find that
the gravitino and two-form gauge potential field equations  imply that
\be
\nabla_M\partial_N \Phi=0
\la{gfeq}
\ee
and
\be
\partial_M\Phi g^{MN} H_{NRS}=0~.
\la{bfeq}
\ee
In the case that $\Phi$ is constant, the dilatino  Killing spinor equation becomes
\be
H_{MNR} \Gamma^{MNR}\epsilon=0~.
\ee
This gives, after using the Jacobi equation, that
\bea
H_{MNR} H^{MNR}=0~.
\eea
Therefore $H$ is null. Since $H$ is null, the spacetime admits at
least eight Killing spinors.

Next let us turn to the case that $\Phi$ is not constant. In such case, (\ref{gfeq}) implies that
 $X^M= g^{MN} \partial_N\Phi$ is parallel.
There are two cases to consider, either $X^2=0$, i.e.~$X$ is null,
or $X^2={\rm const}$ and so $X$ is either timelike or spacelike. In
both cases, using (\ref{bfeq}) and the Jacobi identity for $H$, we
find that the  dilatino Killing spinor equation implies that
\bea
X_M X^M- {1\over24} H_{MNR} H^{MNR}=0~.
\la{lencon}
\eea
Therefore if $X$ is null, then $H$ is also null and the spacetime
admits at least eight Killing spinors which satisfy
$d\Phi\,\epsilon=0$. If $X$ is time-like, the condition
(\ref{bfeq}), also written as $i_X H=0$, implies that $H$ is
spacelike $H^2>0$ and so (\ref{lencon}) cannot be satisfied. There
are no parallelizable    supersymmetric backgrounds, $\hat R=0$, for
which the dilaton can be identified with a time coordinate on the
spacetime. The only remaining possibility is $X$  space-like. Such
supersymmetric backgrounds are known to exist like for example
$\bR^{5,1}\times U(1)\times SU(2)$. The dilaton is identified with
the coordinate along the space-like $U(1)$ direction.

\newsection{Common sector of type II supergravities}

\subsection{Supersymmetric backgrounds}

The Killing spinor equations of the common sector of type II supergravities are
\bea
\nabla^\pm\epsilon_\pm&=&0~,
\cr
(\Gamma^M\partial_M\Phi\mp{1\over12}\Gamma^{MNP} H_{MNP})\epsilon_\pm&=&0~,
\eea
where $\nabla^\pm=\nabla\pm{1\over2} H$
and $\epsilon_\pm$ are Majorana-Weyl spinors of the same (IIB) or opposite (IIA) chiralities.
The field equations are the same as those of the metric and NS$\otimes$NS two-form
gauge potential of the heterotic string. In addition, $dH=0$.

The Killing spinor equations of the common sector resemble those of the gravitino and
 dilatino of the heterotic supergravity. Observe  for example that $\nabla^+=\hat\nabla$.
 The common sector of type II supergravities
 is a consistent truncation of type II supergravities. Therefore it can be thought as a special case of
 eleven-dimensional and IIB supergravities. Because of this, we shall not investigate the Killing spinor
 equations in detail. Instead, we shall focus on some properties of the common sector supersymmetric
 backgrounds that follow from those of the supersymmetric backgrounds of the
 heterotic string that we have presented. In particular, we shall examine the
 common sector of IIB supergravity in which case $\epsilon_\pm$ are both positive chirality spinors.
 It is worth mentioning that the  supersymmetric configurations of the common sector of
  IIA supergravity should be treated
 separately from those of IIB supergravity because there are several differences. To mention one, the spinors
 $1+e_{1234}$ and $e_1-e_{234}$ of IIA supergravity have stability subgroup $G_2\ltimes \bR^8$ in $Spin(9,1)$.
 As we have seen IIB supergravity does not admit  spinors with stability subgroup $G_2\ltimes \bR^8$.
 Therefore, it is expected
 that some of the geometries that appear in the common sector IIA supergravity are different
 from those that appear in IIB.

The relevant spinor bundle of the IIB common sector is $S^+\oplus
S^+$ and the gravitino Killing spinor equation is a parallel
transport equation for the connection $\nabla^+\oplus \nabla^-$.
Thus the gravitino Killing spinor equations are associated with a
$Spin(9,1)\times Spin(9,1)$ connection. However, the gauge group
that preserves the Killing spinor equations is the same as in the
heterotic case, i.e.~it is $Spin(9,1)$. This is the main difference
between the Killing spinor equations of the common sector and those
of the heterotic string. Because the gauge group is the proper
diagonal subgroup of $Spin(9,1)\times Spin(9,1)$, it has many more
orbits in the space of spinors than those of the heterotic string.
As a result there are many more cases to consider.

The Killing spinors $\epsilon$ of the common sector can be written
as $\eta_1\oplus \eta_2$.  To proceed, let $G^+$ and $G^-$  be the
stability subgroups of the parallel spinors $\eta\oplus0$ and
$0\oplus \eta$, respectively, and $G$ be the stability subgroup of
all parallel spinors, $G\subseteq G^+\cap G^-$. It is again the case
that the holonomy of the $\nabla^\pm$ connections should be a
subgroup of the stability subgroups $G^\pm$ of the parallel spinors,
i.e.~${\rm hol}(\nabla^\pm)\subseteq G^\pm$. The general strategy to
analyze the supersymmetric backgrounds of the common sector is to
first choose the parallel spinors of the type $\eta\oplus 0$ as in
the heterotic case and then use the residual gauge symmetry $G^+$ to
simplify the Killing spinors of the type $0\oplus \eta$, or vice
versa. Without loss of generality, we may choose the Killing spinors
$\epsilon\oplus 0$ as those of the heterotic supergravity. If one
requires that there are parallel spinors with stability subgroup
$G^+=\{1\}$ or $G^-=\{1\}$, then either the curvature $R^+=0$ or
$R^-=0$. In such a case,  the spacetime is a metric Lorentzian
group. This follows from the results of the previous section. If
both $G^+=G^-=\{1\}$, then the Riemann curvature of the Levi-Civita
connection vanishes, $R=0$, and the spacetime is locally isometric
to Minkowski space.

On the other hand, if either $\nabla^+$ or $\nabla^-$ does not  admit parallel spinors, then
the analysis of the common sector reduces to that of the heterotic string for the
connection with torsion that admits parallel spinors. In this  case
the heterotic supersymmetric backgrounds are ``embedded'' in the common sector.

It turns out that many common sector supersymmetric backgrounds admit a Killing spinor of the type
  spinor $\epsilon=\eta\oplus \eta$, i.e.~$\eta$ is
 parallel
with respect to both
$\nabla^+$ and $\nabla^-$ connections. If this is the case, then $\eta$ is also parallel
with respect to the Levi-Civita connection. In particular, the Killing spinor equations imply that
that
\bea
\nabla\eta=0~,~~~H_{ABC} \Gamma^{BC}\eta=0~,
\cr
\Gamma^A\partial_A\Phi\eta=0~.
\eea
The last condition follows from the dilatino Killing spinor
equations. Since the stability subgroup of a single spinor  in
$Spin(9,1)$ is $Spin(7)\ltimes\bR^8$,
$G^+=G^-=G=Spin(7)\ltimes\bR^8$. So, ${\rm hol}(\nabla)\subseteq
Spin(7)\ltimes\bR^8$, i.e.~the holonomy of the Levi-Civita
connection is contained in $Spin(7)\ltimes\bR^8$. Furthermore, the
gauge symmetry of the Killing spinor equations can be used to set
$\epsilon=1+e_{1234}$. As a result, one can use the results of $N=1$
heterotic string backgrounds to show that
\bea
&&\Omega_{A,+ B}=0~,~~~\Omega_{A,\a}{}^\a=0~,~~~\Omega_{A,\a\b}={1\over2} \Omega_{A,\g\d} \epsilon_{\a\b}{}^{\g\d}~,
\cr
&&H_{A+ B}=0~,~~~H_{A\a}{}^\a=0~,~~~H_{A\a\b}={1\over2} H_{A\g\d} \epsilon_{\a\b}{}^{\g\d}~,
\cr
&&\partial_+\Phi=\partial_\a\Phi=0~, ~~~\a,\b=1,2,3,4~.
\la{comone}
\eea
These imply that ${\rm hol}(\nabla)\subseteq Spin(7)\ltimes\bR^8$,
as we have already mentioned, $H_{-ij}\in \Lambda^2_{{\bf 21}}$ and
$H_{ijk}=0$. The rotation of the associated null Killing vector
field vanishes. As a result, the spacetime admits Penrose
coordinates. One can also see that  the Lorentzian deformation
family is that of a $Spin(7)$ manifold, i.e.~the Levi-Civita
connection of $B$ has holonomy contained in $Spin(7)$.

Next, we shall use the general results above to investigate common
sector backgrounds with one and two supersymmetries. For $N=1$,
either $\nabla^+$ or $\nabla^-$ admits a parallel spinor $\epsilon$.
This implies that either ${\rm hol}(\nabla^+)\subseteq
Spin(7)\ltimes\bR^8$ or ${\rm hol}(\nabla^-)\subseteq
Spin(7)\ltimes\bR^8$. Suppose that ${\rm hol}(\nabla^+)$ admits the
parallel spinor $\epsilon$. Since $\epsilon$ is Killing by
assumption, it also solves the dilatino Killing spinor equation. The
geometry of spacetime is that described in the case of $N=1$
heterotic string backgrounds. The connection $\nabla^-$ may not
admit parallel spinors, i.e.~the holonomy of $\nabla^-$ is not
contained in $Spin(7)\ltimes\bR^8$. However, if it admits parallel
spinors, they do not solve the dilatino Killing spinor equation.

To examine backgrounds with two supersymmetries, we again use  the results we have derived in the context of
 heterotic string. There are several cases to consider. In the first case both Killing spinors are parallel
 with respect to either the $\nabla^+$ or $\nabla^-$ connection. Without loss of generality, we can assume
 that both spinors are parallel with respect to the $\nabla^+$ connection. There are two such cases to consider with
 stability subgroups $G=G^+=SU(4)\ltimes \bR^8$ and  $G=G^+=G_2$. The geometry of the spacetime
 is that of the $N=2$ heterotic string backgrounds.
 Next suppose that one of the Killing spinors is parallel with respect to the
 $\nabla^+$ connection and the other is parallel with respect to the $\nabla^-$ connection. In this
 case $G^+=G^-=Spin(7)\ltimes\bR^8$ and so the holonomy of both $\nabla^\pm$  connections is contained
  in $Spin(7)\ltimes\bR^8$. The first Killing spinor can always to be chosen as
  $\epsilon_1=f( 1+e_{1234})$. In such a case, one can show that the second Killing spinor can be chosen either
  as $\epsilon_2= g_1 (1+e_{1234})$ or $\epsilon_2= g_1 (1+e_{1234})+ ig_2 (1-e_{1234})$
  or $\epsilon_2= g (e_{15}+e_{2345})$. The argument is similar to the one we have used for the heterotic
  string backgrounds with two supersymmetries. The case for which both Killing spinors point to the same
  direction has already been investigated above and the supersymmetry conditions have been summarized in
  (\ref{comone}). The supersymmetry conditions for the remaining two cases can also be read from
  those of $N=2$ heterotic string  backgrounds. However for the second spinor, one has to alter appropriately
  the sign of the terms containing the flux $H$ in the Killing spinor equations.
  The analysis is routine and we shall not present the results
  here. It is apparent though that one must  consider  many more cases of supersymmetric backgrounds
  in context of the common
  sector than those that appear in  heterotic supergravity. In the table below we summarize the
  stability subgroups of the spinors of $N=2$ IIB common sector backgrounds.

 \begin{table}[h]
 \begin{center}
\begin{tabular}{|c|c|c|c|}\hline
$N=2$&$\mathrm{G^+ }$ & $\mathrm{G^- }$& $\mathrm{G }$
 \\ \hline
 &$SU(4)\ltimes \bR^8 $  &- &$SU(4)\ltimes \bR^8 $ \\
 &$G_2$ &- &$G_2$ \\
&$Spin(7)\ltimes \bR^8$&$Spin(7)\ltimes \bR^8$& $Spin(7)\ltimes \bR^8$ \\
&$Spin(7)\ltimes \bR^8$&$Spin(7)\ltimes \bR^8$&$SU(4)\ltimes \bR^8$ \\
&$Spin(7)\ltimes \bR^8$&$Spin(7)\ltimes \bR^8$&$G_2$
\\ \hline
\end{tabular}
\end{center}
\vskip 0.2cm {\small Table 3:  There are five classes of IIB common
sector backgrounds with two supersymmetries. These are denoted with
the stability subgroups $G^+$, $G^-$ and $G$ of the Killing spinors.
In all cases ${\rm hol}(\nabla^\pm)\subseteq G^\pm$. The entries $-$
denote the  cases for which the sector associated with the
$\nabla^-$ connection  does not admit Killing spinors. }

\end{table}

The parallel forms of string theory backgrounds are associated with conserved currents of the worldvolume action
which described the propagation of (super)strings in such backgrounds \cite{phowegpapas}.
Thus for any of the parallel forms we have presented
in the supersymmetric heterotic and common sector backgrounds, there is an associated conserved current. In particular,
we have shown that  all supersymmetric heterotic and common sector backgrounds admit at least one parallel null vector field.
Without loss of generality let $\nabla^+ \kappa=0$, $\nabla^+=\hat\nabla$. Then the bosonic string
with equations of motion $\nabla^+_+\partial_- Y^M=0$,
where $Y$ is the embedding map of the string worldvolume into spacetime and $\sigma^\pm$ are lightcone worldvolume coordinates,
has a conserve current $\kappa_M \partial_- Y^M$, $\partial_+(\kappa_M \partial_- Y^M)=0$. Therefore
$\kappa_M \partial_- Y^M=f(\sigma^-)$. It is known that the bosonic string action is invariant under the conformal transformations
$\delta Y^M=a(\sigma^+) \partial_+ Y^M+ b(\sigma^-) \partial_- Y^M$, where $a,b$ are the infinitesimal parameters. It is easy
to see that choosing $f(\sigma^-)$ to be constant, one can gauge fix the conformal transformations associated with the parameter
$b(\sigma^-)$. Similarly, if the one-form $\kappa'$ is parallel with respect to the $\nabla^-$ connection,
the current $\kappa'_M\partial_+ Y^M$
is conserved $\partial_-(\kappa'_M\partial_+ Y^M)=0$ and this can be used to gauge fix the conformal transformations
associated with the infinitesimal parameter $a(\sigma^+)$.

\newsection{Conclusions}

We have specified the geometry of the supersymmetric heterotic
string backgrounds (in the lowest order in $\alpha'$) for which all
the parallel spinors of the connection $\hat\nabla$ with torsion
given by the $NS\otimes NS$ three-form field strength
 are also Killing. We have also determined the field equations that are implied
 by the Killing spinor equations in all cases. We have found that there are two classes
 of backgrounds the null and timelike.  The Killing spinors of null backgrounds
 are chiral which respect to a suitable $Spin(8)$ chirality projection or equivalently
 admit a single null $\hat\nabla$-parallel vector field. The stability subgroups of the Killing
 spinors in $Spin(9,1)$ are $K\ltimes\bR^8$ for $K=Spin(7)~ (N=1)$, $K=SU(4)~(N=2)$, $K=Sp(2)~(N=3)$,
 $K=SU(2)\times SU(2)~(N=4)$ and $K=\{1\}~ (N=8)$, where $N$ denotes the number of Killing spinors.
 We have shown that the spacetime is a suitable two-parameter Lorentzian family of an eight-dimensional manifold
 $B$ with a $K$-structure. If the rotation of the null vector field vanishes, then
 $B$ admits a metric connection, $\hat{\tilde \nabla}$ with skew-symmetric torsion on $B$ compatible
 with an integrable conformally balanced $K$-structure on $B$,
 and so in particular ${\rm hol} (\hat{\tilde \nabla})\subseteq K$.

The Killing spinors of timelike backgrounds
 are not chiral which respect to a suitable $Spin(8)$ chirality projection or equivalently
 admit a  time-like $\hat\nabla$-parallel vector field. The stability subgroups of the Killing spinors
 in $Spin(9,1)$ are $G_2~ (N=2)$, $SU(3)~(N=4)$, $SU(2)~(N=8)$ and $\{1\}$ (N=16). Assuming that the
 vector fields constructed from spinor bilinears close under Lie brackets, we have shown that
 the spacetime is locally a principal bundle, $P(\cH, B, \pi)$, whose fibre directions are the orbits
 of the parallel vector fields and the base
 space is a manifold with a $G_2~ (n=7)$ , $SU_c(3)~(n=6)$ and $SU(2)~(n=4)$-structure, respectively,
 where $n={\rm dim}B$. We have described  the geometry of the spacetime of  all supersymmetric backgrounds
 in terms of principal bundle data.

 We also applied some of our results to the supersymmetric configurations of the common sector
 of type II supergravities. We have found that there are some differences between the
 properties of IIA and IIB supersymmetric common sector backgrounds. We also determined the
 conditions for the common sector IIB backgrounds with two supersymmetries.
A consequence of our results is that {\it all} supersymmetric common sector and heterotic string
backgrounds admit a null $\hat\nabla$-parallel vector field. This may be used to lightcone gauge fix
the (super)conformal gauge symmetry of strings propagating in such backgrounds.

We have not investigated in detail the timelike backgrounds for
which the set of the  vector fields constructed from spinor
bilinears  does not close under Lie brackets. However, we have shown
that the commutator vector field $[X,Y]$ of any two
$\hat\nabla$-parallel vectors $X,Y$ is also $\hat\nabla$-parallel.
Therefore there are several possible geometric structures for the
spacetime ranging from a principal bundle, which we have mentioned
above, to a Lorentzian Lie group.

It is well-known the field equations of the heterotic string contain
higher curvature correction terms. These modify the field equations
of the supergravity theory that we have investigated. It has been
shown in \cite{tsimpis} that for certain supersymmetric backgrounds
with $SU(3)$-invariant spinors, these higher order curvature
correction terms are necessary for consistency with the heterotic
anomaly cancelation mechanism. It would be of interest to find
whether this persists to all supersymmetric backgrounds that we have
analyzed.

Another class of supersymmetric heterotic backgrounds that we have
not investigated are those for which the number of Killing spinors
is less than the number of $\hat\nabla$-parallel spinors, i.e.~some
of the $\hat\nabla$-parallel spinors do not solve the dilatino
Killing spinor equation. It is known  such supersymmetric
backgrounds exist. However, the analysis we have done for the
gravitino Killing spinor equation in this paper still applies to
this class of models. In particular, one can determine the stability
subgroup in $Spin(9,1)$ of the parallel spinors and construct the
spacetime form spinor bilinears. Taking a basis in the space of
parallel spinors $\{\eta_i\}$, one then can write the Killing
spinors as $\e_r=f_{ri}\eta_i$ and substitute them in the dilatino
Killing spinors equation. Using the results we have collected in the
appendices, one can derive linear systems  similar to those which
have been found in the context of IIB supergravity \cite{grangpb}.
These linear systems can be solved to find the Killing spinors of
such backgrounds.

\vskip 0.5cm {\bf Acknowledgments:}~G.P. thanks the Erwin
Schr\"odinger International Institute for hospitality during the
``Geometry of pseudo-Riemannian manifolds with applications to
physics'' programme. U.G. has a postdoctoral fellowship funded by
the Research Foundation K.U.~Leuven. In addition U.G.~would like to
acknowledge the support of the Swedish Research Council and the
PPARC grant PPA/G/O/2002/00475 as part of this work was done while
still being a postdoc at King's College.

\setcounter{section}{0}

\appendix{Spinors and forms}\label{Spinors}

\subsection {Spinors from forms}

Spinors can be described in terms of forms. This construction is
explained in, e.g.~\cite{lawson, harvey} and it has been used in
\cite{wang} in the context of manifolds with special holonomy. This
description has been applied  to the the Majorana-Weyl   spinors
 of $Spin(9,1)$ in \cite{grangutowskigp}. Here for completeness, we shall briefly summarize some of the
 aspects of this construction.

Consider the Euclidean space $U=\bR<e_1,\dots,e_5>$, where  $e_1,\dots,e_5$ is an orthonormal
basis.
The space of Dirac spinors of $Spin(9,1)$ is
$\Delta_c=\Lambda^*(U\otimes \bC)$.
The gamma matrices are represented on $\Delta_c$ as
\bea
\Gamma_0\eta&=& -e_5\wedge\eta +e_5\lc\eta~,~~~~ \Gamma_5\eta=
e_5\wedge\eta+e_5\lc \eta~,
\cr
\Gamma_i\eta&=& e_i\wedge \eta+ e_i\lc \eta~,~~~~~~i=1,\dots,4
\cr
\Gamma_{5+i}\eta&=& i e_i\wedge\eta-ie_i\lc\eta~.
\eea
$\Delta_c$ decomposes into two complex chiral representations
according to the degree of the form $\Delta_c^+=\Lambda^{{\rm even}}(U\otimes \bC)$
and  $\Delta_c^-=\Lambda^{{\rm odd}}(U\otimes \bC)$. These are the complex Weyl representations
of $Spin(9,1)$.

The  gamma matrices  $\{\Gamma_i; i=1,\dots, 9\}$  are Hermitian
and $\Gamma_0$ is anti-Hermitian with respect to the
inner product
\be
<z^a e_a, w^b e_b>=\sum_{a=1}^{5} (z^a)^* w^a~,~~~~
\ee
on $U\otimes \bC$  and then extended to $\Delta_c$, where $(z^a)^*$ is the standard complex
conjugate of $z^a$.
The above gamma matrices
satisfy the Clifford algebra relations
$\Gamma_A\Gamma_B+\Gamma_B \Gamma_A=2 \eta_{AB}$ with respect to the
 Lorentzian inner product as expected.

A $Spin(9,1)$  Majorana inner product is
\be
B(\eta,\theta)= <B(\eta^*), \theta>~,~~~~~~~~
\ee
where the map denoted with the same symbol $B=\Gamma_{06789}$. Observe that this Majorana
inner product
is skew-symmetric $B(\eta, \theta)=-B(\theta,\eta)$.
 $B$ pairs the $\Delta^+_c$ and $\Delta^-_c$ representations.
Moreover, both $\Delta^+_c$ and $\Delta^-_c$ are Lagrangian with
respect to $B$, i.e.~$B$ restricted to either $\Delta^+_c$ or
$\Delta^-_c$ vanishes. The $Spin(9,1)$  Dirac inner product is
\be
D(\eta, \theta)=<\Gamma_0\eta, \theta>~.
\ee

It is well-known that $Spin(9,1)$ admits two inequivalent  Majorana-Weyl representations.
So it remains to impose the Majorana condition on the complex Weyl representations we have
constructed above. The Majorana condition  can be chosen as
\be
\eta^*=-\Gamma_0 B(\eta)~,
\ee
or equivalently
\be
\eta^*=\Gamma_{6789}\eta~.
\ee
Observe that this reality condition maps forms of even (odd)-degree
to forms of even (odd)-degree and selects real subspaces
$\Delta^+_{16}$ and $\Delta^-_{16}$ in   $\Delta_c^+$ and
$\Delta^-_c$, respectively. These subspaces are the modules of the
two inequivalent  Majorana-Weyl representations of $Spin(9,1)$. For
example $1$ and $e_{1234}$ are complex Weyl spinors while
$1+e_{1234}$ and $i 1-i e_{1234}$ are Majorana-Weyl, i.e.~real
chiral spinors.

The spacetime form bilinears associated with the spinors $\eta,\theta$.
 are given as
\be
\alpha(\eta, \theta)={1\over k!} B(\eta,\Gamma_{A_1\dots A_k} \theta)
e^{A_1}\wedge\dots\wedge e^{A_k}~,~~~~~~~k=0,\dots, 9~.
\la{forms}
\ee
If both spinors are of the same chirality, then
it is sufficient to compute the forms up to degree $k\leq 5$. This is because the forms with degrees  $k\geq 6$
are related to those with degrees $k\leq 5$ with a Hodge duality operation. The forms of middle dimension
are either self-dual or anti-self-dual. If $\eta, \theta\in \Delta^+_{\bf 16}$, then the non-vanishing forms
are one-forms, three-forms and five-forms. In particular, one finds that $\alpha(\eta, \theta)=\alpha(\theta, \eta)$
for one- and five-forms, and $\alpha(\eta, \theta)=-\alpha(\theta, \eta)$ for three-forms.

In many computations that follow, it is convenient  to use another
 basis in the space of spinors $\Delta_c$. This is an oscillator
basis given in terms of creation and annihilation
 operators. For this, first write
\be
\Gamma_{\bar\a}= {1\over \sqrt {2}}(\Gamma_\a+i
\Gamma_{\a+5})~,~~~~~~~~~ \Gamma_\pm={1\over \sqrt{2}}
(\Gamma_5\pm\Gamma_0) ~,~~~~~~~~~\Gamma_{\a}= {1\over \sqrt
{2}}(\Gamma_\a-i \Gamma_{\a+5})~.
\la{hbasisg}
\ee
 Observe that the
Clifford algebra relations in the above basis are
$\Gamma_A\Gamma_B+\Gamma_B\Gamma_A=2 g_{AB}$,   where the
non-vanishing components of the metric are
$g_{\a\bar\b}=\delta_{\a\bar\b}, g_{+-}=1$. In addition we define
$\Gamma^B=g^{BA} \Gamma_A$. The $1$ spinor is a Clifford  vacuum,
$\Gamma_{\bar\a}1=\Gamma_+ 1=0$ and  the representation $\Delta_c$
can be constructed by acting on $1$ with the creation operators
$\Gamma^{\bar\a}, \Gamma^+$ or equivalently any spinor can be
written as
\be
\eta= \sum_{k=0}^5 {1\over k!}~ \phi_{\bar
a_1\dots \bar a_k}~
 \Gamma^{\bar\a_1\dots\bar\a_k} 1~,~~~~\bar a=\bar\a, +~.
 \la{hbasis}
\ee
This is another manifestation of the relation between spinors and forms.

We conclude this section with our form conventions. A $k$-form
$\alpha$ is denoted as $\alpha={1\over k!} \alpha_{i_1\dots i_k}
e^{i_1}\wedge\dots\wedge e^{i_k}$ and so the components of the
exterior derivative $d\alpha$ of $\alpha$ are $d\alpha_{i_1\dots
i_k}= (k+1) \partial_{[i_1} \alpha_{i_2\dots i_{k+1}]}$. The inner
product of two $k$-forms $\alpha$ and $\beta$ is
$(\alpha,\beta)={1\over k!} \alpha_{i_1\dots i_k} \beta_{j_1\dots
j_k} g^{i_1 j_1}\dots g^{i_k j_k}$, where $g$ is the manifold
metric. The Hodge dual $*\alpha$ of a $k$-form $\alpha$ is defined
as $\alpha\wedge \beta=(*\alpha, \beta)\, d{\rm vol}$, where $\beta$
is a $(n-k)$-form and $d{\rm vol}$ is the volume of the
$n$-dimensional manifold. This gives that $*\alpha_{i_{k+1}\dots
i_n}={1\over k!} \alpha_{j_1\dots j_k} \epsilon^{j_1\dots
j_k}{}_{i_{k+1}\dots i_n}$. The inner derivation $i_I\a$ of a
$k$-form $\a$ with an endomorphism $I$ is
 $i_I\a= {1\over (k-1)!} I^j{}_{i_1} \a_{j i_2\dots
i_k}\, e^{i_1}\wedge\dots\wedge e^{i_k}$.

\subsection{Spacetime forms from spinors}

\subsubsection{The $Spin(7)\ltimes \bR^8$- and $SU(4)\ltimes \bR^8$-invariant spinors}

To compute the spacetime forms that are associated with the
$Spin(7)\ltimes \bR^8$- and $SU(4)\ltimes \bR^8$-invariant spinors, it is
sufficient to know the spacetime forms associated with the $1$ and $e_{1234}$ spinors. This
is because as we have seen $1$ and $e_{1234}$ span the $Spin(7)\ltimes \bR^8$- and
$SU(4)\ltimes \bR^8$-invariant spinors. As a result, the spacetime forms associated
with the $Spin(7)\ltimes \bR^8$- and $SU(4)\ltimes \bR^8$-invariant spinors
are linear combinations of the forms associated with the $1$ and $e_{1234}$
spinors.
Using (\ref{forms}), it is easy to find that the forms associated with the
$1$ and $e_{1234}$ spinors are the following: A one-form
\bea
\kappa(e_{1234},1)= e^0-e^5~,
\eea
a three-form
\bea
\xi(e_{1234}, 1)=-i (e^0-e^5)\wedge \omega~,
\eea
and  five-forms
\bea
&&\tau(1,1)=(e^0-e^5)\wedge \chi~,~~~
\tau(e_{1234}, e_{1234})= (e^0-e^5)\wedge \chi^*~,~~~
\cr
&&\tau(e_{1234}, 1)=- {1\over2} (e^0-e^5)\wedge\omega\wedge \omega~,
\eea
where
\bea
&&\omega= -(e^1\wedge e^6+ e^2\wedge e^7+e^3\wedge e^8+e^4\wedge e^9)~,
\cr
&&\chi=(e^1+i e^6)\wedge
(e^2+i e^7)\wedge (e^3+i e^8)\wedge (e^4+i e^9)~.
\eea

\subsubsection{The $G_2$-invariant spinors}\la{G_2STF}

The $G_2$ invariant spinors are linear combinations of $1$, $e_{1234}$, $e_{15}$
and $e_{2345}$ spinors. The spacetime form bilinears associated with
$1$ and $e_{1234}$ have been given in the previous section. Here we shall compute
the spacetime forms associated with the rest of the spinor bilinears.
In particular, we have the one-forms
\bea
&&\kappa(1, e_{2345})=-e^1-i e^6~,~~~
\kappa(e_{1234}, e_{15})=-e^1+ie^6~,~~~
\cr
&&\kappa(e_{2345}, e_{15})= e^0+e^5~,
\eea
the three-forms
\bea
\xi(1, e_{15})&=& \hat\chi~,
\cr
\xi(1, e_{2345})&=&-i (e^1+ie^6)\wedge \hat\omega+(e^1+ie^6)\wedge e^0\wedge e^5~,
\cr
\xi(e_{1234}, e_{15})&=&i (e^1-ie^6)\wedge \hat\omega+(e^1-ie^6)\wedge e^0\wedge e^5~,
\cr
\xi(e_{1234}, e_{2345})&=&\hat\chi^*~,
\cr
\xi(e_{2345}, e_{15})&=&-i (e^0+ e^5)\wedge \hat\omega- i (e^0+ e^5)\wedge e^1\wedge e^6~,
\eea
where
\bea
\hat\omega&=&-(e^2\wedge e^7+e^3\wedge e^8+e^4\wedge e^9)~,
\cr
\hat\chi&=&(e^2+ie^7)\wedge(e^3+ie^8)\wedge(e^4+ie^9)~,
\eea
and the five-forms
\bea
\tau(1, e_{15})&=&[-e^0\wedge e^5- i e^1\wedge e^6]\wedge\hat\chi~,
\cr
\tau(1, e_{2345})&=& (e^1+i e^6)\wedge [{1\over2} \hat\omega\wedge \hat\omega
+i \hat\omega\wedge e^0\wedge e^5]~,
\cr
\tau(e_{1234}, e_{15})&=& (e^1-ie^6)\wedge [{1\over2}\hat\omega\wedge \hat\omega
-i \hat\omega\wedge e^0\wedge e^5]~,
\cr
\tau(e_{1234}, e_{2345})&=&[- e^0\wedge e^5+ i e^1\wedge e^6]\wedge\hat\chi^*~,
\cr
\tau(e_{2345}, e_{15})&=& (e^0+e^5)\wedge [-{1\over2}\hat\omega\wedge \hat\omega
-\hat\omega\wedge e^1\wedge e^6]~,
\cr
\tau(e_{15},e_{15})&=&-(e^0+e^5)\wedge(e^1-ie^6)\wedge\hat\chi~,
\cr
\tau(e_{2345},e_{2345})&=&-(e^0+e^5)\wedge(e^1+ie^6)\wedge\hat\chi^*~.
\eea

\subsubsection{The $N=3$ case}\la{N=3STF}

The $Sp(2)$ invariant spinors are linear combinations of $1$, $e_{1234}$, $e_{12}$ and $e_{34}$. The spinor
bilinears of  $1$ and $e_{1234}$ have been computed already. Here, we shall give  remaining  forms.
{}For this, let us set
\bea
 \omega_1=-(e^1\wedge e^6+e^2\wedge e^7) ,&&\quad \omega_2=-(e^3\wedge e^8+e^4\wedge e^9)~,
  \cr
 \chi_1=\left(e^1+ie^6\right)\wedge\left(e^2+ie^7\right), &&\quad \chi_2=\left(e^3+ie^8\right)\wedge\left(e^4+ie^9\right)~.
\eea
Note that
\be
 \omega=\omega_1+\omega_2, \quad \chi=\chi_1\wedge\chi_2~.
\ee
Then we find, the one-forms
\be
 \k(e_{12},e_{34})=\k(e_{34},e_{12})=-(e^0-e^5)~,
\ee
the three-forms,
\bea
 \xi(e_{12},1)&=&-(e^0-e^5)\wedge\chi_2~,
 \cr
 \xi(e_{34},1)&=&-(e^0-e^5)\wedge\chi_1~,
 \cr
 \xi(e_{1234},e_{12})&=&-(e^0-e^5)\wedge\chi_1^*~,
 \cr
 \xi(e_{1234},e_{34})&=&-(e^0-e^5)\wedge\chi_2^*~,
 \cr
 \xi(e_{34},e_{12})&=&-i(e^0-e^5)\wedge(\omega_1-\omega_2)~,
\eea
\\
and five-forms
\bea
 \t(e_{12},1)&=&i(e^0-e^5)\wedge\omega_1\wedge\chi_2~,
  \cr
 \t(e_{34},1)&=&i(e^0-e^5)\wedge\omega_2\wedge\chi_1~,
  \cr
 \t(e_{12},e_{1234})&=&i(e^0-e^5)\wedge\omega_2\wedge\chi_1^*~,
  \cr
 \t(e_{34},e_{1234})&=&i(e^0-e^5)\wedge\omega_1\wedge\chi_2^*~,
  \cr
 \t(e_{12},e_{12})&=&(e^0-e^5)\wedge\chi_1^*\wedge\chi_2~,
  \cr
 \t(e_{34},e_{34})&=&(e^0-e^5)\wedge\chi_1\wedge\chi_2^*~,
 \cr
 \t(e_{12},e_{34})&=&\ha(e^0-e^5)\wedge(\omega_1-\omega_2)\wedge(\omega_1-\omega_2)~.
\eea

\subsubsection{$N=4$, $SU(3)$}
The relevant spinors are linear combinations of $1$, $e_{1234}$, $e_{15}$ and $e_{2345}$. The associated
spacetime forms are given in section (\ref{G_2STF}).

\subsubsection{$N=4$, $(SU(2)\times SU(2))\ltimes \bR^8$}
We need to consider linear combinations of $1$, $e_{1234}$, $e_{12}$ and $e_{34}$. The spacetime forms are
given in section (\ref{N=3STF}).

\subsubsection{$N=8$, $SU(2)$} \la{appexsut}
The $SU(2)$-invariant Majorana-Weyl spinors are linear combinations of
$1$, $e_{1234}$, $e_{12}$, $e_{34}$, $e_{15}$, $e_{25}$ $e_{2345}$
and $e_{1345}$. There are five types of spinors that occur in $\Delta^+_{\bf 16}$. These are
\bea
1~,~~~e_{1234}~,~~~e_{ij}~,~~~e_{i5}~,~~~e_{ijk5}~,~~~i,j,k=1,2,3,4~.
\la{fivetypes}
\eea
We have already computed form bilinears of examples of all types.
Because of this, one can compute the form spinor bilinears of the remaining pairs by appropriately
relabelling the indices of the forms of the pairs we have already computed.
We have done the computation but the result is not enlightening. Because of this
we shall not explicitly list all the forms.

\appendix{A linear system}

In order to systematically solve the Killing spinor equations, we
determine the action of the supercovariant derivative on the five
types of  spinors (\ref{fivetypes}) and expand the results in the
basis (\ref{hbasis}). This is similar to the calculation of
IIB-supergravity and M-theory in \cite{grangpb}. We use the
following conventions for the indices: $A,B,C \in
\{+,-,\bar{1},..,\bar{4},1,..,4\}$, $\alpha,\beta\in \{1,..,4\}$ and
$k,l,m,n\in\{1,..,4\}$ with the restriction $k,l,m,n\neq
\alpha,\beta$. The Greek indices are not subject to the sum
convention. In particular, we find
\bea
 \frac{1}{4}\hat{\Om}_{A,BC}\Gamma^{BC}1=&&\ha\left(\hat{\Om}_{A,k}{}^k +\hat{\Om}_{A,-+}\right)1
 +{1\over 4}\hat{\Om}_{A,\bar{k}\bar{l}}\Gamma^{\bar{k}\bar{l}}1+\ha\hat{\Om}_{A,+\bar{k}}\Gamma^{+\bar{k}}1~,
\eea

\bea
 \frac{1}{4}\hat{\Om}_{A,BC}\Gamma^{BC}e_{1234}=&-&{1\over 8}\hat{\Om}_{A,kl}\epsilon^{kl}{}_{\bar{m}\bar{n}}\Gamma^{\bar{m}\bar{n}}1
                                                      +{1\over 24}\hat{\Om}_{A,+k}\epsilon^k{}_{\bar{l}\bar{m}\bar{n}}\Gamma^+\Gamma^{\bar{l}\bar{m}\bar{n}}1
                                         \cr
                                         &+&\ha\left(\hat{\Om}_{A,-+}-\hat{\Om}_{A,k}{}^{k}\right)e_{1234}~,
\eea

\bea
 {1\over 4}\hat{\Om}_{A,BC}\Gamma^{BC} e_{\alpha\beta}=&-&\hat{\Om}_{A,\alpha\beta}1
 +\ha\hat{\Om}_{A,\bar{k}\bar{l}}\epsilon^{\bar{k}\bar{l}\bar{\alpha}\bar{\beta}}e_{1234}
                    \cr
                    &+&{1\over 4}\left(\hat{\Om}_{A,\bar{\alpha}\alpha}
                    +\hat{\Om}_{A,\bar{\beta}\beta}+\hat{\Om}_{A,k}{}^k+\hat{\Om}_{A,-+}\right)\Gamma^{\bar{\alpha}}\Gamma^{\bar{\beta}}1
                    \cr
                    &+&\ha\hat{\Om}_{A,\bar{k}\alpha}\Gamma^{\bar{k}}\Gamma^{\bar{\beta}} 1
                    -\ha\hat{\Om}_{A,\bar{k}\beta}\Gamma^{\bar{k}}\Gamma^{\bar{\alpha}} 1
                    \cr
                    &+&\ha\hat{\Om}_{A,+\alpha}\Gamma^+\Gamma^{\bar{\beta}}1-\ha\hat{\Om}_{A,+\beta}\Gamma^+\Gamma^{\bar{\alpha}}1
                    \cr
                    &+&{1\over 4}\hat{\Om}_{A,+\bar{k}}\Gamma^+\Gamma^{\bar{k}}\Gamma^{\bar{\alpha}\bar{\beta}}1~,
\eea
\bea
 \frac{1}{4}\hat{\Om}_{A,BC}\Gamma^{BC}\Gamma^+\Gamma^{\bar{\alpha}}1=&-&2\hat{\Om}_{A,-\alpha}1
                       +\ha\left(\hat{\Om}_{A,k}{}^k+\hat{\Om}_{A,\bar{\alpha}\alpha}
                       -\hat{\Om}_{A,-+}\right)\Gamma^+\Gamma^{\bar{\alpha}}1
                       \cr
                       &-&\hat{\Om}_{A,-\bar{k}}\Gamma^{\bar{k}}\Gamma^{\bar{\alpha}}1
                       -\hat{\Om}_{A,\alpha\bar{k}}\Gamma^+\Gamma^{\bar{k}}1
                       +{1\over 4}\hat{\Om}_{A,\bar{k}\bar{l}}\Gamma^+\Gamma^{\bar{\alpha}}\Gamma^{\bar{k}\bar{l}}1~,
\eea
\bea
 \frac{1}{4}\hat{\Om}_{A,BC}\Gamma^{BC}\Gamma^+\Gamma^{\alpha}e_{1234}=&-&2\hat{\Om}_{A,-\bar{\alpha}}e_{1234}
                    -{1\over 4}\hat{\Om}_{A,\bar{\alpha}k}\epsilon^k{}_{\bar{\alpha}\bar{l}\bar{m}}\Gamma^+\Gamma^{\bar{\alpha}\bar{l}\bar{m}}1
                    \cr
                    &+&{1\over 24}\left(-\hat{\Om}_{A,k}{}^k+\hat{\Om}_{A,\alpha\bar{\alpha}}
                    -\hat{\Om}_{A,-+}\right)\epsilon^\alpha{}_{\bar{k}\bar{l}\bar{m}}\Gamma^+\Gamma^{\bar{k}\bar{l}\bar{m}}1
                    \cr
                    &-&\ha\hat{\Om}_{A,kl}\epsilon^{\alpha kl}{}_{\bar{m}}\Gamma^+\Gamma^{\bar{m}}1
                     -\ha\hat{\Om}_{A,-k}\epsilon^{\alpha k}{}_{\bar{l}\bar{m}}\Gamma^{\bar{l}\bar{m}}1~.
\eea

 A similar analysis for the dilatino equation yields

\bea
  \left(\del_A\Phi\Gamma^A-{1\over 12} H_{ABC}\Gamma^{ABC}\right)1&=&\left(\del_{\bar{k}}\Phi+\ha H_{\bar{k}}{}^{l}{}_{l}+\ha H_{+-\bar{k}}\right)\Gamma^{\bar{k}}1
            \cr
            &+&\left(\del_+\Phi-\ha H_{+k}{}^k\right)\Gamma^+ 1-{1\over 4}H_{+\bar{k}\bar{l}}\Gamma^+\Gamma^{\bar{k}\bar{l}}1
            \cr
            &-&{1\over 12}H_{\bar{k}\bar{l}\bar{m}}\Gamma^{\bar{k}\bar{l}\bar{m}}1~,
\eea

\bea
  \left(\del_A\Phi\Gamma^A-{1\over 12} H_{ABC}\Gamma^{ABC}\right)e_{1234}=&&{1\over 6}H_{klm}\epsilon^{klm}{}_{\bar{n}}\Gamma^{\bar{n}}1
            \cr
            &+&\left(\del_+\Phi+\ha H_{+k}{}^{k}\right)\Gamma^+ e_{1234}
            \cr
            &+&{1\over 12}\left(\del_k\Phi+\ha H_{kl}{}^{l}+\ha H_{+-k}\right)\epsilon^{k}{}_{\bar{l}\bar{m}\bar{n}}\Gamma^{\bar{l}\bar{m}\bar{n}}1
            \cr
            &+&{1\over 8}H_{+kl}\epsilon^{kl}{}_{\bar{m}\bar{n}}\Gamma^+\Gamma^{\bar{m}\bar{n}}1~,
\eea

\bea
 \left(\del_A\Phi\Gamma^A-\frac{1}{12} H_{ABC}\Gamma^{ABC}\right)e_{\alpha\beta}&=&-
 \ha H_{+\bar{k}\bar{l}}\epsilon^{\bar{k}\bar{l}\bar{\alpha}\bar{\beta}}\Gamma^+ e_{1234}
                     +H_{+\alpha\beta}\Gamma^+1  +H_{\bar{k}\alpha\beta}\Gamma^{\bar{k}}1
                  \cr
                  &+&\left(-\del_\beta \Phi+\ha H_{\bar{\alpha}\alpha\beta}+\ha H_{\beta l}{}^l
                  -\ha H_{+-\beta}\right)\Gamma^{\bar{\alpha}}1
                  \cr
                  &+&\left(\del_\alpha \Phi-\ha H_{\bar{\beta}\beta\alpha}-\ha H_{\alpha l}{}^l
                  +\ha H_{+-\alpha}\right)\Gamma^{\bar{\beta}}1
                  \cr
                  &+&\ha\left(\del_+\Phi-\ha H_{+k}{}^k-\ha H_{+\bar{\alpha}\alpha}-
                  \ha H_{+\bar{\beta}\beta}\right)\Gamma^+\Gamma^{\bar{\alpha}}\Gamma^{\bar{\beta}}1
                  \cr
                  &+&\ha\left(\del_{\bar{k}}\Phi+\ha H_{\alpha\bar{\alpha}\bar{k}}+
                  \ha H_{\beta\bar{\beta}\bar{k}}-\ha H_{\bar{k}l}{}^l+\ha H_{+-\bar{k}}\right)\Gamma^{\bar{k}}\Gamma^{\bar{\alpha}}\Gamma^{\bar{\beta}}1
                  \cr
                  &-&{1\over 4}H_{\alpha\bar{k}\bar{l}}\Gamma^{\bar{k}\bar{l}}\Gamma^{\bar{\beta}}1
                  +{1\over 4}H_{\beta\bar{k}\bar{l}}\Gamma^{\bar{k}\bar{l}}\Gamma^{\bar{\alpha}}1
                  \cr
                  &+&\ha H_{+\alpha\bar{k}}\Gamma^+\Gamma^{\bar{k}}\Gamma^{\bar{\beta}}1
                  -\ha H_{+\beta\bar{k}}\Gamma^+\Gamma^{\bar{k}}\Gamma^{\bar{\alpha}}1~,
\eea

\bea
 \left(\del_A\Phi\Gamma^A-\frac{1}{12} H_{ABC}\Gamma^{ABC}\right)\Gamma^+\Gamma^{\bar{\alpha}}1&=&{1\over3}H_{\bar{k}\bar{l}\bar{m}}\epsilon^{\bar{k}\bar{l}\bar{m}\bar{\alpha}}\Gamma^+ e_{1234}
                  +2H_{-\alpha\bar{k}}\Gamma^{\bar{k}}1
                  \cr
                  &+&\left(2\del_-\Phi-H_{-\bar{\alpha}\alpha}-H_{-k}{}^k\right)\Gamma^{\bar{\alpha}}1
                  \cr
                  &+&\left(-2\del_{\alpha}\Phi+H_{\alpha k}{}^k+H_{+-\alpha}\right)\Gamma^+1
                  \cr
                  &+&\left(\del_{\bar{k}}\Phi+\ha H_{\alpha\bar{\alpha}\bar{k}}-\ha H_{+-\bar{k}}
                  +\ha H_{l\bar{k}}{}^l\right)\Gamma^+\Gamma^{\bar{\alpha}}\Gamma^{\bar{k}}1
                  \cr
                  &+&\ha H_{\alpha\bar{k}\bar{l}}\Gamma^+\Gamma^{\bar{k}\bar{l}}1-
                  \ha H_{-\bar{k}\bar{l}}\Gamma^{\bar{\alpha}}\Gamma^{\bar{k}\bar{l}}1~,
\eea

\bea
 \left(\del_A\Phi\Gamma^A-\frac{1}{12} H_{ABC}\Gamma^{ABC}\right)\Gamma^+\Gamma^{\alpha}e_{1234}&=&{1\over 6}\left(\del_-\Phi+\ha H_{-\bar{\alpha}\alpha}+\ha H_{-k}{}^k \right)\epsilon^{\alpha}{}_{\bar{k}\bar{l}\bar{m}}\Gamma^{\bar{k}\bar{l}\bar{m}}1
                  \cr
                  &-&2\left(\del_{\bar{\alpha}}\Phi-\ha H_{+-\bar{\alpha}}+\ha H_{\bar{\alpha}k}{}^k\right)\Gamma^+ e_{1234}
                  \cr
                  &-&\ha\left(\del_k\Phi+\ha H_{\bar{\alpha}\alpha k}-\ha H_{+-k}+\ha H_{kl}{}^l\right)\epsilon^{\alpha k}{}_{\bar{l}\bar{m}}\Gamma^+\Gamma^{\bar{l}\bar{m}}1
                  \cr
                  &-&\ha H_{\bar{\alpha}kl}\epsilon^{kl}{}_{\bar{\alpha}\bar{m}}\Gamma^+\Gamma^{\bar{\alpha}}\Gamma^{\bar{m}}1
                      +H_{-kl}\epsilon^{\alpha kl}{}_{\bar{m}}\Gamma^{\bar{m}}1
                  \cr
                  &+&\ha H_{-\bar{\alpha}k}\epsilon^k{}_{\bar{\alpha}\bar{l}\bar{m}}\Gamma^{\bar{\alpha}}\Gamma^{\bar{l}\bar{m}}1
                      +{1\over 3}H_{klm}\epsilon^{klm\alpha}\Gamma^+1~.
\eea
The above expressions can be used to construct the linear systems for those backgrounds for which not all parallel
spinors are Killing. We have also used the above equations to determine the conditions on the geometry
of spacetime for the supersymmetric backgrounds we have examined.

\end{document}